\documentclass[sigconf]{acmart}
\AtBeginDocument{%
  }

\usepackage{tabularray}
\usepackage{makecell}

\usepackage{listings}
\usepackage{xcolor}

\setcopyright{acmlicensed} 
\copyrightyear{2025} 
\acmYear{2025} 
\acmConference[KDD '25]{the 31st ACM SIGKDD Conference on Knowledge Discovery and Data Mining}{August 03--07, 2025}{Toronto, Canada} 

\begin{document}

\title[Neurophysiologically Realistic aDBS in Parkinson’s Disease]{Neurophysiologically Realistic Environment for Comparing Adaptive Deep Brain Stimulation Algorithms \\in Parkinson’s Disease}

\author{Ekaterina Kuzmina}
\orcid{0000-0001-5870-6004}
\affiliation{
  \institution{Skoltech}
  \city{Moscow}
  \country{Russian Federation}
}
\affiliation{
  \institution{AIRI}
  \city{Moscow}
  \country{Russian Federation}
}
\email{ekaterina.kuzmina@skoltech.ru}

\author{Dmitrii Kriukov}
\orcid{0000-0001-7629-2466}
\affiliation{%
  \institution{Skoltech}
  \city{Moscow}
  \country{Russian Federation}
}
\affiliation{%
  \institution{AIRI}
  \city{Moscow}
  \country{Russian Federation}
}
\email{dmitrii.kriukov@skoltech.ru}

\author{Mikhail Lebedev}
\affiliation{
  \institution{Lomonosov Moscow State University}
  \city{Moscow}
  \country{Russian Federation} 
}
\email{mikhail.lebedev@math.msu.ru}

\author{Dmitry V. Dylov}
\orcid{0000-0003-2251-3221}
\affiliation{
  \institution{Skoltech}
  \city{Moscow}
  \country{Russian Federation}
}
\affiliation{
  \institution{AIRI}
  \city{Moscow}
  \country{Russian Federation}
}
\email{d.dylov@skoltech.ru}


\renewcommand{\shortauthors}{Kuzmina et al.} 

\begin{abstract}
Adaptive deep brain stimulation (aDBS) has emerged as a promising treatment for Parkinson's disease (PD). In aDBS, a surgically placed electrode sends dynamically altered stimuli to the brain based on neurophysiological feedback: an invasive gadget that limits the amount of data one could collect for optimizing the control offline. As a consequence, a plethora of synthetic models of PD and those of the control algorithms have been proposed. Herein, we introduce the first neurophysiologically realistic benchmark for comparing said models. Specifically, our methodology covers not only conventional basal ganglia circuit dynamics and pathological oscillations, but also captures 15 previously dismissed physiological attributes, such as signal instabilities and noise, neural drift, electrode conductance changes and individual variability – all modeled as spatially distributed and temporally registered features via beta-band activity in the brain and a feedback. Furthermore, we purposely built our framework as a structured environment for training and evaluating deep reinforcement learning (RL) algorithms, opening new possibilities for optimizing aDBS control strategies and inviting the machine learning community to contribute to the emerging field of intelligent neurostimulation interfaces.
\textit{Code repository:} \url{https://github.com/NevVerVer/DBS-Gym}

\end{abstract}





\maketitle

\section{Introduction}

\begin{figure*}
    \centering
    \includegraphics[width=1.0\linewidth]{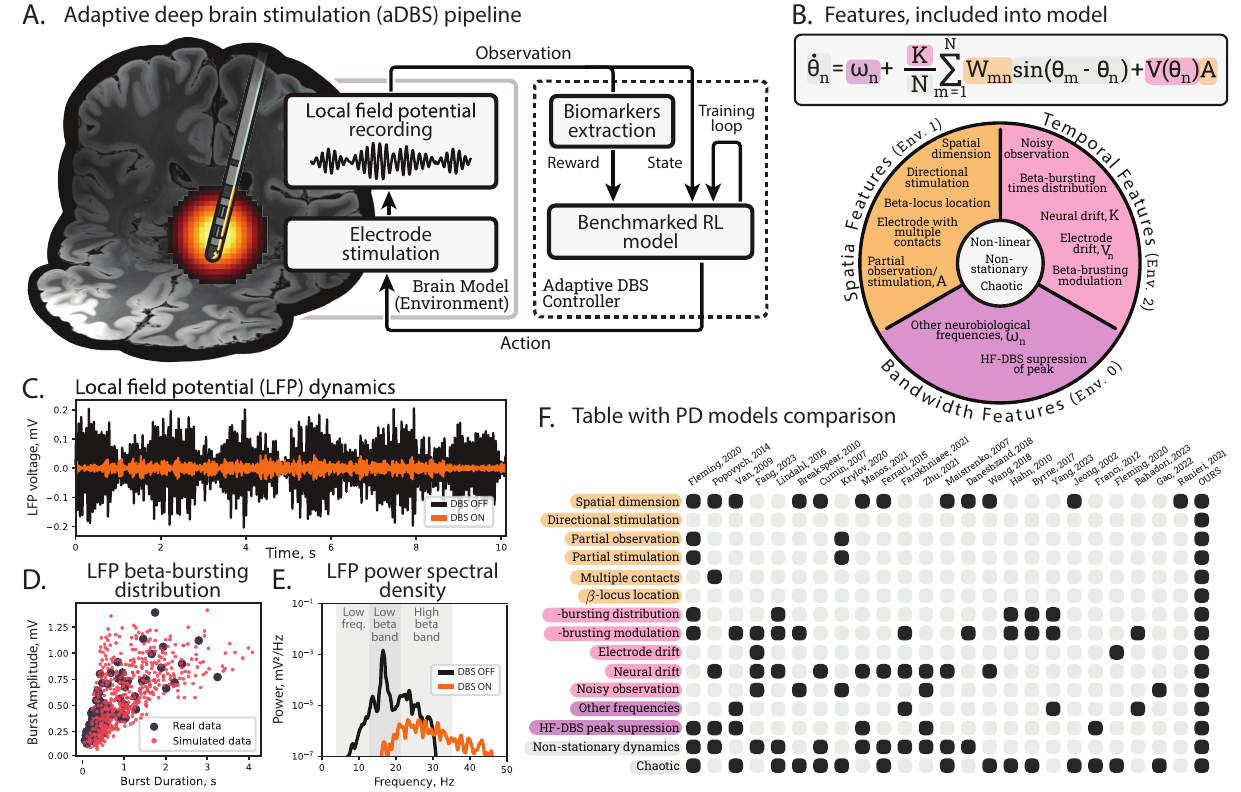}
    \caption{Proposed Environment.
A) A closed-loop electrical brain stimulation system measures neural activity through local field potential (LFP, \textit{`Observation'}) to compute the needed stimuli (\textit{`Action'}) in real time. Then, a chosen algorithm (\textit{e.g.}, RL) should learn to control any undesired signaling in the simulated brain model (\textit{`Environment'}).
B) Three groups of features of the proposed brain model. \textit{Bandwidth features} cover fundamental neurobiological activity as beta oscillations and bursts. \textit{Spatial features} describe spatial relationships between the electrode and surrounding neurons. 
\textit{Temporal features} introduce experimentally relevant noise into learning, simulating neuroplasticity and glial encapsulation and requiring control resilient to the environmental drift. The Kuramoto equations map the corresponding coefficients (color-coded) to the feature groups.
C) An example of LFP dynamics with DBS off and on (a high-frequency HF-DBS with a constant stimulation amplitude is shown).
D) A comparison of beta burst distribution observed in real patients with PD (taken from \cite{tinkhauser2017beta}) and those simulated by the proposed model.
E) The power spectral density of LFP signals from (C). Note the suppression of beta-band power with constant HF-DBS.
F) Modern synthetic Parkinson's disease models and the features they cover (indicated with filled squares).}
    \label{fig:intro}
\end{figure*}

Parkinson's Disease (PD) is a major neurodegenerative disorder. It affects up to 1\% of the population of 60 years of age worldwide \cite{bloem2021parkinson}, leading to severe movement and psychological impairments, including tremors, akinesia, bradykinesia, cognitive decline, depression, and sleep disturbances. 
The commonly-accepted cause of PD is the degeneration of dopamine-producing neurons in the \textit{substantia nigra pars compacta} and, to a lesser extent, the ventral tegmental area, causing abnormal activity in the basal ganglia circuitry \cite{balestrino2020parkinson}.
The standard treatment for PD involves pharmacological interventions, such as L-DOPA; though, their efficacy diminishes as the disease progresses \cite{cao2020dopa}. For patients who no longer respond to medication or suffer from severe side effects, deep brain stimulation (DBS) provides an alternative (Figure \ref{fig:intro}A). 
High-frequency DBS (HF-DBS) delivers electrical pulses, typically between 130 Hz and 185 Hz, via electrodes implanted in the subthalamic nucleus (STN) or globus pallidus internus (GPi). 
DBS is highly effective not only for PD but also for managing essential tremor \cite{houston2019machine, birdno2012stimulus}, major depressive disorder \cite{widge2024closing}, and even sensory-complete spinal cord injury \cite{zhao2021optimization}.
There are different theories of how DBS works. However, there is still no consensus and the underlying mechanisms remain elusive \cite{cagnan2019emerging, neumann2023neurophysiological, rosenbaum2014axonal}.
DBS has shown particular efficacy in alleviating movement-related symptoms of PD. For example, it improves movement velocity and facilitates ballistic movements in PD patients undergoing STN stimulation \cite{rosa2012neurophysiology}. 
Despite its benefits, DBS can induce side effects such as speech impairment, balance problems, mood changes, vision disturbances, and others
\cite{wilkins2023unraveling}.

To enhance DBS's therapeutic outcomes and address its limitations, \textit{adaptive} deep brain stimulation (aDBS) was introduced \cite{little2013adaptive}. aDBS uses a closed-loop system to adjust stimulation parameters automatically based on physiological biomarkers, offering a more tailored intervention than traditional DBS \cite{watts2020machine, oliveira2023machine}. This technique allows the system to respond dynamically to the patient's pathological state, improving efficiency and potentially reducing side effects.
Current aDBS systems use subthalamic nucleus LFPs as feedback, analyzed via power spectral density (PSD) and spectrograms. PSD quantifies frequency power but lacks temporal precision while LFPs capture synchronized neuronal activity \cite{rosa2012neurophysiology}. Excessive beta-band oscillations (12–35 Hz) in PD correlate with symptoms, making their suppression an aDBS’s therapeutic target \cite{oliveira2023machine, wilkins2023unraveling}.  
Early aDBS (2013) used threshold-based control \cite{wilkins2023unraveling, wang2022adaptive}, later adopted clinically \cite{krauss2021technology}. Advanced methods include PI controllers \cite{liu2016closed, su2021model, yang2022beta, zhu2021adaptive, fleming2020self}, fuzzy logic \cite{su2023closed}, phase response curves \cite{holt2016phasic}, and Bayesian optimization \cite{grado2018bayesian}. 

%

Given the rising complexity of DBS systems \cite{rosa2012neurophysiology} and the advances in machine learning (ML), researchers are increasingly exploring ML-based solutions for self-tuning aDBS. These methods range from simple classification algorithms that predict tremor or pre-tremor states \cite{houston2019machine} to more complex neural networks that dynamically adjust stimulation parameters based on LFP data \cite{wang2022adaptive, yu2020review}. Reinforcement learning (RL), in particular, has shown promise in optimizing aDBS control strategies, with several studies successfully demonstrating RL's effectiveness in both simulation \cite{jovanov2018platform, lu2019application, faraji2023adaptive, lu2017desynchronizing, gao2020model, agarwal2023novel, krylov2020reinforcement} and clinical settings \cite{yang2023adaptive}.
As adaptive DBS algorithms advance, they require rigorous validation to ensure autonomous function across diverse patient conditions, necessitating reliable systems that adapt to PD biomarkers and complex basal ganglia dynamics \cite{yu2020review, krauss2021technology}. Last but not least, the invasive nature of the aDBS treatment hinders the collection of datasets of sufficient volume for optimizing the control  strategies offline.

In this work, we tackle the challenge of developing, validating, and comparing aDBS systems in a standardized manner, especially those based on ML and RL. 
Such explorations typically do not require live feedback from a patient and are tested and trained using computational models of neural activity.
The early over-simplified synthetic models (\textit{e.g.}, \cite{so2012relative, schiff2011neural}) could not cover the complex range of all possible neurophysiological responses seen in real patients. The more recent ones, entailing generative approaches, are known to be more computationally intensive \cite{fleming2020self, lu2019application, bahadori2023efficient, kumaravelu2016biophysical}, which has limited their scalability to clinically relevant simulations.

Our contribution is \textit{a comprehensive set of observable pathological features} that could be enabled with simple coefficients in a computationally-efficient set of Kuramoto equations \cite{acebron2005kuramoto}, which will mimic typical LFP patterns in PD patients. We group these as aDBS requirements into spatial, temporal, and bandwidth features (Figure \ref{fig:intro}B, F), allowing for better validation of controllers in realistic, non-linear environments. It fosters the ability of the systems to adapt to signal variations in such real-world scenarios as electrode movement \cite{krauss2021technology} or the plasticity of neurons in response to an interference \cite{manos2021long}.
Lastly, we introduce \textit{a unifying RL environment} encompassing the said features as configurable parameters and \textit{benchmark several modern RL algorithms} as the aDBS controllers.

%



\section{Related work}
\subsection{PD models for aDBS training}

Today, there are many computational models of PD, each based on different hypotheses regarding the dynamics of the disease progression. These models can be broadly categorized into neuron models and mean-field models \cite{yu2020review, pavlides2015computational, carlson2021computational}.

Neuron models, such as extensions of the Hodgkin-Huxley (HH) model \cite{hodgkin1952quantitative}, simulate the spiking behavior of cortical cells by incorporating nonlinear ionic currents to explain low-frequency firing \cite{byrne2017mean, holt2016computational}. The Rubin-Terman model (RTM) is widely used in PD research to explore adaptive DBS (aDBS) dynamics \cite{terman2002activity, so2012relative, bahadori2023efficient, kumaravelu2016biophysical, fleming2020simulation, manos2021long, liu2016closed, popovych2019adaptive, wang2022adaptive}. Mean-field models, on the other hand, are computationally efficient for simulating large-scale cortical networks, allowing the study of statistical neurodynamics without the complexity of biophysically detailed neuron models \cite{van2009mean, yang2022beta, pavlides2015computational, byrne2017mean}.

Initially, these models were developed to interpret experimental findings, investigate the origins of excessive oscillations in PD, and explore DBS effects on the BG circuit \cite{pascual2006computational, romano2020evaluation, cagnan2009frequency}, providing insights into the mechanisms of HF-DBS \cite{wilson2011chaotic}. Although these models have been instrumental in understanding PD, they were primarily designed for conceptual exploration rather than for actual gadget-like training and validation of aDBS. Most PD models aim to: (1) produce oscillations with a pronounced low beta-band peak in the LFP spectrum (Figs. \ref{fig:intro}C, E, black line) and (2) respond realistically to continuous DBS (cDBS), reducing beta power\footnote{The beta-band power usually serves as the main measure of feedback for control \cite{balestrino2020parkinson}.} \cite{grado2018bayesian, lu2019application}. 

For RL-based aDBS, simpler oscillatory mean-field models have been used to train proximal policy algorithms \cite{krylov2020reinforcement2, agarwal2023novel}, Q-learning has been applied to desynchronize Kuramoto model oscillations \cite{lu2017desynchronizing, krylov2020reinforcement, krylov2020reinforcement2}. Variations of the RTM model have also been used for basal ganglia simulations in RL studies \cite{lu2019application, gao2020model, agarwal2023novel, gao2023offline, faraji2023adaptive}.

\subsection{Existing algorithms for aDBS}

Most aDBS algorithms use proportional-integrate-derivative (PID) controllers \cite{daneshzand2018robust, su2023closed, fleming2020self, popovych2014control, schmidt2024home, popovych2019adaptive, liu2016closed}. However, recent research explores neural networks and machine learning for aDBS, with reinforcement learning (RL) showing particular promise due to its adaptability in complex, non-linear environments \cite{agarwal2023novel, wang2022adaptive, swann2018adaptive, liu2020neural, sui2022deep, herron2017cortical, grado2018bayesian, chen2023optimal, houston2019machine}. Online RL methods such as Proximal Policy Optimization (PPO) \cite{gao2020model, faraji2023adaptive, krylov2020reinforcement, krylov2020reinforcement2}, Temporal-difference learning \cite{lu2019application}, and Q-learning \cite{lu2017desynchronizing} have been applied to suppress PD oscillations. Hybrid approaches combining RL and PID, such as PPO to optimize PID parameters, have also been proposed \cite{faraji2022novel}. Model-based RL methods show promise for optimizing aDBS strategies \cite{yang2023adaptive, pan2024coprocessor, gao2023offline}.

\section{Problem setup}  
\subsection{Neural dynamics model description}

Given the importance of computational efficiency for ML-based aDBS systems, we chose the Kuramoto model as the proxy for neural dynamics \cite{acebron2005kuramoto}. This model is computationally efficient and easily extendable with additional features. Moreover, it is widely used in neuroscience to simulate neural activity \cite{nakao2016phase, schmidt2014dynamics, shams2022optimal, hemami2021phase} and to model pathological oscillatory activity \cite{zavaleta2023modification, breakspear2010generative}. 
In general, Kuramoto model captures two basic properties of periodically spiking neurons. First, it can describe regular, bursting, and chaotic oscillatory dynamics via a single set of equations. In the model, the bursting neuron is regarded as a phase oscillator of a given frequency. Second, we can associate its real part with the membrane voltage (a physiologically meaningful output \cite{franci2012desynchronization}). 

Let $\theta \in (\mathbb{S}^1)^N$ be a vector of oscillatory phases of $N$ neurons. A phase of a particular neuron $\theta_n$ is defined via its natural frequency $\omega_n$ term, its interactions with all other neurons, and the effect of DBS electrode, evolving in time as:
\begin{equation}
    \frac{d\theta_n}{dt}=\omega_n+\frac{K}{N} \sum_{m=1}^N W_{mn} \sin \left(\theta_m-\theta_n\right) + V(\theta_n)A, 
\label{eq:kur_alpha}  
\end{equation}
where $K$ refers to a constant coupling parameter; the interaction functions $W_{mn}$ and $V(\theta_n)$ represent neuron-neuron connection and neuron-electrode connection strengths, respectively, and their particular forms define a full connection pattern, allowing for tuning the system to a realistic configuration; $A$ represents the voltage of the electrode stimuli. Herein, we set a network of neurons to be embedded in a cubic grid. The connections are all-to-all, with the interaction function $W_{mn}=\cos(\alpha_{mn})$ depending on a Euclidean distance between a given pair of neurons $\alpha_{mn}$ and the interaction term $\sin(\theta_m-\theta_n)$. The product of both terms reflects the distance and the phase dependent transmission delay between a pair of neurons \cite{breakspear2010generative, ferrari2015phase}. 

The function $V(\theta_n)$, referred to as the neuron-to-electrode interaction function, represents how the phase of a neuron $n$ is influenced by an external deep brain stimulation (DBS) voltage pulse, modulating the effect of the pulse amplitude $A$ on the neuron's phase dynamics. This function is also two-component with one depending on the spatial distance between electrode contact and a neuron and interaction term transforming electrical voltage to a neuron phase:
\begin{equation}
    V(\theta_n) = G(\alpha_{n, el})\cdot PRC(\theta_n). 
\label{eq:kur_prc}
\end{equation}
The spatial component term aims to take into account properties of the extracellular medium and the orientation of target cells which ultimately defines the electrical conductivity between the electrode and a neuron \cite{feng2007optimal}. For the spatial interaction function between neuron $n$ and electrode position $el$ a simple triangular kernel was chosen, i.e. $G=1-\alpha_{n,el}$, if $\alpha_{n,el} \leq 1$ and $G=0$ otherwise. The interaction term which is also called phase response curve (PRC), $PRC(\theta_n)$ \cite{holt2016phasic}, determines the amount that an oscillator $n$ will adjust its phase as a function of its phase value $\theta_n$ upon which the pulse with amplitude $A$ is received \cite{oliveira2023machine, smeal2010phase, danzl2007event, azodi2015phase}. 

Different shapes of stimuli can be used for DBS. In the proposed environment, we implement the simplest square wave pulse and charge-balanced pulse (symmetric biphasic pulses). The latter in real settings allows avoiding tissue damage and produce greater suppression of PD motor symptoms. Increasing evidence suggests that the temporal pattern of stimulation can influence the clinical outcome after DBS, particularly in PD \cite{krauss2021technology}. More details about implementation can be seen in Appendix \ref{app: model_implement_details}.

\subsection{Features description and justification} \label{sec:features}

To accurately model Parkinson's Disease (PD) for adaptive DBS (aDBS) development, key features must be incorporated across three domains: spatial, temporal, and bandwidth (Figure \ref{fig:intro}B, F). Each domain uniquely influences aDBS behavior and suppression efficiency. Detailed justification for feature selection is provided in Appendix \ref{app: features_justif}.

\subsubsection{Bandwidth features.} \label{bf}
Bandwidth features capture neurophysiological biomarkers, particularly beta band activity, which is linked to bradykinesia and rigidity in PD patients \cite{binns2024shared, grado2018bayesian}. However, relying solely on beta oscillations oversimplifies PD dynamics and limits aDBS development \cite{yu2020review}. Recent studies highlight the need to differentiate between low (13–20 Hz) and high (21–35 Hz) beta bands, as they correspond to distinct symptoms and suppression outcomes \cite{yin2021local}.
Our model adjusts power spectral density by controlling the distribution of natural frequencies, $\omega_n \sim P_\omega$. We include low-beta, low-frequency (4–12 Hz), and high-beta bands to create richer, more realistic LFP dynamics \cite{yin2021local}. Frequencies are sampled from a manually designed distribution (Figure \ref{fig:w0_prc}A), enabling flexible control over oscillation strengths for each frequency band. The model also simulates beta peak suppression with high-frequency DBS, consistent with experimental findings \cite{liu2016closed, grado2018bayesian, cagnan2009frequency}. This flexibility makes the model suitable for training and validating ML-based aDBS systems, where diverse frequency features are crucial for robust performance \cite{krauss2021technology}.

\subsubsection{Spatial features.}\label{sf}

Spatial features are essential for accurately modeling neuron interactions and electrode dynamics in aDBS. They introduce non-linearities, time delays, and distance-dependent interactions, impacting aDBS performance. Traditional models often oversimplify these dynamics, limiting aDBS effectiveness. Incorporating spatial features allows for more realistic simulations, essential for advanced neuromodulation techniques like Coordinated Reset \cite{guo2011multi, popovych2014control}. Recent models, including Rubin-Terman and Kuramoto oscillators, have integrated spatial domains to better capture these complexities \cite{ranieri2021data, fleming2020simulation, spiliotis2022deep}. Our model enhances spatial realism by embedding neurons in a three-dimensional grid, with connections defined by $\cos(\alpha_{mn})$, representing excitatory or inhibitory interactions modulated by the coupling parameter $K$. This neurobiologically plausible spatial coupling improves the accuracy of neuron interaction simulations  \cite{ferrari2015phase, breakspear2010generative}.

\begin{figure}[b]
    \centering
    \includegraphics[width=1.0\linewidth]{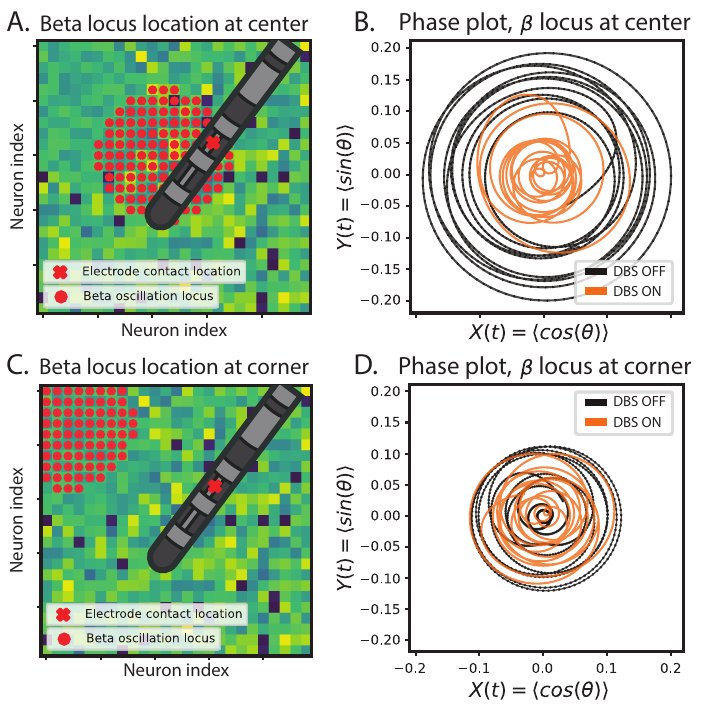}
    \caption{Spatial and phase dynamics of LFP for different placement of the multi-contact electrode relative to beta locus (environment parameters in Table \ref{tab:configs}). A) Beta locus at the center of the signaling neural grid. B) The corresponding phase portrait, shown for no DBS (black) and HF-DBS stimulation (orange). C) The beta locus at the corner. D) The corner-based phase portrait shows decreased synchronization of neurons (note smaller cycle amplitude).}
    \label{fig:phase_main}
\end{figure}

\textit{Electrode location and beta oscillation locus.}
The beta locus refers to the spatial and functional origin of beta-band oscillations in the brain. Electrode placement relative to beta oscillation loci significantly impacts aDBS efficacy. Beta band power is typically highest in the sensorimotor STN, and accurate targeting of these zones enhances therapeutic outcomes \cite{lofredi2019beta, chen2022subthalamic}. Our model replicates this by setting natural frequencies $\omega_{0}$ in specific grid areas to generate localized beta oscillations, simulating spatially localized beta peaks source (Figure \ref{fig:phase_main} and \ref{fig:phase_plot}). This allows testing strategies for optimal electrode placement and stimulation targeting, crucial for adapting to spatially intricate neural activity or electrode shifts \cite{chapelle2021early}.

\textit{Partially observable/stimulated environment.}
Real-world aDBS systems interact locally with neurons, with stimulation efficacy diminishing with distance from the electrode. This spatial constraint creates a partially observable environment, complicating aDBS policy development. Studies show that DBS affects only 20\% to 60\% of STN neurons, emphasizing the need for localized modeling \cite{carlson2021computational, rajamani2024deep}. Our model addresses this by positioning electrodes in a 3D grid and applying a spherical kernel, simulating distance-dependent stimulation and recording. This setup accurately reflects localized neuron interactions, supporting the development of complex aDBS strategies that account for spatial constraints. Due to high nonlinearity, the recorded LFP dynamics change non-uniformly with distance (Figure \ref{fig:electrode_location_appendix}).
\textit{Electrode directed stimulation and multiple contacts.}
Segmented electrodes with directional stimulation allow precise targeting, reducing side effects and enhancing therapeutic effectiveness \cite{yin2021local, krauss2021technology}. These electrodes provide three radial stimulation directions separated by 120 degrees, improving electric field control. This capability supports customized stimulation strategies, critical for symptom-specific aDBS applications \cite{butson2011probabilistic}. Advanced electrodes with multiple contacts enable precise targeting and directional stimulation, improving symptom control and reducing side effects \cite{krauss2021technology}. Our model implements directional stimulation by restricting the stimulation kernel to a $120^{\circ}$ segment using angular constraints in spherical coordinates. Multiple contacts are modeled by setting additional electrode coordinates, automatically adjusting interaction functions and stimulation kernels. This flexible approach facilitates testing of advanced multi-contact aDBS strategies.

\subsubsection{Temporal features}\label{tf}
The temporal dynamics of Parkinson’s disease (PD) are complex and non-stationary, posing challenges for aDBS models and reinforcement learning (RL) algorithms \cite{krylov2020reinforcement, so2012relative, lu2019application}. In real patients, PD-related neural activity fluctuates due to multiple factors and undergoes long-term changes that impact aDBS performance and stability \cite{dovzhenok2013failure, shukla2014modeling, adair2018evolving}. This non-stationarity necessitates adaptive strategies in RL-based aDBS systems \cite{padakandla2020reinforcement, zhangtest}.


\textit{Electrode Drift, Neural Drift, and Noisy Observations}
Temporal signal changes, or drift, affect recorded LFP and neuronal responses. These changes occur across different time scales, from transient artifacts due to head movements to long-term shifts in electrode impedance due to biocompatibility issues \cite{groothuis2014physiological}.

Neural structural plasticity is a major source of drift, influencing synaptic connections and neuronal firing patterns \cite{rubin2017computational, rosa2012neurophysiology}. Our model simulates this by varying the natural oscillator frequencies $\omega_n$ using spike-timing-dependent plasticity (STDP), reflecting changes in neuronal firing rates over time \cite{maistrenko2007multistability, cumin2007generalising}. Electrode drift can result from head movements, stimulation ramping, or implantation inaccuracies, leading to electrode migration within the brain \cite{wilkins2023unraveling, chapelle2021early}. In our model, electrode drift is simulated by periodically changing the coordinates of recording and stimulating contacts coordinate $el$. Chronic electrode encapsulation, caused by foreign body responses like inflammation and gliosis, alters impedance and reduces signal stability \cite{krauss2021technology, little2012brain}. We model this by gradually decreasing electrode conductance, which reduces stimulation effectiveness by shrinking the stimulation kernel $G_{\alpha_{n,el}}$.

\textit{Pathological Beta Bursting and Movement Modulation}
Beta burst amplitude and duration correlate with neural synchronization and symptom severity in PD \cite{tinkhauser2017modulatory, yin2021local, tinkhauser2017beta, yu2021parkinsonism, vinding2020reduction}. Unlike other models that explicitly simulate beta bursts \cite{fleming2020simulation, zhu2021adaptive}, our Kuramoto model naturally exhibits beta bursting through neural synchronization, producing bursts of 400–1500 ms. This distribution is shown in Figure \ref{fig:intro}C, D. Adjusting the coupling parameter $K$ controls burst dynamics, allowing precise tuning of LFP patterns.

Basal ganglia activity is modulated by patient behavior, including sleep, emotional states, motor intentions, and medication intake, particularly during movement \cite{fang2023robust, binns2024shared, hahn2010modeling, gilron2021sleep, krauss2021technology, tinkhauser2017beta, tinkhauser2017modulatory, cassidy2002movement}. Our model simulates movement modulation by randomly increasing the coupling parameter $K$, emulating movement preparation. The frequency and intensity of these changes can be manually adjusted, allowing the model to capture dynamic neural responses to behavioral states.

\subsection{DBS control problem}

The proposed brain model, incorporating three essential feature groups, requires an interface for agent interaction. Given the effectiveness of RL in control of nonlinear systems, we present a \textit{gymnasium environment} designed for training and evaluating various control strategies. 

\subsubsection{Observation}
As mentioned, an implanted electrode records local field potentials (LFPs), calculated by averaging the real part of neuronal phases. To model partial observability, the environment incorporates a conductance factor based on the distance between the electrode contact and neurons. Then, the expression for LFP reads as follows:
\begin{equation}
    LFP(t) = \frac{1}{N}\sum_{n=1}^N\cos(\theta_n(t)) \cdot G(\alpha_{n,el}).
\label{eq:lfp}
\end{equation}
Since a number of previous studies proposed to consider fully observed LFP (a case of $G(\alpha_{n,el})=1, \forall n,el$), we provide flexibility in our environment for adjusting this parameter. The observation vector length in an RL environment depends on the chosen control strategy, making it a user-defined hyperparameter: $y(t) = [LFP(t-k), \ldots, LFP(t)]^T \in \mathbb{R}^k$. In practice, we recommend a 1.2-second LFP window, implemented as a sliding window and updated step-by-step. To remove high-frequency components (>35 Hz) (irrelevant for control), the LFP output is filtered using the beta-band filter function before being sent to the agent.

\subsubsection{Action space}

In our environment, the DBS control problem involves regulating electrode contact voltages and stimulation direction. The action space is defined as $\mathcal{A} = \mathbb{R}^p \times \{0^\circ, 120^\circ, 240^\circ\}^p$, where $p$ is the number of electrode contacts, each with a continuous voltage amplitude $A_i \in \mathbb{R}$, and stimulation direction is a discrete variable $d \in \{0^\circ, 120^\circ, 240^\circ\}$. Each contact influences neurons through the neuron-to-electrode interaction function (\ref{eq:kur_prc}), with spatial parameters (conductances) dependent on the selected stimulation direction $d_i$ (refer to Figure ~\ref{fig:electrode_location_appendix} in Appendix).

\subsubsection{Evaluation metrics and reward function.} \label{section:metrics}

Formally, the DBS control problem targets a trade-off: maximizing efficiency while minimizing energy sent to the brain\footnote{Unlike HF-DBS that maximizes efficiency, disregarding the energy costs}.

Hence, the efficiency is the first metric, calculated as the value of the integral under the curve of the low beta-band spectrum in the 13–21 Hz range, known to be the source of pathological oscillations (see Section \ref{sec:features}). 
Specifically, $\beta_{BP} = \int_{13}^{21} S_y(f)df$, where $S_y(f)$ is the power spectral density of the recorded signal $y(t)$ (see Eq. (\ref{eq:lfp})). 
The second metric is the energy cost, calculated as the summation of all DBS pulses sent over the time step of the environment $T_{step}$, \textit{i.e.}, $E_A =  \int_{0}^{T_{step}}\sum^p_{i=1} A_i(t)dt$. 
In practice, we set the $T_{step}$ to be equal to the duration of one DBS pulse, which is equal to 90 ms. Per-episode metrics are averaged over and the total energy is summed over all stimulating contacts and steps. 

The corresponding reward functions address the wanted control problem for the trade-off between suppression efficiency $\beta_{BP}$ and the energy cost $E_A$ simultaneously and are widely used in prior studies \cite{lu2019application, faraji2023adaptive, fleming2020self, faraji2022novel, lu2017desynchronizing}:
\begin{equation}
    r_1 = -\lambda_{r_1}\beta_{BP} - \kappa_{r_1}E_A.
\label{eq:R1}
\end{equation}
Weights $\lambda_{r_1}$ and $\kappa_{r_1}$ are to adjust the balance between efficiency and energy during training. The other two standard reward functions, used to train RL agents, can be found in Appendix \ref{app: rewards}.

\subsubsection{Complexity levels of proposed environment}
Initial experiments indicated that the full-scale control problem was too complex for effective policy learning. Additionally, each feature domain introduces unique challenges, making it essential to assess their individual impact on aDBS performance. To address this, we structured the environment into three levels of increasing complexity, corresponding to the three feature domains (refer to Figure \ref{fig:intro}).  

\textbf{Env0.} Baseline environment incorporating bandwidth features. It includes a complex PSD, complete suppression of the pathological peak with HF-DBS, and basic spatial interactions between neurons. The electrode is placed within the beta locus, eliminating the need to account for electrode–beta locus distance.  

\textbf{Env1.} Includes Env0 features and adds the spatial features, including interactions between the electrode, neurons, and the beta locus. Partial observability is introduced through medium conductance, and stimulation effectiveness is reduced based on the distance between electrode contacts and the beta locus.  

\textbf{Env2.} Incorporates spatio-temporal dynamics, extending Env1 with the neural drift (modeled as drift of $\omega_n$), electrode drift (random electrode displacement and encapsulation), and beta burstings modulation (coupling parameter $K$ changes to reflect beta oscillation growth prior to the movement, see Section \ref{tf}).

\subsection{Baseline (\textit{i.e.}, non-adaptive) DBS agents} 

The baseline high-frequency DBS (HF-DBS) was configured with parameters close to the clinical practice: 110 Hz frequency, 90 microseconds pulse duration, and a maximum voltage of 5 V \cite{krauss2021technology}. 
Typical DBS settings for movement disorders range from 2 to 4 V amplitude, 60–450 microseconds pulse width, and 130–185 Hz frequency \cite{kuncel2004selection}. 
In all environments, HF-DBS fully suppressed low beta-band oscillations (Figure \ref{fig:sup_agents_performance}A).

The second baseline is Random DBS, which samples amplitude from $\mathcal{U}[-5, 5]$ at each step. Under this policy, the environment showed up to a 15\% reduction in low beta-band power in Env0 with half the energy cost compared to HF-DBS (Figure \ref{fig:sup_agents_performance}B, Table \ref{tab:exp_results_main_PERCENT}). However, its effectiveness was minimal in Env2, making it a valuable starting point for further exploring optimal policies.

\section{Experiments}
\begin{table*}
\centering
\caption{Performance of baseline aDBS algorithms in three environment levels, using $r_1$ (eq. \ref{eq:R1}) reward function. Details on the evaluation environment parameters are provided in Appendix \ref{app: model_implement_details}. Two key metrics were used for assessment: low beta-band power and the total stimulation energy. Evaluations were conducted six times for Env0, Env1 and 25 times for Env2. The best-performing model is highlighted in bold as a trade-off between the beta-band power and the stimuli energy. HF-DBS represents the case of maximum possible energy, while the DBS OFF condition indicates the highest beta-band power, reflecting base neural activity without stimulation.}
\label{tab:exp_results_main_PERCENT}

\begin{tabular}{lllllll}
\toprule
& \multicolumn{2}{c|}{Env0 (Basic model)} & \multicolumn{2}{c|}{Env1 (Spatial features)} & \multicolumn{2}{c|}{Env2 (Temporal features)} \\
\midrule
&  \multicolumn{1}{p{2.2cm}|}{Beta-band power, \%} 
&  \multicolumn{1}{p{2.2cm}|}{Stimulation energy, \%} 
&  \multicolumn{1}{p{2.2cm}|}{Beta-band power, \%} 
&  \multicolumn{1}{p{2.2cm}|}{Stimulation energy, \%} 
&  \multicolumn{1}{p{2.2cm}|}{Beta-band power, \%} 
&  \multicolumn{1}{p{2.2cm}|}{Stimulation energy, \%} \\
\midrule
DBS OFF &                   100 $\pm$27 &                             0 &                   100 $\pm$27 &                             0 &                 100 $\pm$13.4 &                             0 \\
HF-DBS &                 19.8 $\pm$1.9 &                       100 $\pm$0 &                 34.0 $\pm$2.5 &                       100 $\pm$0 &                 30.0 $\pm$5.7 &                       100 $\pm$0 \\
PI &                 18.6 $\pm$1.7 &                       100 $\pm$0 &                 33.6 $\pm$1.9 &                       100 $\pm$0 &                 31.8 $\pm$5.8 &                       100 $\pm$0 \\
PID &                 18.6 $\pm$1.7 &                       100 $\pm$0 &                 33.0 $\pm$2.2 &                       100 $\pm$0 &                43.4 $\pm$14.3 &                       100 $\pm$0 \\
Random policy &                85.8 $\pm$19.4 &                    49.5 $\pm$0.3 &                88.9 $\pm$23.8 &                    49.9 $\pm$0.6 &                97.2 $\pm$14.9 &                    49.9 $\pm$0.2 \\
PPO \cite{schulman2017proximal} &                66.4 $\pm$15.2 &                    94.9 $\pm$0.4 &                71.8 $\pm$10.2 &                      93 $\pm$0.4 &                86.0 $\pm$16.2 &                    99.2 $\pm$0.1 \\
SAC \cite{haarnoja2018soft} &                 \textbf{27.4 $\pm$5.9} &                    \textbf{87.8 $\pm$0.5} &                 \textbf{36.4 $\pm$1.9} &                   \textbf{86.5 $\pm$0.5}  &                 \textbf{36.9 $\pm$5.7} &                    \textbf{88.7 $\pm$0.7} \\
DDPG \cite{silver2014deterministic} &                 26.9 $\pm$4.9 &                    99.8 $\pm$0.1 &                 36.9 $\pm$4.2 &                    80.5 $\pm$1.0 &                93.9 $\pm$14.6 &                    89.6 $\pm$2.7 \\
IQL \cite{kostrikov2021iql} &                 30.4 $\pm$5.8 &                    99.2 $\pm$0.5 &                   39 $\pm$5.5 &                    99.8 $\pm$0.2 &                 26.6 $\pm$1.8 &                    99.9 $\pm$0.0 \\
CQL-SAC \cite{pan2024coprocessor} &                  97.2 $\pm$27 &                       7 $\pm$0.2 &                97.8 $\pm$20.9 &                     5.3 $\pm$0.1 &               107.9 $\pm$25.6 &                     4.8 $\pm$0.3 \\
\bottomrule
\end{tabular}
\end{table*}

While our environment supports all described features and action space dimensions, below, we focus on evaluating policies using the simplest single-contact, non-directional DBS across different environment levels for clarity. We compared traditional control methods, such as proportional–integral (PI) and proportional–integral–derivative (PID) controllers, with modern RL-based approaches. For RL agents, we selected well-established online and offline algorithms used in previous aDBS studies: PPO \cite{schulman2017proximal}, Deep Deterministic Policy Gradient (DDPG) \cite{silver2014deterministic}, and Soft Actor-Critic (SAC) \cite{haarnoja2018soft} for online RL, and Implicit Q-learning (IQL) \cite{kostrikov2021iql} and Conservative Q-learning (CQL) \cite{pan2024coprocessor} for offline RL.

\subsection{Experimental setup}
\subsubsection{Parameters choice and the training procedure.}
In the single-contact non-direction setting, the only control parameter is the amplitude of stimulation. The frequency and pulse duration of the stimulation was set as for the case of HF-DBS. 
For all aDBS algorithms, we tested 3 reward functions, described above.
For PI/PID controller, the error term was equal to a negative reward. All online agents were trained for 2.4 millions steps. The neural dynamics model was implemented with JAX library \cite{jax2018github} and created as a gymnasium environment class. It enables parallel computations on both GPU and CPU. We utilized the Diffrax library \cite{kidger2021on} for differential equation solving, selecting the Runge-Kutta45 solver with a step size of $5 \times 10^{-4}$ sec. To balance computational efficiency and dynamic complexity, we set $N=512$ with a grid size of $8\times8\times8$, ensuring sufficient richness in dynamics while maintaining manageable training times for RL agents. 
The process of training the agent is set as: 
\begin{lstlisting}[caption={Basic use of SpatialKuramoto environment.}, captionpos=b]
from neurokuramoto.model_v1 import SpatialKuramoto
from stable_baselines3 import PPO

# Define environment with default parameters
env = SpatialKuramoto(params_dict=params_dict_train)

# Define aDBS agent and start training  
agent = PPO("MlpPolicy", env, n_steps=2**7)
agent.learn(total_timesteps=2e6)
\end{lstlisting}
Our primary objective was not to optimize the performance of all tested aDBS algorithms but to evaluate whether RL could effectively learn in our environment and identify challenging features. To ensure a fair comparison, all algorithms were used without modification, with identical reward function weights. Further details on RL agent training are provided in Appendix \ref{app:agents}.  

\subsubsection{Evaluation Procedure}

The evaluation used the same environment parameters as training. For Env0 and Env1, 10 episodes of 1500 steps each (13 seconds per episode, 130 seconds total) were run with varying initializations. For Env2, which included temporal drifts, 5 environments were run for 25 episodes (1687 seconds total).

In Env0, only the random seed was varied. In Env1, spatial parameters were adjusted, including the positions of the beta locus, recording, and stimulating contacts. The distance between the electrode and beta locus was fixed at 2-3 grid cells, and placement near grid boundaries was prohibited for consistency. Env2 included temporal drift events: electrode shifts, electrode encapsulation, and neural drift. Electrode shifts occurred every 7th episode, moving randomly within 20\% of the grid's length while avoiding grid boundaries. 
Electrode encapsulation reduced conductance by 5\% every 5th episode, totaling a 25\% reduction by the end of evaluation. Neural drift occurred in each episode, shifting natural frequencies by 1\%, leading to a cumulative 25\% change. Complete evaluation results are shown in Table \ref{tab:vertical_results}, and the percentage gains relative to HF-DBS and DBS OFF are shown in Table \ref{tab:exp_results_main_PERCENT}.

\subsection{Results: comparison of RL algorithms}

Table \ref{tab:exp_results_main_PERCENT} and Figure \ref{fig:barplots} show the performance of algorithms across three levels of environment complexity. The metrics are presented as percentage gains: average beta band power relative to the maximum power observed in the OFF DBS condition, and the stimulation energy consumption relative to the maximum energy used by the HF-DBS. The absolute values are shown in Appendix (Table \ref{tab:vertical_results}). All algorithms were trained using the $r_1$ reward function (Eq. \ref{eq:R1}) with consistent weighting for efficiency and energy.

\textbf{Env0 results}: PI and PID controllers reduced beta power to about 20\% but used as much energy as HF-DBS. The random policy conserved 50\% of energy but only reduced beta power by 15\%. Online RL agents generally balanced efficiency and energy savings, with SAC performing best by reducing oscillations fourfold while saving 13\% of stimulation energy. Among offline RL methods, IQL behaved similarly to HF-DBS, while CQL-SAC failed to learn an effective policy.

\textbf{Env1 results}: Introducing spatial separation between the beta locus and electrode contacts reduced the performance of all algorithms, with HF-DBS only suppressing beta power to 34\%. SAC and DDPG were most effective in reducing beta oscillations, with DDPG using slightly less energy but showing more variability (standard deviation of 4.2). CQL-SAC performed poorly, sometimes worsening oscillatory dynamics.

\textbf{Env2 results}: Temporal drifts severely impacted Random DBS, CQL-SAC, and DDPG, with DDPG consuming 90\% of energy but reducing beta power by only 6\%. However, SAC maintained effectiveness, reducing beta power by 63\%, close to the HF-DBS’s 70\% reduction.

\begin{figure*}
    \centering
    \includegraphics[width=1.0\linewidth]{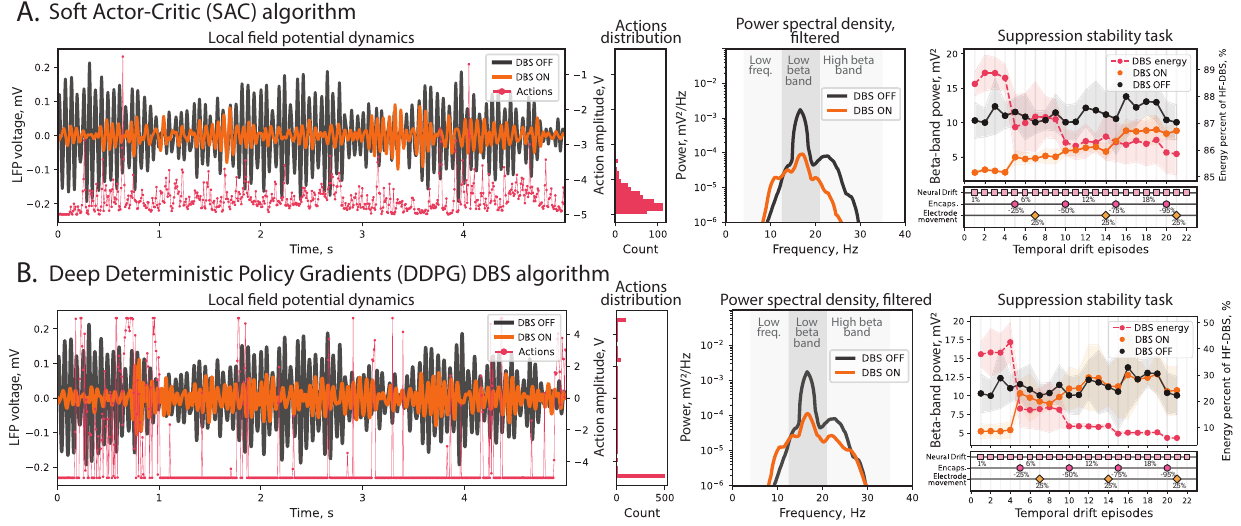}
    \caption{Performance of selected DBS Algorithms in proposed \textit{suppression stability task}: A) SAC-DBS and B) DDPG-DBS. The left column shows beta oscillation suppression (black: No DBS, orange: DBS), with red markers indicating DBS pulse amplitudes. The second column depicts the DBS amplitude distribution. The third column displays LFP power spectral densities, highlighting beta-band suppression. The right column illustrates control stability performance mimicking real-life `experimental' issues encoded as an event sequence: electrode movement (diamonds, simulating random shifts in the observation position), neural drift (squares, reflecting changes in the natural frequencies of neurons), encapsulation (hexagons, representing degradation in conductance between the DBS electrode and surrounding neurons). Beta power per episode (black and orange) and total stimulation energy (red points) are also shown.  
    }
    \label{fig:sac_performance_fig3}
\end{figure*}

Figure \ref{fig:sac_performance_fig3} illustrates the performance of the top-performing algorithms, SAC and DDPG, showing LFP dynamics, action distributions, and PSD spectrograms for both on and off DBS states\footnote{In clinical protocols, the goal of DBS is not to remove 
these LFP oscillations completely \cite{tinkhauser2017beta}, but to suppress excessive oscillations above some threshold.}. SAC and DDPG adopted different suppression strategies: DDPG primarily fired at the maximum negative voltage of $-5$ V, occasionally at $5$ V and $2.1$ V, while SAC consistently used negative amplitudes around $-4.7$ V. Both methods effectively suppressed excessive low beta band oscillations.

\textbf{Suppression stability task.} We evaluated the stability of algorithms in Env2 under strong temporal drift, as shown in the right part of Figure \ref{fig:sac_performance_fig3}. This task tests aDBS stability under worsening conditions. Temporal drift occurred every episode, with timing and intensity marked in the event plot (Figure \ref{fig:sac_performance_fig3} and Figure \ref{fig:sup_agents_performance}, rightmost panel). The drift was more severe than in the evaluation phase, progressively worsening the conditions for the aDBS to keep up with. Additional stability evaluations and other models' details are shown in the Appendix \ref{app:stability}.

Figure \ref{fig:sac_performance_fig3}A shows that SAC gradually allowed beta-band power to increase, maintaining suppression until the 14th episode. Conversely, DDPG failed to maintain low beta power after the first encapsulation event. Interestingly, SAC and DDP reduced stimulation energy in response to encapsulation drift, which was counterintuitive. Results for the other algorithms (PPO, HF-DBS, Random DBS, IQL) are shown in Figure \ref{fig:sup_agents_performance}.

\section{Discussion}

DBS offers an alternative to pharmacological treatment, effectively alleviating tremor, akinesia, bradykinesia, depressive symptoms, and other disease manifestations \cite{krauss2021technology}. Beyond its therapeutic benefits, DBS presents a compelling control problem due to the non-stationary nature of the neural dynamics. The complexity of the underlying system makes it a challenging task, even for modern reinforcement learning (RL) approaches, emphasizing the need for robust adaptive control strategies \cite{merk2022machine, watts2020machine}.

Developing an RL-based adaptive deep brain stimulation (aDBS) environment requires balancing biological relevance and computational feasibility. While several computational models of Parkinson’s disease exist, only one RL framework has been proposed \cite{krylov2020reinforcement}, all remaining incomplete in terms of implemented features (Figure \ref{fig:intro}F). We introduce a comprehensive RL training environment that integrates spatial, temporal, and frequency-domain features within a Kuramoto oscillator model, capturing essential aspects of neural synchrony for aDBS and brain-computer interfaces (BCIs). This neurophysiological relevance enhances the ability to suppress pathological beta oscillations and expands the potential for developing and evaluating aDBS strategies. These features have long been recognized, with numerous studies highlighting their importance \cite{yin2021local}; yet, to the authors' knowledge this is the first effort to standardize the evaluation within a unified synthetic environment.

Current closed-loop DBS systems rely on simple control mechanisms, typically limited to on/off switching or proportional scaling. While extensively tested clinically \cite{krauss2021technology}, future DBS systems are expected to become more adaptive, state-dependent, and energy-saving. We frame DBS control as a bicriterial optimization problem—maximizing efficiency while minimizing energy consumption—contrasting with HF-DBS, which prioritizes efficiency regardless of energy cost. To support this, we define an evaluation procedure and reward functions for training.
%
We evaluated various control strategies, including online RL, offline RL, and traditional PID controllers. SAC outperformed other algorithms across different environment levels, but the bicriterial task appeared quite complex, leaving room for further optimization and improved control strategies.

Our approach complements patient-based DBS studies while addressing a key gap: the need for a controlled pre-training environment. RL-based aDBS studies often rely solely on patient data, limiting generalizability. Our simulation-based training enables RL agents to pre-train in a structured, noise-resilient setting before fine-tuning on patient-specific data.

In summary, our RL-based aDBS framework provides a flexible, computationally efficient approach for training and evaluating adaptive stimulation strategies. By integrating multiple feature dimensions and leveraging a simulation-based pre-training pipeline, we offer a robust methodology that enhances generalizability and stability. These advancements contribute to the broader goal of making aDBS systems more adaptive, interpretable, and applicable across diverse neural disorders.

\bibliographystyle{ACM-Reference-Format}
\bibliography{main}


\begin{thebibliography}{134}


\ifx \showCODEN    \undefined \def \showCODEN     #1{\unskip}     \fi
\ifx \showISBNx    \undefined \def \showISBNx     #1{\unskip}     \fi
\ifx \showISBNxiii \undefined \def \showISBNxiii  #1{\unskip}     \fi
\ifx \showISSN     \undefined \def \showISSN      #1{\unskip}     \fi
\ifx \showLCCN     \undefined \def \showLCCN      #1{\unskip}     \fi
\ifx \shownote     \undefined \def \shownote      #1{#1}          \fi
\ifx \showarticletitle \undefined \def \showarticletitle #1{#1}   \fi
\ifx \showURL      \undefined \def \showURL       {\relax}        \fi
\providecommand\bibfield[2]{#2}
\providecommand\bibinfo[2]{#2}
\providecommand\natexlab[1]{#1}
\providecommand\showeprint[2][]{arXiv:#2}

\bibitem[Acebr{\'o}n et~al\mbox{.}(2005)]%
        {acebron2005kuramoto}
\bibfield{author}{\bibinfo{person}{Juan~A Acebr{\'o}n}, \bibinfo{person}{Luis~L Bonilla}, \bibinfo{person}{Conrad~J P{\'e}rez~Vicente}, \bibinfo{person}{F{\'e}lix Ritort}, {and} \bibinfo{person}{Renato Spigler}.} \bibinfo{year}{2005}\natexlab{}.
\newblock \showarticletitle{The Kuramoto model: A simple paradigm for synchronization phenomena}.
\newblock \bibinfo{journal}{\emph{Reviews of modern physics}} \bibinfo{volume}{77}, \bibinfo{number}{1} (\bibinfo{year}{2005}), \bibinfo{pages}{137--185}.
\newblock


\bibitem[Adair et~al\mbox{.}(2018)]%
        {adair2018evolving}
\bibfield{author}{\bibinfo{person}{Jason Adair}, \bibinfo{person}{Alexander Brownlee}, \bibinfo{person}{Fabio Daolio}, {and} \bibinfo{person}{Gabriela Ochoa}.} \bibinfo{year}{2018}\natexlab{}.
\newblock \showarticletitle{Evolving training sets for improved transfer learning in brain computer interfaces}. In \bibinfo{booktitle}{\emph{Machine Learning, Optimization, and Big Data: Third International Conference, MOD 2017, Volterra, Italy, September 14--17, 2017, Revised Selected Papers 3}}. Springer, \bibinfo{pages}{186--197}.
\newblock


\bibitem[Agarwal and Rathore(2023)]%
        {agarwal2023novel}
\bibfield{author}{\bibinfo{person}{Harsh Agarwal} {and} \bibinfo{person}{Heena Rathore}.} \bibinfo{year}{2023}\natexlab{}.
\newblock \showarticletitle{Novel Reinforcement Learning Algorithm for Suppressing Synchronization in Closed Loop Deep Brain Stimulators}. In \bibinfo{booktitle}{\emph{2023 11th International IEEE/EMBS Conference on Neural Engineering (NER)}}. IEEE, \bibinfo{pages}{1--5}.
\newblock


\bibitem[Akiba et~al\mbox{.}(2019)]%
        {akiba2019optuna}
\bibfield{author}{\bibinfo{person}{Takuya Akiba}, \bibinfo{person}{Shotaro Sano}, \bibinfo{person}{Toshihiko Yanase}, \bibinfo{person}{Takeru Ohta}, {and} \bibinfo{person}{Masanori Koyama}.} \bibinfo{year}{2019}\natexlab{}.
\newblock \showarticletitle{Optuna: A next-generation hyperparameter optimization framework}. In \bibinfo{booktitle}{\emph{Proceedings of the 25th ACM SIGKDD international conference on knowledge discovery \& data mining}}. \bibinfo{pages}{2623--2631}.
\newblock


\bibitem[Aman et~al\mbox{.}(2020)]%
        {aman2020directional}
\bibfield{author}{\bibinfo{person}{Joshua~E Aman}, \bibinfo{person}{Luke~A Johnson}, \bibinfo{person}{David~Escobar Sanabria}, \bibinfo{person}{Jing Wang}, \bibinfo{person}{Remi Patriat}, \bibinfo{person}{Meghan Hill}, \bibinfo{person}{Ethan Marshall}, \bibinfo{person}{Colum~D MacKinnon}, \bibinfo{person}{Scott~E Cooper}, \bibinfo{person}{Lauren~E Schrock}, {et~al\mbox{.}}} \bibinfo{year}{2020}\natexlab{}.
\newblock \showarticletitle{Directional deep brain stimulation leads reveal spatially distinct oscillatory activity in the globus pallidus internus of Parkinson's disease patients}.
\newblock \bibinfo{journal}{\emph{Neurobiology of disease}}  \bibinfo{volume}{139} (\bibinfo{year}{2020}), \bibinfo{pages}{104819}.
\newblock


\bibitem[Anderson et~al\mbox{.}(2020)]%
        {anderson2020novel}
\bibfield{author}{\bibinfo{person}{RW Anderson}, \bibinfo{person}{YM Kehnemouyi}, \bibinfo{person}{RS Neuville}, \bibinfo{person}{KB Wilkins}, \bibinfo{person}{CM Anidi}, \bibinfo{person}{MN Petrucci}, \bibinfo{person}{JE Parker}, \bibinfo{person}{A Velisar}, {and} \bibinfo{person}{HM Bront{\"e}-Stewart}.} \bibinfo{year}{2020}\natexlab{}.
\newblock \showarticletitle{A novel method for calculating beta band burst durations in Parkinson’s disease using a physiological baseline}.
\newblock \bibinfo{journal}{\emph{Journal of neuroscience methods}}  \bibinfo{volume}{343} (\bibinfo{year}{2020}), \bibinfo{pages}{108811}.
\newblock


\bibitem[Azodi-Avval and Gharabaghi(2015)]%
        {azodi2015phase}
\bibfield{author}{\bibinfo{person}{Ramin Azodi-Avval} {and} \bibinfo{person}{Alireza Gharabaghi}.} \bibinfo{year}{2015}\natexlab{}.
\newblock \showarticletitle{Phase-dependent modulation as a novel approach for therapeutic brain stimulation}.
\newblock \bibinfo{journal}{\emph{Frontiers in computational neuroscience}}  \bibinfo{volume}{9} (\bibinfo{year}{2015}), \bibinfo{pages}{26}.
\newblock


\bibitem[Bahadori-Jahromi et~al\mbox{.}(2023)]%
        {bahadori2023efficient}
\bibfield{author}{\bibinfo{person}{Fatemeh Bahadori-Jahromi}, \bibinfo{person}{Sina Salehi}, \bibinfo{person}{Mojtaba Madadi~Asl}, {and} \bibinfo{person}{Alireza Valizadeh}.} \bibinfo{year}{2023}\natexlab{}.
\newblock \showarticletitle{Efficient suppression of parkinsonian beta oscillations in a closed-loop model of deep brain stimulation with amplitude modulation}.
\newblock \bibinfo{journal}{\emph{Frontiers in human neuroscience}}  \bibinfo{volume}{16} (\bibinfo{year}{2023}), \bibinfo{pages}{1013155}.
\newblock


\bibitem[Balestrino and Schapira(2020)]%
        {balestrino2020parkinson}
\bibfield{author}{\bibinfo{person}{Roberta Balestrino} {and} \bibinfo{person}{AHV Schapira}.} \bibinfo{year}{2020}\natexlab{}.
\newblock \showarticletitle{Parkinson disease}.
\newblock \bibinfo{journal}{\emph{European journal of neurology}} \bibinfo{volume}{27}, \bibinfo{number}{1} (\bibinfo{year}{2020}), \bibinfo{pages}{27--42}.
\newblock


\bibitem[Binns et~al\mbox{.}(2024)]%
        {binns2024shared}
\bibfield{author}{\bibinfo{person}{Thomas~S Binns}, \bibinfo{person}{Richard~M K{\"o}hler}, \bibinfo{person}{Jonathan Vanhoecke}, \bibinfo{person}{Meera Chikermane}, \bibinfo{person}{Moritz Gerster}, \bibinfo{person}{Timon Merk}, \bibinfo{person}{Franziska Pellegrini}, \bibinfo{person}{Johannes~L Busch}, \bibinfo{person}{Jeroen~GV Habets}, \bibinfo{person}{Allesia Cavallo}, {et~al\mbox{.}}} \bibinfo{year}{2024}\natexlab{}.
\newblock \showarticletitle{Shared pathway-specific network mechanisms of dopamine and deep brain stimulation for the treatment of Parkinson's disease}.
\newblock \bibinfo{journal}{\emph{bioRxiv}} (\bibinfo{year}{2024}), \bibinfo{pages}{2024--04}.
\newblock


\bibitem[Birdno et~al\mbox{.}(2012)]%
        {birdno2012stimulus}
\bibfield{author}{\bibinfo{person}{Merrill~J Birdno}, \bibinfo{person}{Alexis~M Kuncel}, \bibinfo{person}{Alan~D Dorval}, \bibinfo{person}{Dennis~A Turner}, \bibinfo{person}{Robert~E Gross}, {and} \bibinfo{person}{Warren~M Grill}.} \bibinfo{year}{2012}\natexlab{}.
\newblock \showarticletitle{Stimulus features underlying reduced tremor suppression with temporally patterned deep brain stimulation}.
\newblock \bibinfo{journal}{\emph{Journal of neurophysiology}} \bibinfo{volume}{107}, \bibinfo{number}{1} (\bibinfo{year}{2012}), \bibinfo{pages}{364--383}.
\newblock


\bibitem[Bloem et~al\mbox{.}(2021)]%
        {bloem2021parkinson}
\bibfield{author}{\bibinfo{person}{Bastiaan~R Bloem}, \bibinfo{person}{Michael~S Okun}, {and} \bibinfo{person}{Christine Klein}.} \bibinfo{year}{2021}\natexlab{}.
\newblock \showarticletitle{Parkinson's disease}.
\newblock \bibinfo{journal}{\emph{The Lancet}} \bibinfo{volume}{397}, \bibinfo{number}{10291} (\bibinfo{year}{2021}), \bibinfo{pages}{2284--2303}.
\newblock


\bibitem[Bradbury et~al\mbox{.}(2018)]%
        {jax2018github}
\bibfield{author}{\bibinfo{person}{James Bradbury}, \bibinfo{person}{Roy Frostig}, \bibinfo{person}{Peter Hawkins}, \bibinfo{person}{Matthew~James Johnson}, \bibinfo{person}{Chris Leary}, \bibinfo{person}{Dougal Maclaurin}, \bibinfo{person}{George Necula}, \bibinfo{person}{Adam Paszke}, \bibinfo{person}{Jake Vander{P}las}, \bibinfo{person}{Skye Wanderman-{M}ilne}, {and} \bibinfo{person}{Qiao Zhang}.} \bibinfo{year}{2018}\natexlab{}.
\newblock \bibinfo{booktitle}{\emph{{JAX}: composable transformations of {P}ython+{N}um{P}y programs}}.
\newblock
\urldef\tempurl%
\url{http://github.com/jax-ml/jax}
\showURL{%
\tempurl}


\bibitem[Breakspear et~al\mbox{.}(2010)]%
        {breakspear2010generative}
\bibfield{author}{\bibinfo{person}{Michael Breakspear}, \bibinfo{person}{Stewart Heitmann}, {and} \bibinfo{person}{Andreas Daffertshofer}.} \bibinfo{year}{2010}\natexlab{}.
\newblock \showarticletitle{Generative models of cortical oscillations: neurobiological implications of the Kuramoto model}.
\newblock \bibinfo{journal}{\emph{Frontiers in human neuroscience}}  \bibinfo{volume}{4} (\bibinfo{year}{2010}), \bibinfo{pages}{190}.
\newblock


\bibitem[Butson et~al\mbox{.}(2011)]%
        {butson2011probabilistic}
\bibfield{author}{\bibinfo{person}{Christopher~R Butson}, \bibinfo{person}{Scott~E Cooper}, \bibinfo{person}{Jaimie~M Henderson}, \bibinfo{person}{Barbara Wolgamuth}, {and} \bibinfo{person}{Cameron~C McIntyre}.} \bibinfo{year}{2011}\natexlab{}.
\newblock \showarticletitle{Probabilistic analysis of activation volumes generated during deep brain stimulation}.
\newblock \bibinfo{journal}{\emph{Neuroimage}} \bibinfo{volume}{54}, \bibinfo{number}{3} (\bibinfo{year}{2011}), \bibinfo{pages}{2096--2104}.
\newblock


\bibitem[Byrne et~al\mbox{.}(2017)]%
        {byrne2017mean}
\bibfield{author}{\bibinfo{person}{Aine Byrne}, \bibinfo{person}{Matthew~J Brookes}, {and} \bibinfo{person}{Stephen Coombes}.} \bibinfo{year}{2017}\natexlab{}.
\newblock \showarticletitle{A mean field model for movement induced changes in the beta rhythm}.
\newblock \bibinfo{journal}{\emph{Journal of computational neuroscience}}  \bibinfo{volume}{43} (\bibinfo{year}{2017}), \bibinfo{pages}{143--158}.
\newblock


\bibitem[Cagnan et~al\mbox{.}(2019)]%
        {cagnan2019emerging}
\bibfield{author}{\bibinfo{person}{Hayriye Cagnan}, \bibinfo{person}{Timothy Denison}, \bibinfo{person}{Cameron McIntyre}, {and} \bibinfo{person}{Peter Brown}.} \bibinfo{year}{2019}\natexlab{}.
\newblock \showarticletitle{Emerging technologies for improved deep brain stimulation}.
\newblock \bibinfo{journal}{\emph{Nature biotechnology}} \bibinfo{volume}{37}, \bibinfo{number}{9} (\bibinfo{year}{2019}), \bibinfo{pages}{1024--1033}.
\newblock


\bibitem[Cagnan et~al\mbox{.}(2009)]%
        {cagnan2009frequency}
\bibfield{author}{\bibinfo{person}{Hayriye Cagnan}, \bibinfo{person}{Hil~GE Meijer}, \bibinfo{person}{Stephan~A Van~Gils}, \bibinfo{person}{Martin Krupa}, \bibinfo{person}{Tjitske Heida}, \bibinfo{person}{Michelle Rudolph}, \bibinfo{person}{Wytse~J Wadman}, {and} \bibinfo{person}{Hubert~CF Martens}.} \bibinfo{year}{2009}\natexlab{}.
\newblock \showarticletitle{Frequency-selectivity of a thalamocortical relay neuron during Parkinson’s disease and deep brain stimulation: a computational study}.
\newblock \bibinfo{journal}{\emph{European Journal of Neuroscience}} \bibinfo{volume}{30}, \bibinfo{number}{7} (\bibinfo{year}{2009}), \bibinfo{pages}{1306--1317}.
\newblock


\bibitem[Campbell and Wu(2018)]%
        {campbell2018chronically}
\bibfield{author}{\bibinfo{person}{Andrew Campbell} {and} \bibinfo{person}{Chengyuan Wu}.} \bibinfo{year}{2018}\natexlab{}.
\newblock \showarticletitle{Chronically implanted intracranial electrodes: tissue reaction and electrical changes}.
\newblock \bibinfo{journal}{\emph{Micromachines}} \bibinfo{volume}{9}, \bibinfo{number}{9} (\bibinfo{year}{2018}), \bibinfo{pages}{430}.
\newblock


\bibitem[Cao et~al\mbox{.}(2020)]%
        {cao2020dopa}
\bibfield{author}{\bibinfo{person}{Chunyan Cao}, \bibinfo{person}{Dianyou Li}, \bibinfo{person}{Shikun Zhan}, \bibinfo{person}{Chencheng Zhang}, \bibinfo{person}{Bomin Sun}, {and} \bibinfo{person}{Vladimir Litvak}.} \bibinfo{year}{2020}\natexlab{}.
\newblock \showarticletitle{L-dopa treatment increases oscillatory power in the motor cortex of Parkinson's disease patients}.
\newblock \bibinfo{journal}{\emph{NeuroImage: Clinical}}  \bibinfo{volume}{26} (\bibinfo{year}{2020}), \bibinfo{pages}{102255}.
\newblock


\bibitem[Carlson et~al\mbox{.}(2021)]%
        {carlson2021computational}
\bibfield{author}{\bibinfo{person}{Kristen Carlson}, \bibinfo{person}{Jay~L Shils}, \bibinfo{person}{Sahil Patel}, \bibinfo{person}{Longzhi Mei}, {and} \bibinfo{person}{Jeffrey Arle}.} \bibinfo{year}{2021}\natexlab{}.
\newblock \showarticletitle{Computational Modelling of Deep Brain Stimulation for Parkinson's Disease: A Critical Review}.
\newblock \bibinfo{journal}{\emph{OBM Neurobiology}} \bibinfo{volume}{5}, \bibinfo{number}{2} (\bibinfo{year}{2021}), \bibinfo{pages}{1--31}.
\newblock


\bibitem[Cassidy et~al\mbox{.}(2002)]%
        {cassidy2002movement}
\bibfield{author}{\bibinfo{person}{Michael Cassidy}, \bibinfo{person}{Paolo Mazzone}, \bibinfo{person}{Antonio Oliviero}, \bibinfo{person}{Angelo Insola}, \bibinfo{person}{Pietro Tonali}, \bibinfo{person}{Vincenzo~Di Lazzaro}, {and} \bibinfo{person}{Peter Brown}.} \bibinfo{year}{2002}\natexlab{}.
\newblock \showarticletitle{Movement-related changes in synchronization in the human basal ganglia}.
\newblock \bibinfo{journal}{\emph{Brain}} \bibinfo{volume}{125}, \bibinfo{number}{6} (\bibinfo{year}{2002}), \bibinfo{pages}{1235--1246}.
\newblock


\bibitem[Chapelle et~al\mbox{.}(2021)]%
        {chapelle2021early}
\bibfield{author}{\bibinfo{person}{Fr{\'e}d{\'e}ric Chapelle}, \bibinfo{person}{Lucie Manciet}, \bibinfo{person}{Bruno Pereira}, \bibinfo{person}{Anna Sontheimer}, \bibinfo{person}{J{\'e}r{\^o}me Coste}, \bibinfo{person}{Youssef El~Ouadih}, \bibinfo{person}{Ruxandra Cimpeanu}, \bibinfo{person}{Dimitri Gouot}, \bibinfo{person}{Yuri Lapusta}, \bibinfo{person}{B{\'e}atrice Claise}, {et~al\mbox{.}}} \bibinfo{year}{2021}\natexlab{}.
\newblock \showarticletitle{Early deformation of deep brain stimulation electrodes following surgical implantation: intracranial, brain, and electrode mechanics}.
\newblock \bibinfo{journal}{\emph{Frontiers in Bioengineering and Biotechnology}}  \bibinfo{volume}{9} (\bibinfo{year}{2021}), \bibinfo{pages}{657875}.
\newblock


\bibitem[Chen et~al\mbox{.}(2022)]%
        {chen2022subthalamic}
\bibfield{author}{\bibinfo{person}{Po-Lin Chen}, \bibinfo{person}{Yi-Chieh Chen}, \bibinfo{person}{Po-Hsun Tu}, \bibinfo{person}{Tzu-Chi Liu}, \bibinfo{person}{Min-Chi Chen}, \bibinfo{person}{Hau-Tieng Wu}, \bibinfo{person}{Mun-Chun Yeap}, \bibinfo{person}{Chih-Hua Yeh}, \bibinfo{person}{Chin-Song Lu}, {and} \bibinfo{person}{Chiung-Chu Chen}.} \bibinfo{year}{2022}\natexlab{}.
\newblock \showarticletitle{Subthalamic high-beta oscillation informs the outcome of deep brain stimulation in patients with Parkinson's disease}.
\newblock \bibinfo{journal}{\emph{Frontiers in Human Neuroscience}}  \bibinfo{volume}{16} (\bibinfo{year}{2022}), \bibinfo{pages}{958521}.
\newblock


\bibitem[Chen et~al\mbox{.}(2023)]%
        {chen2023optimal}
\bibfield{author}{\bibinfo{person}{Ziqin Chen}, \bibinfo{person}{Timothy Anglea}, \bibinfo{person}{Yuanzhao Zhang}, {and} \bibinfo{person}{Yongqiang Wang}.} \bibinfo{year}{2023}\natexlab{}.
\newblock \showarticletitle{Optimal synchronization in pulse-coupled oscillator networks using reinforcement learning}.
\newblock \bibinfo{journal}{\emph{PNAS nexus}} \bibinfo{volume}{2}, \bibinfo{number}{4} (\bibinfo{year}{2023}), \bibinfo{pages}{pgad102}.
\newblock


\bibitem[Cumin and Unsworth(2007)]%
        {cumin2007generalising}
\bibfield{author}{\bibinfo{person}{David Cumin} {and} \bibinfo{person}{CP2296240 Unsworth}.} \bibinfo{year}{2007}\natexlab{}.
\newblock \showarticletitle{Generalising the Kuramoto model for the study of neuronal synchronisation in the brain}.
\newblock \bibinfo{journal}{\emph{Physica D: Nonlinear Phenomena}} \bibinfo{volume}{226}, \bibinfo{number}{2} (\bibinfo{year}{2007}), \bibinfo{pages}{181--196}.
\newblock


\bibitem[Daneshzand et~al\mbox{.}(2018)]%
        {daneshzand2018robust}
\bibfield{author}{\bibinfo{person}{Mohammad Daneshzand}, \bibinfo{person}{Miad Faezipour}, {and} \bibinfo{person}{Buket~D Barkana}.} \bibinfo{year}{2018}\natexlab{}.
\newblock \showarticletitle{Robust desynchronization of Parkinson’s disease pathological oscillations by frequency modulation of delayed feedback deep brain stimulation}.
\newblock \bibinfo{journal}{\emph{PloS one}} \bibinfo{volume}{13}, \bibinfo{number}{11} (\bibinfo{year}{2018}), \bibinfo{pages}{e0207761}.
\newblock


\bibitem[Danzl and Moehlis(2007)]%
        {danzl2007event}
\bibfield{author}{\bibinfo{person}{Per Danzl} {and} \bibinfo{person}{Jeff Moehlis}.} \bibinfo{year}{2007}\natexlab{}.
\newblock \showarticletitle{Event-based feedback control of nonlinear oscillators using phase response curves}. In \bibinfo{booktitle}{\emph{2007 46th IEEE Conference on Decision and Control}}. IEEE, \bibinfo{pages}{5806--5811}.
\newblock


\bibitem[Degenhart et~al\mbox{.}(2020)]%
        {degenhart2020stabilization}
\bibfield{author}{\bibinfo{person}{Alan~D Degenhart}, \bibinfo{person}{William~E Bishop}, \bibinfo{person}{Emily~R Oby}, \bibinfo{person}{Elizabeth~C Tyler-Kabara}, \bibinfo{person}{Steven~M Chase}, \bibinfo{person}{Aaron~P Batista}, {and} \bibinfo{person}{Byron~M Yu}.} \bibinfo{year}{2020}\natexlab{}.
\newblock \showarticletitle{Stabilization of a brain--computer interface via the alignment of low-dimensional spaces of neural activity}.
\newblock \bibinfo{journal}{\emph{Nature biomedical engineering}} \bibinfo{volume}{4}, \bibinfo{number}{7} (\bibinfo{year}{2020}), \bibinfo{pages}{672--685}.
\newblock


\bibitem[Dovzhenok et~al\mbox{.}(2013)]%
        {dovzhenok2013failure}
\bibfield{author}{\bibinfo{person}{Andrey Dovzhenok}, \bibinfo{person}{Choongseok Park}, \bibinfo{person}{Robert~M Worth}, {and} \bibinfo{person}{Leonid~L Rubchinsky}.} \bibinfo{year}{2013}\natexlab{}.
\newblock \showarticletitle{Failure of delayed feedback deep brain stimulation for intermittent pathological synchronization in Parkinson’s disease}.
\newblock \bibinfo{journal}{\emph{PLoS One}} \bibinfo{volume}{8}, \bibinfo{number}{3} (\bibinfo{year}{2013}), \bibinfo{pages}{e58264}.
\newblock


\bibitem[Fang et~al\mbox{.}(2023)]%
        {fang2023robust}
\bibfield{author}{\bibinfo{person}{Hao Fang}, \bibinfo{person}{Stephen~A Berman}, \bibinfo{person}{Yueming Wang}, {and} \bibinfo{person}{Yuxiao Yang}.} \bibinfo{year}{2023}\natexlab{}.
\newblock \showarticletitle{Robust adaptive deep brain stimulation control of non-stationary cortex-basal ganglia-thalamus network models in parkinson’s disease}.
\newblock \bibinfo{journal}{\emph{bioRxiv}} (\bibinfo{year}{2023}), \bibinfo{pages}{2023--08}.
\newblock


\bibitem[Faraji et~al\mbox{.}(2022)]%
        {faraji2022novel}
\bibfield{author}{\bibinfo{person}{Behnam Faraji}, \bibinfo{person}{Korosh Rouhollahi}, \bibinfo{person}{Akram Nezhadi}, {and} \bibinfo{person}{Zahra Jamalpoor}.} \bibinfo{year}{2022}\natexlab{}.
\newblock \showarticletitle{A novel closed-loop deep brain stimulation technique for Parkinson’s patients rehabilitation utilizing machine learning}.
\newblock \bibinfo{journal}{\emph{IEEE Sensors Journal}} \bibinfo{volume}{23}, \bibinfo{number}{3} (\bibinfo{year}{2022}), \bibinfo{pages}{2914--2921}.
\newblock


\bibitem[Faraji et~al\mbox{.}(2023)]%
        {faraji2023adaptive}
\bibfield{author}{\bibinfo{person}{Behnam Faraji}, \bibinfo{person}{Korosh Rouhollahi}, \bibinfo{person}{Saeed~Mollahoseini Paghaleh}, \bibinfo{person}{Meysam Gheisarnejad}, {and} \bibinfo{person}{Mohammad-Hassan Khooban}.} \bibinfo{year}{2023}\natexlab{}.
\newblock \showarticletitle{Adaptive multi symptoms control of Parkinson's disease by deep reinforcement learning}.
\newblock \bibinfo{journal}{\emph{Biomedical Signal Processing and Control}}  \bibinfo{volume}{80} (\bibinfo{year}{2023}), \bibinfo{pages}{104410}.
\newblock


\bibitem[Farokhniaee and Lowery(2021)]%
        {farokhniaee2021cortical}
\bibfield{author}{\bibinfo{person}{AmirAli Farokhniaee} {and} \bibinfo{person}{Madeleine~M Lowery}.} \bibinfo{year}{2021}\natexlab{}.
\newblock \showarticletitle{Cortical network effects of subthalamic deep brain stimulation in a thalamo-cortical microcircuit model}.
\newblock \bibinfo{journal}{\emph{Journal of Neural Engineering}} \bibinfo{volume}{18}, \bibinfo{number}{5} (\bibinfo{year}{2021}), \bibinfo{pages}{056006}.
\newblock


\bibitem[Feng et~al\mbox{.}(2007)]%
        {feng2007optimal}
\bibfield{author}{\bibinfo{person}{Xiao-Jiang Feng}, \bibinfo{person}{Eric Shea-Brown}, \bibinfo{person}{Brian Greenwald}, \bibinfo{person}{Robert Kosut}, {and} \bibinfo{person}{Herschel Rabitz}.} \bibinfo{year}{2007}\natexlab{}.
\newblock \showarticletitle{Optimal deep brain stimulation of the subthalamic nucleus—a computational study}.
\newblock \bibinfo{journal}{\emph{Journal of computational neuroscience}}  \bibinfo{volume}{23} (\bibinfo{year}{2007}), \bibinfo{pages}{265--282}.
\newblock


\bibitem[Ferrari et~al\mbox{.}(2015)]%
        {ferrari2015phase}
\bibfield{author}{\bibinfo{person}{Fabiano Alan~Serafim Ferrari}, \bibinfo{person}{Ricardo~L Viana}, \bibinfo{person}{Sergio~Roberto Lopes}, {and} \bibinfo{person}{Ruedi Stoop}.} \bibinfo{year}{2015}\natexlab{}.
\newblock \showarticletitle{Phase synchronization of coupled bursting neurons and the generalized Kuramoto model}.
\newblock \bibinfo{journal}{\emph{Neural Networks}}  \bibinfo{volume}{66} (\bibinfo{year}{2015}), \bibinfo{pages}{107--118}.
\newblock


\bibitem[Fleming et~al\mbox{.}(2020a)]%
        {fleming2020simulation}
\bibfield{author}{\bibinfo{person}{John~E Fleming}, \bibinfo{person}{Eleanor Dunn}, {and} \bibinfo{person}{Madeleine~M Lowery}.} \bibinfo{year}{2020}\natexlab{a}.
\newblock \showarticletitle{Simulation of closed-loop deep brain stimulation control schemes for suppression of pathological beta oscillations in Parkinson’s disease}.
\newblock \bibinfo{journal}{\emph{Frontiers in neuroscience}}  \bibinfo{volume}{14} (\bibinfo{year}{2020}), \bibinfo{pages}{520710}.
\newblock


\bibitem[Fleming et~al\mbox{.}(2020b)]%
        {fleming2020self}
\bibfield{author}{\bibinfo{person}{John~E Fleming}, \bibinfo{person}{Jakub Or{\l}owski}, \bibinfo{person}{Madeleine~M Lowery}, {and} \bibinfo{person}{Antoine Chaillet}.} \bibinfo{year}{2020}\natexlab{b}.
\newblock \showarticletitle{Self-tuning deep brain stimulation controller for suppression of beta oscillations: analytical derivation and numerical validation}.
\newblock \bibinfo{journal}{\emph{Frontiers in neuroscience}}  \bibinfo{volume}{14} (\bibinfo{year}{2020}), \bibinfo{pages}{639}.
\newblock


\bibitem[Franci et~al\mbox{.}(2012)]%
        {franci2012desynchronization}
\bibfield{author}{\bibinfo{person}{Alessio Franci}, \bibinfo{person}{Antoine Chaillet}, \bibinfo{person}{Elena Panteley}, {and} \bibinfo{person}{Fran{\c{c}}oise Lamnabhi-Lagarrigue}.} \bibinfo{year}{2012}\natexlab{}.
\newblock \showarticletitle{Desynchronization and inhibition of Kuramoto oscillators by scalar mean-field feedback}.
\newblock \bibinfo{journal}{\emph{Mathematics of Control, Signals, and Systems}} \bibinfo{volume}{24}, \bibinfo{number}{1} (\bibinfo{year}{2012}), \bibinfo{pages}{169--217}.
\newblock


\bibitem[Fung et~al\mbox{.}(2013)]%
        {fung2013neural}
\bibfield{author}{\bibinfo{person}{PK Fung}, \bibinfo{person}{AL Haber}, {and} \bibinfo{person}{PA Robinson}.} \bibinfo{year}{2013}\natexlab{}.
\newblock \showarticletitle{Neural field theory of plasticity in the cerebral cortex}.
\newblock \bibinfo{journal}{\emph{Journal of Theoretical Biology}}  \bibinfo{volume}{318} (\bibinfo{year}{2013}), \bibinfo{pages}{44--57}.
\newblock


\bibitem[Gao et~al\mbox{.}(2020)]%
        {gao2020model}
\bibfield{author}{\bibinfo{person}{Qitong Gao}, \bibinfo{person}{Michael Naumann}, \bibinfo{person}{Ilija Jovanov}, \bibinfo{person}{Vuk Lesi}, \bibinfo{person}{Karthik Kamaravelu}, \bibinfo{person}{Warren~M Grill}, {and} \bibinfo{person}{Miroslav Pajic}.} \bibinfo{year}{2020}\natexlab{}.
\newblock \showarticletitle{Model-based design of closed loop deep brain stimulation controller using reinforcement learning}. In \bibinfo{booktitle}{\emph{2020 ACM/IEEE 11th International Conference on Cyber-Physical Systems (ICCPS)}}. IEEE, \bibinfo{pages}{108--118}.
\newblock


\bibitem[Gao et~al\mbox{.}(2023)]%
        {gao2023offline}
\bibfield{author}{\bibinfo{person}{Qitong Gao}, \bibinfo{person}{Stephen~L Schmidt}, \bibinfo{person}{Afsana Chowdhury}, \bibinfo{person}{Guangyu Feng}, \bibinfo{person}{Jennifer~J Peters}, \bibinfo{person}{Katherine Genty}, \bibinfo{person}{Warren~M Grill}, \bibinfo{person}{Dennis~A Turner}, {and} \bibinfo{person}{Miroslav Pajic}.} \bibinfo{year}{2023}\natexlab{}.
\newblock \showarticletitle{Offline learning of closed-loop deep brain stimulation controllers for parkinson disease treatment}. In \bibinfo{booktitle}{\emph{Proceedings of the ACM/IEEE 14th International Conference on Cyber-Physical Systems (with CPS-IoT Week 2023)}}. \bibinfo{pages}{44--55}.
\newblock


\bibitem[Gilron et~al\mbox{.}(2021a)]%
        {gilron2021long}
\bibfield{author}{\bibinfo{person}{Ro’ee Gilron}, \bibinfo{person}{Simon Little}, \bibinfo{person}{Randy Perrone}, \bibinfo{person}{Robert Wilt}, \bibinfo{person}{Coralie de Hemptinne}, \bibinfo{person}{Maria~S Yaroshinsky}, \bibinfo{person}{Caroline~A Racine}, \bibinfo{person}{Sarah~S Wang}, \bibinfo{person}{Jill~L Ostrem}, \bibinfo{person}{Paul~S Larson}, {et~al\mbox{.}}} \bibinfo{year}{2021}\natexlab{a}.
\newblock \showarticletitle{Long-term wireless streaming of neural recordings for circuit discovery and adaptive stimulation in individuals with Parkinson’s disease}.
\newblock \bibinfo{journal}{\emph{Nature biotechnology}} \bibinfo{volume}{39}, \bibinfo{number}{9} (\bibinfo{year}{2021}), \bibinfo{pages}{1078--1085}.
\newblock


\bibitem[Gilron et~al\mbox{.}(2021b)]%
        {gilron2021sleep}
\bibfield{author}{\bibinfo{person}{Ro’ee Gilron}, \bibinfo{person}{Simon Little}, \bibinfo{person}{Robert Wilt}, \bibinfo{person}{Randy Perrone}, \bibinfo{person}{Juan Anso}, {and} \bibinfo{person}{Philip~A Starr}.} \bibinfo{year}{2021}\natexlab{b}.
\newblock \showarticletitle{Sleep-aware adaptive deep brain stimulation control: chronic use at home with dual independent linear discriminate detectors}.
\newblock \bibinfo{journal}{\emph{Frontiers in Neuroscience}}  \bibinfo{volume}{15} (\bibinfo{year}{2021}), \bibinfo{pages}{732499}.
\newblock


\bibitem[Grado et~al\mbox{.}(2018)]%
        {grado2018bayesian}
\bibfield{author}{\bibinfo{person}{Logan~L Grado}, \bibinfo{person}{Matthew~D Johnson}, {and} \bibinfo{person}{Theoden~I Netoff}.} \bibinfo{year}{2018}\natexlab{}.
\newblock \showarticletitle{Bayesian adaptive dual control of deep brain stimulation in a computational model of Parkinson’s disease}.
\newblock \bibinfo{journal}{\emph{PLoS computational biology}} \bibinfo{volume}{14}, \bibinfo{number}{12} (\bibinfo{year}{2018}), \bibinfo{pages}{e1006606}.
\newblock


\bibitem[Groothuis et~al\mbox{.}(2014)]%
        {groothuis2014physiological}
\bibfield{author}{\bibinfo{person}{Jitte Groothuis}, \bibinfo{person}{Nick~F Ramsey}, \bibinfo{person}{Geert~MJ Ramakers}, {and} \bibinfo{person}{Geoffrey van~der Plasse}.} \bibinfo{year}{2014}\natexlab{}.
\newblock \showarticletitle{Physiological challenges for intracortical electrodes}.
\newblock \bibinfo{journal}{\emph{Brain stimulation}} \bibinfo{volume}{7}, \bibinfo{number}{1} (\bibinfo{year}{2014}), \bibinfo{pages}{1--6}.
\newblock


\bibitem[Guo et~al\mbox{.}(2013)]%
        {guo2013subthalamic}
\bibfield{author}{\bibinfo{person}{Song Guo}, \bibinfo{person}{Ping Zhuang}, \bibinfo{person}{Mark Hallett}, \bibinfo{person}{Zhe Zheng}, \bibinfo{person}{Yuqing Zhang}, \bibinfo{person}{Jianyu Li}, {and} \bibinfo{person}{Yongjie Li}.} \bibinfo{year}{2013}\natexlab{}.
\newblock \showarticletitle{Subthalamic deep brain stimulation for Parkinson's disease: correlation between locations of oscillatory activity and optimal site of stimulation}.
\newblock \bibinfo{journal}{\emph{Parkinsonism \& Related Disorders}} \bibinfo{volume}{19}, \bibinfo{number}{1} (\bibinfo{year}{2013}), \bibinfo{pages}{109--114}.
\newblock


\bibitem[Guo and Rubin(2011)]%
        {guo2011multi}
\bibfield{author}{\bibinfo{person}{Yixin Guo} {and} \bibinfo{person}{Jonathan~E Rubin}.} \bibinfo{year}{2011}\natexlab{}.
\newblock \showarticletitle{Multi-site stimulation of subthalamic nucleus diminishes thalamocortical relay errors in a biophysical network model}.
\newblock \bibinfo{journal}{\emph{Neural Networks}} \bibinfo{volume}{24}, \bibinfo{number}{6} (\bibinfo{year}{2011}), \bibinfo{pages}{602--616}.
\newblock


\bibitem[Haarnoja et~al\mbox{.}(2018)]%
        {haarnoja2018soft}
\bibfield{author}{\bibinfo{person}{Tuomas Haarnoja}, \bibinfo{person}{Aurick Zhou}, \bibinfo{person}{Kristian Hartikainen}, \bibinfo{person}{George Tucker}, \bibinfo{person}{Sehoon Ha}, \bibinfo{person}{Jie Tan}, \bibinfo{person}{Vikash Kumar}, \bibinfo{person}{Henry Zhu}, \bibinfo{person}{Abhishek Gupta}, \bibinfo{person}{Pieter Abbeel}, {et~al\mbox{.}}} \bibinfo{year}{2018}\natexlab{}.
\newblock \showarticletitle{Soft actor-critic algorithms and applications}.
\newblock \bibinfo{journal}{\emph{arXiv preprint arXiv:1812.05905}} (\bibinfo{year}{2018}).
\newblock


\bibitem[Hahn and McIntyre(2010)]%
        {hahn2010modeling}
\bibfield{author}{\bibinfo{person}{Philip~J Hahn} {and} \bibinfo{person}{Cameron~C McIntyre}.} \bibinfo{year}{2010}\natexlab{}.
\newblock \showarticletitle{Modeling shifts in the rate and pattern of subthalamopallidal network activity during deep brain stimulation}.
\newblock \bibinfo{journal}{\emph{Journal of computational neuroscience}}  \bibinfo{volume}{28} (\bibinfo{year}{2010}), \bibinfo{pages}{425--441}.
\newblock


\bibitem[Hansel et~al\mbox{.}(1993)]%
        {hansel1993phase}
\bibfield{author}{\bibinfo{person}{David Hansel}, \bibinfo{person}{German Mato}, {and} \bibinfo{person}{Claude Meunier}.} \bibinfo{year}{1993}\natexlab{}.
\newblock \showarticletitle{Phase dynamics for weakly coupled Hodgkin-Huxley neurons}.
\newblock \bibinfo{journal}{\emph{Europhysics Letters}} \bibinfo{volume}{23}, \bibinfo{number}{5} (\bibinfo{year}{1993}), \bibinfo{pages}{367}.
\newblock


\bibitem[Hemami et~al\mbox{.}(2021)]%
        {hemami2021phase}
\bibfield{author}{\bibinfo{person}{Mohammad Hemami}, \bibinfo{person}{Jamal~Amani Rad}, {and} \bibinfo{person}{Kourosh Parand}.} \bibinfo{year}{2021}\natexlab{}.
\newblock \showarticletitle{Phase distribution control of neural oscillator populations using local radial basis function meshfree technique with application in epileptic seizures: A numerical simulation approach}.
\newblock \bibinfo{journal}{\emph{Communications in Nonlinear Science and Numerical Simulation}}  \bibinfo{volume}{103} (\bibinfo{year}{2021}), \bibinfo{pages}{105961}.
\newblock


\bibitem[Herron et~al\mbox{.}(2017)]%
        {herron2017cortical}
\bibfield{author}{\bibinfo{person}{Jeffrey~A Herron}, \bibinfo{person}{Margaret~C Thompson}, \bibinfo{person}{Timothy Brown}, \bibinfo{person}{Howard~Jay Chizeck}, \bibinfo{person}{Jeffrey~G Ojemann}, {and} \bibinfo{person}{Andrew~L Ko}.} \bibinfo{year}{2017}\natexlab{}.
\newblock \showarticletitle{Cortical brain--computer interface for closed-loop deep brain stimulation}.
\newblock \bibinfo{journal}{\emph{IEEE Transactions on Neural Systems and Rehabilitation Engineering}} \bibinfo{volume}{25}, \bibinfo{number}{11} (\bibinfo{year}{2017}), \bibinfo{pages}{2180--2187}.
\newblock


\bibitem[Hodgkin and Huxley(1952)]%
        {hodgkin1952quantitative}
\bibfield{author}{\bibinfo{person}{Alan~L Hodgkin} {and} \bibinfo{person}{Andrew~F Huxley}.} \bibinfo{year}{1952}\natexlab{}.
\newblock \showarticletitle{A quantitative description of membrane current and its application to conduction and excitation in nerve}.
\newblock \bibinfo{journal}{\emph{The Journal of physiology}} \bibinfo{volume}{117}, \bibinfo{number}{4} (\bibinfo{year}{1952}), \bibinfo{pages}{500}.
\newblock


\bibitem[Holt and Netoff(2016)]%
        {holt2016computational}
\bibfield{author}{\bibinfo{person}{Abbey~B Holt} {and} \bibinfo{person}{Theoden~I Netoff}.} \bibinfo{year}{2016}\natexlab{}.
\newblock \showarticletitle{Computational modeling to advance deep brain stimulation for the treatment of Parkinson’s disease}.
\newblock \bibinfo{journal}{\emph{Drug Discovery Today: Disease Models}}  \bibinfo{volume}{19} (\bibinfo{year}{2016}), \bibinfo{pages}{31--36}.
\newblock


\bibitem[Holt et~al\mbox{.}(2016)]%
        {holt2016phasic}
\bibfield{author}{\bibinfo{person}{Abbey~B Holt}, \bibinfo{person}{Dan Wilson}, \bibinfo{person}{Max Shinn}, \bibinfo{person}{Jeff Moehlis}, {and} \bibinfo{person}{Theoden~I Netoff}.} \bibinfo{year}{2016}\natexlab{}.
\newblock \showarticletitle{Phasic burst stimulation: a closed-loop approach to tuning deep brain stimulation parameters for Parkinson’s disease}.
\newblock \bibinfo{journal}{\emph{PLoS computational biology}} \bibinfo{volume}{12}, \bibinfo{number}{7} (\bibinfo{year}{2016}), \bibinfo{pages}{e1005011}.
\newblock


\bibitem[Houston et~al\mbox{.}(2019)]%
        {houston2019machine}
\bibfield{author}{\bibinfo{person}{Brady Houston}, \bibinfo{person}{Margaret Thompson}, \bibinfo{person}{Andrew Ko}, {and} \bibinfo{person}{Howard Chizeck}.} \bibinfo{year}{2019}\natexlab{}.
\newblock \showarticletitle{A machine-learning approach to volitional control of a closed-loop deep brain stimulation system}.
\newblock \bibinfo{journal}{\emph{Journal of neural engineering}} \bibinfo{volume}{16}, \bibinfo{number}{1} (\bibinfo{year}{2019}), \bibinfo{pages}{016004}.
\newblock


\bibitem[Johnson et~al\mbox{.}(2016)]%
        {johnson2016closed}
\bibfield{author}{\bibinfo{person}{Luke~A Johnson}, \bibinfo{person}{Shane~D Nebeck}, \bibinfo{person}{Abirami Muralidharan}, \bibinfo{person}{Matthew~D Johnson}, \bibinfo{person}{Kenneth~B Baker}, {and} \bibinfo{person}{Jerrold~L Vitek}.} \bibinfo{year}{2016}\natexlab{}.
\newblock \showarticletitle{Closed-loop deep brain stimulation effects on parkinsonian motor symptoms in a non-human primate--is beta enough?}
\newblock \bibinfo{journal}{\emph{Brain stimulation}} \bibinfo{volume}{9}, \bibinfo{number}{6} (\bibinfo{year}{2016}), \bibinfo{pages}{892--896}.
\newblock


\bibitem[Jovanov et~al\mbox{.}(2018)]%
        {jovanov2018platform}
\bibfield{author}{\bibinfo{person}{Ilija Jovanov}, \bibinfo{person}{Michael Naumann}, \bibinfo{person}{Karthik Kumaravelu}, \bibinfo{person}{Warren~M Grill}, {and} \bibinfo{person}{Miroslav Pajic}.} \bibinfo{year}{2018}\natexlab{}.
\newblock \showarticletitle{Platform for model-based design and testing for deep brain stimulation}. In \bibinfo{booktitle}{\emph{2018 ACM/IEEE 9th International Conference on Cyber-Physical Systems (ICCPS)}}. IEEE, \bibinfo{pages}{263--274}.
\newblock


\bibitem[Kidger(2021)]%
        {kidger2021on}
\bibfield{author}{\bibinfo{person}{Patrick Kidger}.} \bibinfo{year}{2021}\natexlab{}.
\newblock \emph{\bibinfo{title}{{O}n {N}eural {D}ifferential {E}quations}}.
\newblock \bibinfo{thesistype}{Ph.\,D. Dissertation}. \bibinfo{school}{University of Oxford}.
\newblock


\bibitem[Kostrikov et~al\mbox{.}(2021)]%
        {kostrikov2021iql}
\bibfield{author}{\bibinfo{person}{Ilya Kostrikov}, \bibinfo{person}{Ashvin Nair}, {and} \bibinfo{person}{Sergey Levine}.} \bibinfo{year}{2021}\natexlab{}.
\newblock \showarticletitle{Offline Reinforcement Learning with Implicit Q-Learning}.
\newblock  (\bibinfo{year}{2021}).
\newblock


\bibitem[Krauss et~al\mbox{.}(2021)]%
        {krauss2021technology}
\bibfield{author}{\bibinfo{person}{Joachim~K Krauss}, \bibinfo{person}{Nir Lipsman}, \bibinfo{person}{Tipu Aziz}, \bibinfo{person}{Alexandre Boutet}, \bibinfo{person}{Peter Brown}, \bibinfo{person}{Jin~Woo Chang}, \bibinfo{person}{Benjamin Davidson}, \bibinfo{person}{Warren~M Grill}, \bibinfo{person}{Marwan~I Hariz}, \bibinfo{person}{Andreas Horn}, {et~al\mbox{.}}} \bibinfo{year}{2021}\natexlab{}.
\newblock \showarticletitle{Technology of deep brain stimulation: current status and future directions}.
\newblock \bibinfo{journal}{\emph{Nature Reviews Neurology}} \bibinfo{volume}{17}, \bibinfo{number}{2} (\bibinfo{year}{2021}), \bibinfo{pages}{75--87}.
\newblock


\bibitem[Krylov et~al\mbox{.}(2020a)]%
        {krylov2020reinforcement2}
\bibfield{author}{\bibinfo{person}{Dmitrii Krylov}, \bibinfo{person}{Dmitry~V Dylov}, {and} \bibinfo{person}{Michael Rosenblum}.} \bibinfo{year}{2020}\natexlab{a}.
\newblock \showarticletitle{Reinforcement learning for suppression of collective activity in oscillatory ensembles}.
\newblock \bibinfo{journal}{\emph{Chaos: An Interdisciplinary Journal of Nonlinear Science}} \bibinfo{volume}{30}, \bibinfo{number}{3} (\bibinfo{year}{2020}).
\newblock


\bibitem[Krylov et~al\mbox{.}(2020b)]%
        {krylov2020reinforcement}
\bibfield{author}{\bibinfo{person}{Dmitrii Krylov}, \bibinfo{person}{Remi Tachet}, \bibinfo{person}{Romain Laroche}, \bibinfo{person}{Michael Rosenblum}, {and} \bibinfo{person}{Dmitry~V Dylov}.} \bibinfo{year}{2020}\natexlab{b}.
\newblock \showarticletitle{Reinforcement learning framework for deep brain stimulation study}.
\newblock \bibinfo{journal}{\emph{arXiv preprint arXiv:2002.10948}} (\bibinfo{year}{2020}).
\newblock


\bibitem[Kumaravelu et~al\mbox{.}(2016)]%
        {kumaravelu2016biophysical}
\bibfield{author}{\bibinfo{person}{Karthik Kumaravelu}, \bibinfo{person}{David~T Brocker}, {and} \bibinfo{person}{Warren~M Grill}.} \bibinfo{year}{2016}\natexlab{}.
\newblock \showarticletitle{A biophysical model of the cortex-basal ganglia-thalamus network in the 6-OHDA lesioned rat model of Parkinson’s disease}.
\newblock \bibinfo{journal}{\emph{Journal of computational neuroscience}}  \bibinfo{volume}{40} (\bibinfo{year}{2016}), \bibinfo{pages}{207--229}.
\newblock


\bibitem[Kuncel and Grill(2004)]%
        {kuncel2004selection}
\bibfield{author}{\bibinfo{person}{Alexis~M Kuncel} {and} \bibinfo{person}{Warren~M Grill}.} \bibinfo{year}{2004}\natexlab{}.
\newblock \showarticletitle{Selection of stimulus parameters for deep brain stimulation}.
\newblock \bibinfo{journal}{\emph{Clinical neurophysiology}} \bibinfo{volume}{115}, \bibinfo{number}{11} (\bibinfo{year}{2004}), \bibinfo{pages}{2431--2441}.
\newblock


\bibitem[Lindahl and Kotaleski(2016)]%
        {lindahl2016untangling}
\bibfield{author}{\bibinfo{person}{Mikael Lindahl} {and} \bibinfo{person}{Jeanette~Hellgren Kotaleski}.} \bibinfo{year}{2016}\natexlab{}.
\newblock \showarticletitle{Untangling basal ganglia network dynamics and function: Role of dopamine depletion and inhibition investigated in a spiking network model}.
\newblock \bibinfo{journal}{\emph{eneuro}} \bibinfo{volume}{3}, \bibinfo{number}{6} (\bibinfo{year}{2016}).
\newblock


\bibitem[Little and Brown(2012)]%
        {little2012brain}
\bibfield{author}{\bibinfo{person}{Simon Little} {and} \bibinfo{person}{Peter Brown}.} \bibinfo{year}{2012}\natexlab{}.
\newblock \showarticletitle{Brain Stimulation in Neurology and Psychiatry}.
\newblock \bibinfo{journal}{\emph{Annals of the New York Academy of Sciences}} \bibinfo{volume}{1265}, \bibinfo{number}{1} (\bibinfo{year}{2012}), \bibinfo{pages}{9}.
\newblock


\bibitem[Little et~al\mbox{.}(2013)]%
        {little2013adaptive}
\bibfield{author}{\bibinfo{person}{Simon Little}, \bibinfo{person}{Alex Pogosyan}, \bibinfo{person}{Spencer Neal}, \bibinfo{person}{Baltazar Zavala}, \bibinfo{person}{Ludvic Zrinzo}, \bibinfo{person}{Marwan Hariz}, \bibinfo{person}{Thomas Foltynie}, \bibinfo{person}{Patricia Limousin}, \bibinfo{person}{Keyoumars Ashkan}, \bibinfo{person}{James FitzGerald}, {et~al\mbox{.}}} \bibinfo{year}{2013}\natexlab{}.
\newblock \showarticletitle{Adaptive deep brain stimulation in advanced Parkinson disease}.
\newblock \bibinfo{journal}{\emph{Annals of neurology}} \bibinfo{volume}{74}, \bibinfo{number}{3} (\bibinfo{year}{2013}), \bibinfo{pages}{449--457}.
\newblock


\bibitem[Liu et~al\mbox{.}(2016)]%
        {liu2016closed}
\bibfield{author}{\bibinfo{person}{Chen Liu}, \bibinfo{person}{Jiang Wang}, \bibinfo{person}{Huiyan Li}, \bibinfo{person}{Meili Lu}, \bibinfo{person}{Bin Deng}, \bibinfo{person}{Haitao Yu}, \bibinfo{person}{Xile Wei}, \bibinfo{person}{Chris Fietkiewicz}, {and} \bibinfo{person}{Kenneth~A Loparo}.} \bibinfo{year}{2016}\natexlab{}.
\newblock \showarticletitle{Closed-loop modulation of the pathological disorders of the basal ganglia network}.
\newblock \bibinfo{journal}{\emph{IEEE Transactions on Neural Networks and Learning Systems}} \bibinfo{volume}{28}, \bibinfo{number}{2} (\bibinfo{year}{2016}), \bibinfo{pages}{371--382}.
\newblock


\bibitem[Liu et~al\mbox{.}(2020)]%
        {liu2020neural}
\bibfield{author}{\bibinfo{person}{Chen Liu}, \bibinfo{person}{Ge Zhao}, \bibinfo{person}{Jiang Wang}, \bibinfo{person}{Hao Wu}, \bibinfo{person}{Huiyan Li}, \bibinfo{person}{Chris Fietkiewicz}, {and} \bibinfo{person}{Kenneth~A Loparo}.} \bibinfo{year}{2020}\natexlab{}.
\newblock \showarticletitle{Neural network-based closed-loop deep brain stimulation for modulation of pathological oscillation in Parkinson’s disease}.
\newblock \bibinfo{journal}{\emph{Ieee Access}}  \bibinfo{volume}{8} (\bibinfo{year}{2020}), \bibinfo{pages}{161067--161079}.
\newblock


\bibitem[Lofredi et~al\mbox{.}(2019)]%
        {lofredi2019beta}
\bibfield{author}{\bibinfo{person}{Roxanne Lofredi}, \bibinfo{person}{Huiling Tan}, \bibinfo{person}{Wolf-Julian Neumann}, \bibinfo{person}{Chien-Hung Yeh}, \bibinfo{person}{Gerd-Helge Schneider}, \bibinfo{person}{Andrea~A K{\"u}hn}, {and} \bibinfo{person}{Peter Brown}.} \bibinfo{year}{2019}\natexlab{}.
\newblock \showarticletitle{Beta bursts during continuous movements accompany the velocity decrement in Parkinson's disease patients}.
\newblock \bibinfo{journal}{\emph{Neurobiology of disease}}  \bibinfo{volume}{127} (\bibinfo{year}{2019}), \bibinfo{pages}{462--471}.
\newblock


\bibitem[Lu and Wei(2017)]%
        {lu2017desynchronizing}
\bibfield{author}{\bibinfo{person}{Meili Lu} {and} \bibinfo{person}{Xile Wei}.} \bibinfo{year}{2017}\natexlab{}.
\newblock \showarticletitle{Desynchronizing of noisy neuron networks using reinforcement learning}. In \bibinfo{booktitle}{\emph{2017 8th International IEEE/EMBS Conference on Neural Engineering (NER)}}. IEEE, \bibinfo{pages}{296--299}.
\newblock


\bibitem[Lu et~al\mbox{.}(2019)]%
        {lu2019application}
\bibfield{author}{\bibinfo{person}{Meili Lu}, \bibinfo{person}{Xile Wei}, \bibinfo{person}{Yanqiu Che}, \bibinfo{person}{Jiang Wang}, {and} \bibinfo{person}{Kenneth~A Loparo}.} \bibinfo{year}{2019}\natexlab{}.
\newblock \showarticletitle{Application of reinforcement learning to deep brain stimulation in a computational model of Parkinson’s disease}.
\newblock \bibinfo{journal}{\emph{IEEE Transactions on Neural Systems and Rehabilitation Engineering}} \bibinfo{volume}{28}, \bibinfo{number}{1} (\bibinfo{year}{2019}), \bibinfo{pages}{339--349}.
\newblock


\bibitem[Ma et~al\mbox{.}(2023)]%
        {ma2023using}
\bibfield{author}{\bibinfo{person}{Xuan Ma}, \bibinfo{person}{Fabio Rizzoglio}, \bibinfo{person}{Kevin~L Bodkin}, \bibinfo{person}{Eric Perreault}, \bibinfo{person}{Lee~E Miller}, {and} \bibinfo{person}{Ann Kennedy}.} \bibinfo{year}{2023}\natexlab{}.
\newblock \showarticletitle{Using adversarial networks to extend brain computer interface decoding accuracy over time}.
\newblock \bibinfo{journal}{\emph{elife}}  \bibinfo{volume}{12} (\bibinfo{year}{2023}), \bibinfo{pages}{e84296}.
\newblock


\bibitem[Maistrenko et~al\mbox{.}(2007)]%
        {maistrenko2007multistability}
\bibfield{author}{\bibinfo{person}{Yuri~L Maistrenko}, \bibinfo{person}{Borys Lysyansky}, \bibinfo{person}{Christian Hauptmann}, \bibinfo{person}{Oleksandr Burylko}, {and} \bibinfo{person}{Peter~A Tass}.} \bibinfo{year}{2007}\natexlab{}.
\newblock \showarticletitle{Multistability in the Kuramoto model with synaptic plasticity}.
\newblock \bibinfo{journal}{\emph{Physical Review E—Statistical, Nonlinear, and Soft Matter Physics}} \bibinfo{volume}{75}, \bibinfo{number}{6} (\bibinfo{year}{2007}), \bibinfo{pages}{066207}.
\newblock


\bibitem[Manos et~al\mbox{.}(2021)]%
        {manos2021long}
\bibfield{author}{\bibinfo{person}{Thanos Manos}, \bibinfo{person}{Sandra Diaz-Pier}, {and} \bibinfo{person}{Peter~A Tass}.} \bibinfo{year}{2021}\natexlab{}.
\newblock \showarticletitle{Long-term desynchronization by coordinated reset stimulation in a neural network model with synaptic and structural plasticity}.
\newblock \bibinfo{journal}{\emph{Frontiers in physiology}}  \bibinfo{volume}{12} (\bibinfo{year}{2021}), \bibinfo{pages}{716556}.
\newblock


\bibitem[Merk et~al\mbox{.}(2022)]%
        {merk2022machine}
\bibfield{author}{\bibinfo{person}{Timon Merk}, \bibinfo{person}{Victoria Peterson}, \bibinfo{person}{Richard K{\"o}hler}, \bibinfo{person}{Stefan Haufe}, \bibinfo{person}{R~Mark Richardson}, {and} \bibinfo{person}{Wolf-Julian Neumann}.} \bibinfo{year}{2022}\natexlab{}.
\newblock \showarticletitle{Machine learning based brain signal decoding for intelligent adaptive deep brain stimulation}.
\newblock \bibinfo{journal}{\emph{Experimental Neurology}}  \bibinfo{volume}{351} (\bibinfo{year}{2022}), \bibinfo{pages}{113993}.
\newblock


\bibitem[Nakao(2016)]%
        {nakao2016phase}
\bibfield{author}{\bibinfo{person}{Hiroya Nakao}.} \bibinfo{year}{2016}\natexlab{}.
\newblock \showarticletitle{Phase reduction approach to synchronisation of nonlinear oscillators}.
\newblock \bibinfo{journal}{\emph{Contemporary Physics}} \bibinfo{volume}{57}, \bibinfo{number}{2} (\bibinfo{year}{2016}), \bibinfo{pages}{188--214}.
\newblock


\bibitem[Neumann et~al\mbox{.}(2023)]%
        {neumann2023neurophysiological}
\bibfield{author}{\bibinfo{person}{Wolf-Julian Neumann}, \bibinfo{person}{Leon~A Steiner}, {and} \bibinfo{person}{Luka Milosevic}.} \bibinfo{year}{2023}\natexlab{}.
\newblock \showarticletitle{Neurophysiological mechanisms of deep brain stimulation across spatiotemporal resolutions}.
\newblock \bibinfo{journal}{\emph{Brain}} \bibinfo{volume}{146}, \bibinfo{number}{11} (\bibinfo{year}{2023}), \bibinfo{pages}{4456--4468}.
\newblock


\bibitem[Oliveira et~al\mbox{.}(2023)]%
        {oliveira2023machine}
\bibfield{author}{\bibinfo{person}{Andreia~M Oliveira}, \bibinfo{person}{Luis Coelho}, \bibinfo{person}{Eduardo Carvalho}, \bibinfo{person}{Manuel~J Ferreira-Pinto}, \bibinfo{person}{Rui Vaz}, {and} \bibinfo{person}{Paulo Aguiar}.} \bibinfo{year}{2023}\natexlab{}.
\newblock \showarticletitle{Machine learning for adaptive deep brain stimulation in Parkinson’s disease: closing the loop}.
\newblock \bibinfo{journal}{\emph{Journal of Neurology}} \bibinfo{volume}{270}, \bibinfo{number}{11} (\bibinfo{year}{2023}), \bibinfo{pages}{5313--5326}.
\newblock


\bibitem[Padakandla et~al\mbox{.}(2020)]%
        {padakandla2020reinforcement}
\bibfield{author}{\bibinfo{person}{Sindhu Padakandla}, \bibinfo{person}{Prabuchandran KJ}, {and} \bibinfo{person}{Shalabh Bhatnagar}.} \bibinfo{year}{2020}\natexlab{}.
\newblock \showarticletitle{Reinforcement learning algorithm for non-stationary environments}.
\newblock \bibinfo{journal}{\emph{Applied Intelligence}} \bibinfo{volume}{50}, \bibinfo{number}{11} (\bibinfo{year}{2020}), \bibinfo{pages}{3590--3606}.
\newblock


\bibitem[Pan et~al\mbox{.}(2024)]%
        {pan2024coprocessor}
\bibfield{author}{\bibinfo{person}{Michelle Pan}, \bibinfo{person}{Mariah Schrum}, \bibinfo{person}{Vivek Myers}, \bibinfo{person}{Erdem B{\i}y{\i}k}, {and} \bibinfo{person}{Anca Dragan}.} \bibinfo{year}{2024}\natexlab{}.
\newblock \showarticletitle{Coprocessor Actor Critic: A Model-Based Reinforcement Learning Approach For Adaptive Brain Stimulation}.
\newblock \bibinfo{journal}{\emph{arXiv preprint arXiv:2406.06714}} (\bibinfo{year}{2024}).
\newblock


\bibitem[Pascual et~al\mbox{.}(2006)]%
        {pascual2006computational}
\bibfield{author}{\bibinfo{person}{Alejandro Pascual}, \bibinfo{person}{Julien Modolo}, {and} \bibinfo{person}{Anne Beuter}.} \bibinfo{year}{2006}\natexlab{}.
\newblock \showarticletitle{Is a computational model useful to understand the effect of deep brain stimulation in Parkinson's disease?}
\newblock \bibinfo{journal}{\emph{Journal of integrative neuroscience}} \bibinfo{volume}{5}, \bibinfo{number}{04} (\bibinfo{year}{2006}), \bibinfo{pages}{541--559}.
\newblock


\bibitem[Paszke et~al\mbox{.}(2017)]%
        {paszke2017automatic}
\bibfield{author}{\bibinfo{person}{Adam Paszke}, \bibinfo{person}{Sam Gross}, \bibinfo{person}{Soumith Chintala}, \bibinfo{person}{Gregory Chanan}, \bibinfo{person}{Edward Yang}, \bibinfo{person}{Zachary DeVito}, \bibinfo{person}{Zeming Lin}, \bibinfo{person}{Alban Desmaison}, \bibinfo{person}{Luca Antiga}, {and} \bibinfo{person}{Adam Lerer}.} \bibinfo{year}{2017}\natexlab{}.
\newblock \showarticletitle{Automatic differentiation in PyTorch}.
\newblock  (\bibinfo{year}{2017}).
\newblock


\bibitem[Pavlides et~al\mbox{.}(2015)]%
        {pavlides2015computational}
\bibfield{author}{\bibinfo{person}{Alex Pavlides}, \bibinfo{person}{S~John Hogan}, {and} \bibinfo{person}{Rafal Bogacz}.} \bibinfo{year}{2015}\natexlab{}.
\newblock \showarticletitle{Computational models describing possible mechanisms for generation of excessive beta oscillations in Parkinson’s disease}.
\newblock \bibinfo{journal}{\emph{PLoS computational biology}} \bibinfo{volume}{11}, \bibinfo{number}{12} (\bibinfo{year}{2015}), \bibinfo{pages}{e1004609}.
\newblock


\bibitem[Popovych and Tass(2014)]%
        {popovych2014control}
\bibfield{author}{\bibinfo{person}{Oleksandr~V Popovych} {and} \bibinfo{person}{Peter~A Tass}.} \bibinfo{year}{2014}\natexlab{}.
\newblock \showarticletitle{Control of abnormal synchronization in neurological disorders}.
\newblock \bibinfo{journal}{\emph{Frontiers in neurology}}  \bibinfo{volume}{5} (\bibinfo{year}{2014}), \bibinfo{pages}{268}.
\newblock


\bibitem[Popovych and Tass(2019)]%
        {popovych2019adaptive}
\bibfield{author}{\bibinfo{person}{Oleksandr~V Popovych} {and} \bibinfo{person}{Peter~A Tass}.} \bibinfo{year}{2019}\natexlab{}.
\newblock \showarticletitle{Adaptive delivery of continuous and delayed feedback deep brain stimulation-a computational study}.
\newblock \bibinfo{journal}{\emph{Scientific Reports}} \bibinfo{volume}{9}, \bibinfo{number}{1} (\bibinfo{year}{2019}), \bibinfo{pages}{10585}.
\newblock


\bibitem[Raffin et~al\mbox{.}(2021)]%
        {stable-baselines3}
\bibfield{author}{\bibinfo{person}{Antonin Raffin}, \bibinfo{person}{Ashley Hill}, \bibinfo{person}{Adam Gleave}, \bibinfo{person}{Anssi Kanervisto}, \bibinfo{person}{Maximilian Ernestus}, {and} \bibinfo{person}{Noah Dormann}.} \bibinfo{year}{2021}\natexlab{}.
\newblock \showarticletitle{Stable-Baselines3: Reliable Reinforcement Learning Implementations}.
\newblock \bibinfo{journal}{\emph{Journal of Machine Learning Research}} \bibinfo{volume}{22}, \bibinfo{number}{268} (\bibinfo{year}{2021}), \bibinfo{pages}{1--8}.
\newblock
\urldef\tempurl%
\url{http://jmlr.org/papers/v22/20-1364.html}
\showURL{%
\tempurl}


\bibitem[Rajamani et~al\mbox{.}(2024)]%
        {rajamani2024deep}
\bibfield{author}{\bibinfo{person}{Nanditha Rajamani}, \bibinfo{person}{Helen Friedrich}, \bibinfo{person}{Konstantin Butenko}, \bibinfo{person}{Till Dembek}, \bibinfo{person}{Florian Lange}, \bibinfo{person}{Pavel Navr{\'a}til}, \bibinfo{person}{Patricia Zvarova}, \bibinfo{person}{Barbara Hollunder}, \bibinfo{person}{Rob~MA de Bie}, \bibinfo{person}{Vincent~JJ Odekerken}, {et~al\mbox{.}}} \bibinfo{year}{2024}\natexlab{}.
\newblock \showarticletitle{Deep brain stimulation of symptom-specific networks in Parkinson’s disease}.
\newblock \bibinfo{journal}{\emph{Nature Communications}} \bibinfo{volume}{15}, \bibinfo{number}{1} (\bibinfo{year}{2024}), \bibinfo{pages}{4662}.
\newblock


\bibitem[Ranieri et~al\mbox{.}(2021)]%
        {ranieri2021data}
\bibfield{author}{\bibinfo{person}{Caetano~M Ranieri}, \bibinfo{person}{Jhielson~M Pimentel}, \bibinfo{person}{Marcelo~R Romano}, \bibinfo{person}{Leonardo~A Elias}, \bibinfo{person}{Roseli~AF Romero}, \bibinfo{person}{Michael~A Lones}, \bibinfo{person}{Mariana~FP Araujo}, \bibinfo{person}{Patricia~A Vargas}, {and} \bibinfo{person}{Renan~C Moioli}.} \bibinfo{year}{2021}\natexlab{}.
\newblock \showarticletitle{A data-driven biophysical computational model of Parkinson’s disease based on marmoset monkeys}.
\newblock \bibinfo{journal}{\emph{IEEE access}}  \bibinfo{volume}{9} (\bibinfo{year}{2021}), \bibinfo{pages}{122548--122567}.
\newblock


\bibitem[Romano et~al\mbox{.}(2020)]%
        {romano2020evaluation}
\bibfield{author}{\bibinfo{person}{Marcelo~R Romano}, \bibinfo{person}{Renan~C Moioli}, {and} \bibinfo{person}{Leonardo~A Elias}.} \bibinfo{year}{2020}\natexlab{}.
\newblock \showarticletitle{Evaluation of frequency-dependent effects of deep brain stimulation in a cortex-basal ganglia-thalamus network model of Parkinson’s disease}. In \bibinfo{booktitle}{\emph{2020 42nd Annual International Conference of the IEEE Engineering in Medicine \& Biology Society (EMBC)}}. IEEE, \bibinfo{pages}{3638--3641}.
\newblock


\bibitem[Rosa et~al\mbox{.}(2012)]%
        {rosa2012neurophysiology}
\bibfield{author}{\bibinfo{person}{Manuela Rosa}, \bibinfo{person}{Gaia Giannicola}, \bibinfo{person}{Sara Marceglia}, \bibinfo{person}{Manuela Fumagalli}, \bibinfo{person}{Sergio Barbieri}, {and} \bibinfo{person}{Alberto Priori}.} \bibinfo{year}{2012}\natexlab{}.
\newblock \showarticletitle{Neurophysiology of deep brain stimulation}.
\newblock \bibinfo{journal}{\emph{International review of neurobiology}}  \bibinfo{volume}{107} (\bibinfo{year}{2012}), \bibinfo{pages}{23--55}.
\newblock


\bibitem[Rosenbaum et~al\mbox{.}(2014)]%
        {rosenbaum2014axonal}
\bibfield{author}{\bibinfo{person}{Robert Rosenbaum}, \bibinfo{person}{Andrew Zimnik}, \bibinfo{person}{Fang Zheng}, \bibinfo{person}{Robert~S Turner}, \bibinfo{person}{Christian Alzheimer}, \bibinfo{person}{Brent Doiron}, {and} \bibinfo{person}{Jonathan~E Rubin}.} \bibinfo{year}{2014}\natexlab{}.
\newblock \showarticletitle{Axonal and synaptic failure suppress the transfer of firing rate oscillations, synchrony and information during high frequency deep brain stimulation}.
\newblock \bibinfo{journal}{\emph{Neurobiology of disease}}  \bibinfo{volume}{62} (\bibinfo{year}{2014}), \bibinfo{pages}{86--99}.
\newblock


\bibitem[Rubin(2017)]%
        {rubin2017computational}
\bibfield{author}{\bibinfo{person}{Jonathan~E Rubin}.} \bibinfo{year}{2017}\natexlab{}.
\newblock \showarticletitle{Computational models of basal ganglia dysfunction: the dynamics is in the details}.
\newblock \bibinfo{journal}{\emph{Current opinion in neurobiology}}  \bibinfo{volume}{46} (\bibinfo{year}{2017}), \bibinfo{pages}{127--135}.
\newblock


\bibitem[Rule et~al\mbox{.}(2019)]%
        {rule2019causes}
\bibfield{author}{\bibinfo{person}{Michael~E Rule}, \bibinfo{person}{Timothy O’Leary}, {and} \bibinfo{person}{Christopher~D Harvey}.} \bibinfo{year}{2019}\natexlab{}.
\newblock \showarticletitle{Causes and consequences of representational drift}.
\newblock \bibinfo{journal}{\emph{Current opinion in neurobiology}}  \bibinfo{volume}{58} (\bibinfo{year}{2019}), \bibinfo{pages}{141--147}.
\newblock


\bibitem[Schiff(2011)]%
        {schiff2011neural}
\bibfield{author}{\bibinfo{person}{Steven~J Schiff}.} \bibinfo{year}{2011}\natexlab{}.
\newblock \bibinfo{booktitle}{\emph{Neural control engineering: the emerging intersection between control theory and neuroscience}}.
\newblock \bibinfo{publisher}{MIT Press}.
\newblock


\bibitem[Schmidt et~al\mbox{.}(2014)]%
        {schmidt2014dynamics}
\bibfield{author}{\bibinfo{person}{Helmut Schmidt}, \bibinfo{person}{George Petkov}, \bibinfo{person}{Mark~P Richardson}, {and} \bibinfo{person}{John~R Terry}.} \bibinfo{year}{2014}\natexlab{}.
\newblock \showarticletitle{Dynamics on networks: the role of local dynamics and global networks on the emergence of hypersynchronous neural activity}.
\newblock \bibinfo{journal}{\emph{PLoS computational biology}} \bibinfo{volume}{10}, \bibinfo{number}{11} (\bibinfo{year}{2014}), \bibinfo{pages}{e1003947}.
\newblock


\bibitem[Schmidt et~al\mbox{.}(2024)]%
        {schmidt2024home}
\bibfield{author}{\bibinfo{person}{Stephen~L Schmidt}, \bibinfo{person}{Afsana~H Chowdhury}, \bibinfo{person}{Kyle~T Mitchell}, \bibinfo{person}{Jennifer~J Peters}, \bibinfo{person}{Qitong Gao}, \bibinfo{person}{Hui-Jie Lee}, \bibinfo{person}{Katherine Genty}, \bibinfo{person}{Shein-Chung Chow}, \bibinfo{person}{Warren~M Grill}, \bibinfo{person}{Miroslav Pajic}, {et~al\mbox{.}}} \bibinfo{year}{2024}\natexlab{}.
\newblock \showarticletitle{At home adaptive dual target deep brain stimulation in Parkinson’s disease with proportional control}.
\newblock \bibinfo{journal}{\emph{Brain}} \bibinfo{volume}{147}, \bibinfo{number}{3} (\bibinfo{year}{2024}), \bibinfo{pages}{911--922}.
\newblock


\bibitem[Schroll et~al\mbox{.}(2014)]%
        {schroll2014dysfunctional}
\bibfield{author}{\bibinfo{person}{Henning Schroll}, \bibinfo{person}{Julien Vitay}, {and} \bibinfo{person}{Fred~H Hamker}.} \bibinfo{year}{2014}\natexlab{}.
\newblock \showarticletitle{Dysfunctional and compensatory synaptic plasticity in P arkinson's disease}.
\newblock \bibinfo{journal}{\emph{European Journal of Neuroscience}} \bibinfo{volume}{39}, \bibinfo{number}{4} (\bibinfo{year}{2014}), \bibinfo{pages}{688--702}.
\newblock


\bibitem[Schulman et~al\mbox{.}(2017)]%
        {schulman2017proximal}
\bibfield{author}{\bibinfo{person}{John Schulman}, \bibinfo{person}{Filip Wolski}, \bibinfo{person}{Prafulla Dhariwal}, \bibinfo{person}{Alec Radford}, {and} \bibinfo{person}{Oleg Klimov}.} \bibinfo{year}{2017}\natexlab{}.
\newblock \showarticletitle{Proximal policy optimization algorithms}.
\newblock \bibinfo{journal}{\emph{arXiv preprint arXiv:1707.06347}} (\bibinfo{year}{2017}).
\newblock


\bibitem[Shams et~al\mbox{.}(2022)]%
        {shams2022optimal}
\bibfield{author}{\bibinfo{person}{Siavash Shams}, \bibinfo{person}{Sana Motallebi}, {and} \bibinfo{person}{Mohammad~Javad Yazdanpanah}.} \bibinfo{year}{2022}\natexlab{}.
\newblock \showarticletitle{An Optimal Data-Driven Method for Controlling Epileptic Seizures}. In \bibinfo{booktitle}{\emph{2022 29th National and 7th International Iranian Conference on Biomedical Engineering (ICBME)}}. IEEE, \bibinfo{pages}{250--255}.
\newblock


\bibitem[Shukla et~al\mbox{.}(2014)]%
        {shukla2014modeling}
\bibfield{author}{\bibinfo{person}{Pitamber Shukla}, \bibinfo{person}{Ishita Basu}, \bibinfo{person}{Daniela Tuninetti}, \bibinfo{person}{Daniel Graupe}, {and} \bibinfo{person}{Konstantin~V Slavin}.} \bibinfo{year}{2014}\natexlab{}.
\newblock \showarticletitle{On modeling the neuronal activity in movement disorder patients by using the Ornstein Uhlenbeck Process}. In \bibinfo{booktitle}{\emph{2014 36th Annual International Conference of the IEEE Engineering in Medicine and Biology Society}}. IEEE, \bibinfo{pages}{2609--2612}.
\newblock


\bibitem[Silver et~al\mbox{.}(2014)]%
        {silver2014deterministic}
\bibfield{author}{\bibinfo{person}{David Silver}, \bibinfo{person}{Guy Lever}, \bibinfo{person}{Nicolas Heess}, \bibinfo{person}{Thomas Degris}, \bibinfo{person}{Daan Wierstra}, {and} \bibinfo{person}{Martin Riedmiller}.} \bibinfo{year}{2014}\natexlab{}.
\newblock \showarticletitle{Deterministic policy gradient algorithms}. In \bibinfo{booktitle}{\emph{International conference on machine learning}}. Pmlr, \bibinfo{pages}{387--395}.
\newblock


\bibitem[Smeal et~al\mbox{.}(2010)]%
        {smeal2010phase}
\bibfield{author}{\bibinfo{person}{Roy~M Smeal}, \bibinfo{person}{G~Bard Ermentrout}, {and} \bibinfo{person}{John~A White}.} \bibinfo{year}{2010}\natexlab{}.
\newblock \showarticletitle{Phase-response curves and synchronized neural networks}.
\newblock \bibinfo{journal}{\emph{Philosophical Transactions of the Royal Society B: Biological Sciences}} \bibinfo{volume}{365}, \bibinfo{number}{1551} (\bibinfo{year}{2010}), \bibinfo{pages}{2407--2422}.
\newblock


\bibitem[So et~al\mbox{.}(2012)]%
        {so2012relative}
\bibfield{author}{\bibinfo{person}{Rosa~Q So}, \bibinfo{person}{Alexander~R Kent}, {and} \bibinfo{person}{Warren~M Grill}.} \bibinfo{year}{2012}\natexlab{}.
\newblock \showarticletitle{Relative contributions of local cell and passing fiber activation and silencing to changes in thalamic fidelity during deep brain stimulation and lesioning: a computational modeling study}.
\newblock \bibinfo{journal}{\emph{Journal of computational neuroscience}} \bibinfo{volume}{32}, \bibinfo{number}{3} (\bibinfo{year}{2012}), \bibinfo{pages}{499--519}.
\newblock


\bibitem[Spiliotis et~al\mbox{.}(2022)]%
        {spiliotis2022deep}
\bibfield{author}{\bibinfo{person}{Konstantinos Spiliotis}, \bibinfo{person}{Jens Starke}, \bibinfo{person}{Denise Franz}, \bibinfo{person}{Angelika Richter}, {and} \bibinfo{person}{R{\"u}diger K{\"o}hling}.} \bibinfo{year}{2022}\natexlab{}.
\newblock \showarticletitle{Deep brain stimulation for movement disorder treatment: exploring frequency-dependent efficacy in a computational network model}.
\newblock \bibinfo{journal}{\emph{Biological Cybernetics}} \bibinfo{volume}{116}, \bibinfo{number}{1} (\bibinfo{year}{2022}), \bibinfo{pages}{93--116}.
\newblock


\bibitem[Su et~al\mbox{.}(2021)]%
        {su2021model}
\bibfield{author}{\bibinfo{person}{Fei Su}, \bibinfo{person}{Min Chen}, \bibinfo{person}{Linlu Zu}, \bibinfo{person}{Shanshan Li}, {and} \bibinfo{person}{Huiyan Li}.} \bibinfo{year}{2021}\natexlab{}.
\newblock \showarticletitle{Model-based closed-loop suppression of parkinsonian beta band oscillations through origin analysis}.
\newblock \bibinfo{journal}{\emph{IEEE Transactions on Neural Systems and Rehabilitation Engineering}}  \bibinfo{volume}{29} (\bibinfo{year}{2021}), \bibinfo{pages}{450--457}.
\newblock


\bibitem[Su et~al\mbox{.}(2023)]%
        {su2023closed}
\bibfield{author}{\bibinfo{person}{Fei Su}, \bibinfo{person}{Hong Wang}, \bibinfo{person}{Linlu Zu}, {and} \bibinfo{person}{Yan Chen}.} \bibinfo{year}{2023}\natexlab{}.
\newblock \showarticletitle{Closed-loop modulation of model parkinsonian beta oscillations based on CAR-fuzzy control algorithm}.
\newblock \bibinfo{journal}{\emph{Cognitive Neurodynamics}} \bibinfo{volume}{17}, \bibinfo{number}{5} (\bibinfo{year}{2023}), \bibinfo{pages}{1185--1199}.
\newblock


\bibitem[Sui et~al\mbox{.}(2022)]%
        {sui2022deep}
\bibfield{author}{\bibinfo{person}{Yanan Sui}, \bibinfo{person}{Huiling Yu}, \bibinfo{person}{Chen Zhang}, \bibinfo{person}{Yue Chen}, \bibinfo{person}{Changqing Jiang}, {and} \bibinfo{person}{Luming Li}.} \bibinfo{year}{2022}\natexlab{}.
\newblock \showarticletitle{Deep brain--machine interfaces: sensing and modulating the human deep brain}.
\newblock \bibinfo{journal}{\emph{National Science Review}} \bibinfo{volume}{9}, \bibinfo{number}{10} (\bibinfo{year}{2022}), \bibinfo{pages}{nwac212}.
\newblock


\bibitem[Swann et~al\mbox{.}(2018)]%
        {swann2018adaptive}
\bibfield{author}{\bibinfo{person}{Nicole~C Swann}, \bibinfo{person}{Coralie De~Hemptinne}, \bibinfo{person}{Margaret~C Thompson}, \bibinfo{person}{Svjetlana Miocinovic}, \bibinfo{person}{Andrew~M Miller}, \bibinfo{person}{Jill~L Ostrem}, \bibinfo{person}{Howard~J Chizeck}, \bibinfo{person}{Philip~A Starr}, {et~al\mbox{.}}} \bibinfo{year}{2018}\natexlab{}.
\newblock \showarticletitle{Adaptive deep brain stimulation for Parkinson’s disease using motor cortex sensing}.
\newblock \bibinfo{journal}{\emph{Journal of neural engineering}} \bibinfo{volume}{15}, \bibinfo{number}{4} (\bibinfo{year}{2018}), \bibinfo{pages}{046006}.
\newblock


\bibitem[Terman et~al\mbox{.}(2002)]%
        {terman2002activity}
\bibfield{author}{\bibinfo{person}{David Terman}, \bibinfo{person}{Jonathan~E Rubin}, \bibinfo{person}{AC Yew}, {and} \bibinfo{person}{CJ Wilson}.} \bibinfo{year}{2002}\natexlab{}.
\newblock \showarticletitle{Activity patterns in a model for the subthalamopallidal network of the basal ganglia}.
\newblock \bibinfo{journal}{\emph{Journal of Neuroscience}} \bibinfo{volume}{22}, \bibinfo{number}{7} (\bibinfo{year}{2002}), \bibinfo{pages}{2963--2976}.
\newblock


\bibitem[Tinkhauser et~al\mbox{.}(2017a)]%
        {tinkhauser2017modulatory}
\bibfield{author}{\bibinfo{person}{Gerd Tinkhauser}, \bibinfo{person}{Alek Pogosyan}, \bibinfo{person}{Simon Little}, \bibinfo{person}{Martijn Beudel}, \bibinfo{person}{Damian~M Herz}, \bibinfo{person}{Huiling Tan}, {and} \bibinfo{person}{Peter Brown}.} \bibinfo{year}{2017}\natexlab{a}.
\newblock \showarticletitle{The modulatory effect of adaptive deep brain stimulation on beta bursts in Parkinson’s disease}.
\newblock \bibinfo{journal}{\emph{Brain}} \bibinfo{volume}{140}, \bibinfo{number}{4} (\bibinfo{year}{2017}), \bibinfo{pages}{1053--1067}.
\newblock


\bibitem[Tinkhauser et~al\mbox{.}(2017b)]%
        {tinkhauser2017beta}
\bibfield{author}{\bibinfo{person}{Gerd Tinkhauser}, \bibinfo{person}{Alek Pogosyan}, \bibinfo{person}{Huiling Tan}, \bibinfo{person}{Damian~M Herz}, \bibinfo{person}{Andrea~A K{\"u}hn}, {and} \bibinfo{person}{Peter Brown}.} \bibinfo{year}{2017}\natexlab{b}.
\newblock \showarticletitle{Beta burst dynamics in Parkinson’s disease OFF and ON dopaminergic medication}.
\newblock \bibinfo{journal}{\emph{Brain}} \bibinfo{volume}{140}, \bibinfo{number}{11} (\bibinfo{year}{2017}), \bibinfo{pages}{2968--2981}.
\newblock


\bibitem[van Albada and Robinson(2009)]%
        {van2009mean}
\bibfield{author}{\bibinfo{person}{Sacha~Jennifer van Albada} {and} \bibinfo{person}{Peter~A Robinson}.} \bibinfo{year}{2009}\natexlab{}.
\newblock \showarticletitle{Mean-field modeling of the basal ganglia-thalamocortical system. I: Firing rates in healthy and parkinsonian states}.
\newblock \bibinfo{journal}{\emph{Journal of theoretical biology}} \bibinfo{volume}{257}, \bibinfo{number}{4} (\bibinfo{year}{2009}), \bibinfo{pages}{642--663}.
\newblock


\bibitem[Van~Hartevelt et~al\mbox{.}(2014)]%
        {van2014neural}
\bibfield{author}{\bibinfo{person}{Tim~J Van~Hartevelt}, \bibinfo{person}{Joana Cabral}, \bibinfo{person}{Gustavo Deco}, \bibinfo{person}{Arne M{\o}ller}, \bibinfo{person}{Alexander~L Green}, \bibinfo{person}{Tipu~Z Aziz}, {and} \bibinfo{person}{Morten~L Kringelbach}.} \bibinfo{year}{2014}\natexlab{}.
\newblock \showarticletitle{Neural plasticity in human brain connectivity: the effects of long term deep brain stimulation of the subthalamic nucleus in Parkinson’s disease}.
\newblock \bibinfo{journal}{\emph{PloS one}} \bibinfo{volume}{9}, \bibinfo{number}{1} (\bibinfo{year}{2014}), \bibinfo{pages}{e86496}.
\newblock


\bibitem[Vinding et~al\mbox{.}(2020)]%
        {vinding2020reduction}
\bibfield{author}{\bibinfo{person}{Mikkel~C Vinding}, \bibinfo{person}{Panagiota Tsitsi}, \bibinfo{person}{Josefine Waldthaler}, \bibinfo{person}{Robert Oostenveld}, \bibinfo{person}{Martin Ingvar}, \bibinfo{person}{Per Svenningsson}, {and} \bibinfo{person}{Daniel Lundqvist}.} \bibinfo{year}{2020}\natexlab{}.
\newblock \showarticletitle{Reduction of spontaneous cortical beta bursts in Parkinson’s disease is linked to symptom severity}.
\newblock \bibinfo{journal}{\emph{Brain Communications}} \bibinfo{volume}{2}, \bibinfo{number}{1} (\bibinfo{year}{2020}), \bibinfo{pages}{fcaa052}.
\newblock


\bibitem[Wang et~al\mbox{.}(2018)]%
        {wang2018suppressing}
\bibfield{author}{\bibinfo{person}{JiaYi Wang}, \bibinfo{person}{XiaoLi Yang}, {and} \bibinfo{person}{ZhongKui Sun}.} \bibinfo{year}{2018}\natexlab{}.
\newblock \showarticletitle{Suppressing bursting synchronization in a modular neuronal network with synaptic plasticity}.
\newblock \bibinfo{journal}{\emph{Cognitive neurodynamics}}  \bibinfo{volume}{12} (\bibinfo{year}{2018}), \bibinfo{pages}{625--636}.
\newblock


\bibitem[Wang et~al\mbox{.}(2022)]%
        {wang2022adaptive}
\bibfield{author}{\bibinfo{person}{Kuanchuan Wang}, \bibinfo{person}{Jiang Wang}, \bibinfo{person}{Yulin Zhu}, \bibinfo{person}{Huiyan Li}, \bibinfo{person}{Chen Liu}, \bibinfo{person}{Chris Fietkiewicz}, {and} \bibinfo{person}{Kenneth~A Loparo}.} \bibinfo{year}{2022}\natexlab{}.
\newblock \showarticletitle{Adaptive closed-loop control strategy inhibiting pathological basal ganglia oscillations}.
\newblock \bibinfo{journal}{\emph{Biomedical Signal Processing and Control}}  \bibinfo{volume}{77} (\bibinfo{year}{2022}), \bibinfo{pages}{103776}.
\newblock


\bibitem[Watts et~al\mbox{.}(2020)]%
        {watts2020machine}
\bibfield{author}{\bibinfo{person}{Jeremy Watts}, \bibinfo{person}{Anahita Khojandi}, \bibinfo{person}{Oleg Shylo}, {and} \bibinfo{person}{Ritesh~A Ramdhani}.} \bibinfo{year}{2020}\natexlab{}.
\newblock \showarticletitle{Machine learning’s application in deep brain stimulation for Parkinson’s disease: A review}.
\newblock \bibinfo{journal}{\emph{Brain Sciences}} \bibinfo{volume}{10}, \bibinfo{number}{11} (\bibinfo{year}{2020}), \bibinfo{pages}{809}.
\newblock


\bibitem[Wei et~al\mbox{.}(2022)]%
        {wei2022parkinsonian}
\bibfield{author}{\bibinfo{person}{Xi-Le Wei}, \bibinfo{person}{Yu-Lin Bai}, \bibinfo{person}{Jiang Wang}, \bibinfo{person}{Si-Yuan Chang}, {and} \bibinfo{person}{Chen Liu}.} \bibinfo{year}{2022}\natexlab{}.
\newblock \showarticletitle{Parkinsonian oscillations and their suppression by closed-loop deep brain stimulation based on fuzzy concept}.
\newblock \bibinfo{journal}{\emph{Chinese Physics B}} \bibinfo{volume}{31}, \bibinfo{number}{12} (\bibinfo{year}{2022}), \bibinfo{pages}{128701}.
\newblock


\bibitem[Widge(2024)]%
        {widge2024closing}
\bibfield{author}{\bibinfo{person}{Alik~S Widge}.} \bibinfo{year}{2024}\natexlab{}.
\newblock \showarticletitle{Closing the loop in psychiatric deep brain stimulation: physiology, psychometrics, and plasticity}.
\newblock \bibinfo{journal}{\emph{Neuropsychopharmacology}} \bibinfo{volume}{49}, \bibinfo{number}{1} (\bibinfo{year}{2024}), \bibinfo{pages}{138--149}.
\newblock


\bibitem[Wilkins et~al\mbox{.}(2023)]%
        {wilkins2023unraveling}
\bibfield{author}{\bibinfo{person}{Kevin~B Wilkins}, \bibinfo{person}{Jillian~A Melbourne}, \bibinfo{person}{Pranav Akella}, {and} \bibinfo{person}{Helen~M Bronte-Stewart}.} \bibinfo{year}{2023}\natexlab{}.
\newblock \showarticletitle{Unraveling the complexities of programming neural adaptive deep brain stimulation in Parkinson’s disease}.
\newblock \bibinfo{journal}{\emph{Frontiers in Human Neuroscience}}  \bibinfo{volume}{17} (\bibinfo{year}{2023}).
\newblock


\bibitem[Wilson et~al\mbox{.}(2011)]%
        {wilson2011chaotic}
\bibfield{author}{\bibinfo{person}{Charles~J Wilson}, \bibinfo{person}{Bryce Beverlin}, {and} \bibinfo{person}{Theoden Netoff}.} \bibinfo{year}{2011}\natexlab{}.
\newblock \showarticletitle{Chaotic desynchronization as the therapeutic mechanism of deep brain stimulation}.
\newblock \bibinfo{journal}{\emph{Frontiers in systems neuroscience}}  \bibinfo{volume}{5} (\bibinfo{year}{2011}), \bibinfo{pages}{50}.
\newblock


\bibitem[Yang and Che(2022)]%
        {yang2022beta}
\bibfield{author}{\bibinfo{person}{Jingyu Yang} {and} \bibinfo{person}{Yanqiu Che}.} \bibinfo{year}{2022}\natexlab{}.
\newblock \showarticletitle{Beta Oscillations Suppression in a Population Model of Parkinson’s Disease by Linear Delayed Feedback Control}. In \bibinfo{booktitle}{\emph{2022 37th Youth Academic Annual Conference of Chinese Association of Automation (YAC)}}. IEEE, \bibinfo{pages}{974--978}.
\newblock


\bibitem[Yang et~al\mbox{.}(2023)]%
        {yang2023adaptive}
\bibfield{author}{\bibinfo{person}{Ming Yang}, \bibinfo{person}{Jiang Wang}, \bibinfo{person}{Shanshan Li}, \bibinfo{person}{Kuanchuan Wang}, \bibinfo{person}{Wei Yue}, {and} \bibinfo{person}{Chen Liu}.} \bibinfo{year}{2023}\natexlab{}.
\newblock \showarticletitle{Adaptive closed-loop paradigm of electrophysiology for neuron models}.
\newblock \bibinfo{journal}{\emph{Neural Networks}}  \bibinfo{volume}{165} (\bibinfo{year}{2023}), \bibinfo{pages}{406--419}.
\newblock


\bibitem[Yin et~al\mbox{.}(2021)]%
        {yin2021local}
\bibfield{author}{\bibinfo{person}{Zixiao Yin}, \bibinfo{person}{Guanyu Zhu}, \bibinfo{person}{Baotian Zhao}, \bibinfo{person}{Yutong Bai}, \bibinfo{person}{Yin Jiang}, \bibinfo{person}{Wolf-Julian Neumann}, \bibinfo{person}{Andrea~A K{\"u}hn}, {and} \bibinfo{person}{Jianguo Zhang}.} \bibinfo{year}{2021}\natexlab{}.
\newblock \showarticletitle{Local field potentials in Parkinson's disease: a frequency-based review}.
\newblock \bibinfo{journal}{\emph{Neurobiology of Disease}}  \bibinfo{volume}{155} (\bibinfo{year}{2021}), \bibinfo{pages}{105372}.
\newblock


\bibitem[Yu et~al\mbox{.}(2021)]%
        {yu2021parkinsonism}
\bibfield{author}{\bibinfo{person}{Ying Yu}, \bibinfo{person}{David~Escobar Sanabria}, \bibinfo{person}{Jing Wang}, \bibinfo{person}{Claudia~M Hendrix}, \bibinfo{person}{Jianyu Zhang}, \bibinfo{person}{Shane~D Nebeck}, \bibinfo{person}{Alexia~M Amundson}, \bibinfo{person}{Zachary~B Busby}, \bibinfo{person}{Devyn~L Bauer}, \bibinfo{person}{Matthew~D Johnson}, {et~al\mbox{.}}} \bibinfo{year}{2021}\natexlab{}.
\newblock \showarticletitle{Parkinsonism alters Beta burst dynamics across the basal ganglia--motor cortical network}.
\newblock \bibinfo{journal}{\emph{Journal of Neuroscience}} \bibinfo{volume}{41}, \bibinfo{number}{10} (\bibinfo{year}{2021}), \bibinfo{pages}{2274--2286}.
\newblock


\bibitem[Yu et~al\mbox{.}(2020)]%
        {yu2020review}
\bibfield{author}{\bibinfo{person}{Ying Yu}, \bibinfo{person}{Xiaomin Wang}, \bibinfo{person}{Qishao Wang}, {and} \bibinfo{person}{Qingyun Wang}.} \bibinfo{year}{2020}\natexlab{}.
\newblock \showarticletitle{A review of computational modeling and deep brain stimulation: applications to Parkinson’s disease}.
\newblock \bibinfo{journal}{\emph{Applied mathematics and mechanics}} \bibinfo{volume}{41}, \bibinfo{number}{12} (\bibinfo{year}{2020}), \bibinfo{pages}{1747--1768}.
\newblock


\bibitem[Zavaleta-Viveros et~al\mbox{.}(2023)]%
        {zavaleta2023modification}
\bibfield{author}{\bibinfo{person}{Jos{\'e}~Alfredo Zavaleta-Viveros}, \bibinfo{person}{Porfirio Toledo}, \bibinfo{person}{Martha~Lorena Avenda{\~n}o-Garrido}, \bibinfo{person}{Jes{\'u}s~Enrique Escalante-Mart{\'\i}nez}, \bibinfo{person}{Mar{\'\i}a-Leonor L{\'o}pez-Meraz}, {and} \bibinfo{person}{Karen~Paola Ramos-Riera}.} \bibinfo{year}{2023}\natexlab{}.
\newblock \showarticletitle{A modification to the Kuramoto model to simulate epileptic seizures as synchronization}.
\newblock \bibinfo{journal}{\emph{Journal of Mathematical Biology}} \bibinfo{volume}{87}, \bibinfo{number}{1} (\bibinfo{year}{2023}), \bibinfo{pages}{9}.
\newblock


\bibitem[Zhang et~al\mbox{.}({[n.\,d.]})]%
        {zhangtest}
\bibfield{author}{\bibinfo{person}{Zhen-Yu Zhang}, \bibinfo{person}{Zhiyu Xie}, \bibinfo{person}{Huaxiu Yao}, {and} \bibinfo{person}{Masashi Sugiyama}.} \bibinfo{year}{[n.\,d.]}\natexlab{}.
\newblock \showarticletitle{Test-time Adaptation in Non-stationary Environments via Adaptive Representation Alignment}. In \bibinfo{booktitle}{\emph{The Thirty-eighth Annual Conference on Neural Information Processing Systems}}.
\newblock


\bibitem[Zhao et~al\mbox{.}(2021)]%
        {zhao2021optimization}
\bibfield{author}{\bibinfo{person}{Zixi Zhao}, \bibinfo{person}{Aliya Ahmadi}, \bibinfo{person}{Caleb Hoover}, \bibinfo{person}{Logan Grado}, \bibinfo{person}{Nicholas Peterson}, \bibinfo{person}{Xinran Wang}, \bibinfo{person}{David Freeman}, \bibinfo{person}{Thomas Murray}, \bibinfo{person}{Andrew Lamperski}, \bibinfo{person}{David Darrow}, {et~al\mbox{.}}} \bibinfo{year}{2021}\natexlab{}.
\newblock \showarticletitle{Optimization of spinal cord stimulation using bayesian preference learning and its validation}.
\newblock \bibinfo{journal}{\emph{IEEE Transactions on Neural Systems and Rehabilitation Engineering}}  \bibinfo{volume}{29} (\bibinfo{year}{2021}), \bibinfo{pages}{1987--1997}.
\newblock


\bibitem[Zhou et~al\mbox{.}(2017)]%
        {zhou2017adaptive}
\bibfield{author}{\bibinfo{person}{Shijie Zhou}, \bibinfo{person}{Peng Ji}, \bibinfo{person}{Qing Zhou}, \bibinfo{person}{Jianfeng Feng}, \bibinfo{person}{J{\"u}rgen Kurths}, {and} \bibinfo{person}{Wei Lin}.} \bibinfo{year}{2017}\natexlab{}.
\newblock \showarticletitle{Adaptive elimination of synchronization in coupled oscillator}.
\newblock \bibinfo{journal}{\emph{New Journal of Physics}} \bibinfo{volume}{19}, \bibinfo{number}{8} (\bibinfo{year}{2017}), \bibinfo{pages}{083004}.
\newblock


\bibitem[Zhu et~al\mbox{.}(2021)]%
        {zhu2021adaptive}
\bibfield{author}{\bibinfo{person}{Yulin Zhu}, \bibinfo{person}{Jiang Wang}, \bibinfo{person}{Huiyan Li}, \bibinfo{person}{Chen Liu}, {and} \bibinfo{person}{Warren~M Grill}.} \bibinfo{year}{2021}\natexlab{}.
\newblock \showarticletitle{Adaptive parameter modulation of deep brain stimulation based on improved supervisory algorithm}.
\newblock \bibinfo{journal}{\emph{Frontiers in neuroscience}}  \bibinfo{volume}{15} (\bibinfo{year}{2021}), \bibinfo{pages}{750806}.
\newblock


\end{thebibliography}

\appendix

\section{Appendix} \label{apdx}

\setcounter{figure}{0}
\setcounter{table}{0}
\renewcommand{\thefigure}{A\arabic{figure}}
\renewcommand{\thetable}{A\arabic{table}}

\subsection{Code accessibility}
The code will be open-access upon publication.

\subsection{Limitations}

To achieve better performance metrics, further training and hyperparameter optimization are necessary for online agents. Additionally, larger synthetic datasets would enhance the performance of the offline RL methods used. Careful selection of reward functions is crucial, as they significantly influence RL behavior, as shown in Table \ref{tab:vertical_results}, which compares the outcomes of different reward functions. For PI and PID controllers, constructing a separate error function is required, as no ablation study was conducted to investigate their underperformance in our environment.

Experiments with different stimulation pulse shapes, such as the widely used charge-balanced pulse, were not conducted. However, our environment is adaptable enough to support the implementation of such pulses if needed.

Previous modeling studies have demonstrated that tissue volume activation depends on various factors, including electrode interface voltage drop, electrode and tissue capacitance, tissue encapsulation, and tissue heterogeneity (anisotropy) \cite{rosa2012neurophysiology}. These factors were not fully incorporated into our model.

Additionally, not all features of our model were presented here. For example, train and evaluate RL models for directional stimulation represents a distinct challenge significantly extending the action space that was beyond the scope of this study.

Alternative metrics could also be considered for evaluating aDBS algorithm performance, depending on how pathological neural activity is defined.

\subsection{Extended justification for selected PD features} \label{app: features_justif}

\subsubsection{Bandwidth features}
Beta band activity is the most commonly used biomarker for adaptive deep brain stimulation (aDBS) in Parkinson's Disease (PD), as suppressing beta oscillations (24.5–28.5 Hz) has been shown to alleviate bradykinesia and rigidity \cite{binns2024shared, grado2018bayesian}. Consequently, many studies focus exclusively on beta power \cite{yu2020review}. However, this narrow focus oversimplifies PD dynamics, potentially compromising aDBS testing and development. Increasing evidence highlights the importance of other frequency features and their combinations in accurately modeling PD symptoms \cite{yin2021local, wilkins2023unraveling}. For example, beta activity can be further divided into low (13–20 Hz) and high (21–35 Hz) sub-bands, each associated with different PD symptoms and suppression outcomes \cite{yin2021local}. In real PD patients, local field potential power spectral density (PSD) exhibits broad beta peaks, with distinct low and high beta band peaks that differentiate them from healthy controls (Figure \ref{fig:intro}E) \cite{yin2021local}.

Beyond beta oscillations, other neurophysiological biomarkers have been proposed, including phase-amplitude coupling (PAC), evoked compound action potentials (ECAPs), and the ratio between low-field oscillations (LF) and beta power \cite{yin2021local}. Additionally, kinematic feedback, such as tremor, rigidity, and bradykinesia metrics, can enhance aDBS models \cite{grado2018bayesian, wilkins2023unraveling}. For example, tremor-dominant PD patients often exhibit excessive gamma band activity, rendering beta suppression ineffective \cite{gilron2021long, swann2018adaptive}. In such cases, multiple control loops targeting different symptoms may be necessary, especially for patients with frequent motor fluctuations or breakthrough tremor during beta-based aDBS \cite{krauss2021technology}. Theta oscillations also correlate with tremor and bradykinesia, further highlighting the need for a multi-frequency approach \cite{yin2021local}. Relying solely on beta band activity poses challenges, particularly in tremor-dominant patients or cases where “good” beta oscillations are present, leading to suboptimal aDBS performance \cite{johnson2016closed, yin2021local}.

This complexity is particularly problematic for machine learning-based aDBS systems, where limited frequency features or poorly designed reward functions can degrade algorithm performance \cite{krylov2020reinforcement, krylov2020reinforcement2}. Therefore, incorporating a broader range of frequencies and flexible spectrograms is essential for robust aDBS training and validation. Some computational models already simulate neurobiologically plausible PSD shapes within the beta band \cite{farokhniaee2021cortical, fleming2020simulation, van2009mean, grado2018bayesian} and include other frequencies such as theta and gamma \cite{wei2022parkinsonian, su2021model}.

In our Kuramoto model, spectrogram flexibility is achieved by adjusting the natural frequencies of oscillators, distributed according to a manually defined probability density $P_{\omega}$, allowing for complex PSD shapes. We also decided to include other frequencies (low frequencies band from 4 to 12 Hz and high-beta band from 21 to 35 Hz) into the model, to ensure richer and more realistic LFP dynamics \cite{yin2021local}. Therefore, we sampled $\omega_n$ from a manually designed distribution (Figure \ref{fig:w0_prc}A). This approach provides flexibility in setting model dynamics and allows adjustment of each frequency band's oscillation strength as needed. This design enables the simulation of broad beta peaks, capturing the nuanced spectral characteristics observed in PD patients. Additionally, our model supports the community-standard feature of beta peak suppression through high-frequency DBS, consistent with existing models of PD (\ref{fig:intro}F) \cite{liu2016closed, grado2018bayesian, cagnan2009frequency}.

\subsubsection{Spatial features}
Including spatial features in the model is crucial because it adds complexity to neuron interactions, both among neurons and between the electrode and neurons \cite{ferrari2015phase}.

Our model incorporates spatial domain as a 3-dimensional grid of interconnected neurons (the same can be applied to 2-dimensional grid), parameterized by $\cos(\alpha_{mn})$, determining the nature (inhibitory or excitatory) of connections and the strength of connection with coupling parameter $K$ (Figure \ref{fig:phase_main}). 

In our model the interaction (i.e. coupling) function is neurobiologically plausible, using a cosine kernel to represent spatial coupling, though other functions can be used \cite{ferrari2015phase, breakspear2010generative}. Our model supports all-to-all connectivity but allows flexible coupling functions for customized simulations. This spatial embedding improves the realism of neural dynamics, facilitating more robust aDBS testing and development.

\textit{Electrode location and Beta oscillation locus.}
Current advancements in electrode technology, which include electrodes with multiple contacts capable of both stimulating and recording, necessitate the integration of spatial dynamics into the model. This technology enables more precise stimulation, such as targeting specific contacts or creating directed stimulation \cite{krauss2021technology}. The multiple contacts (on one electrode) can create different shape of electric field \cite{krauss2021technology}.
It has been shown that the highest beta band power often coincides with locations in the sensorimotor STN \cite{lofredi2019beta}. And it was shown that the relative distance between an electrode and oscillation locus affects the efficiency of DBS to suppress PD symptoms, as well as the direction of stimulation \cite{guo2013subthalamic}.
It also has been reported that the contact pair with maximal STN beta power is very likely to induce the best symptom control and has the widest therapeutic window \cite{yin2021local}.

\begin{figure*}
    \centering
    \includegraphics[width=1.0\linewidth]{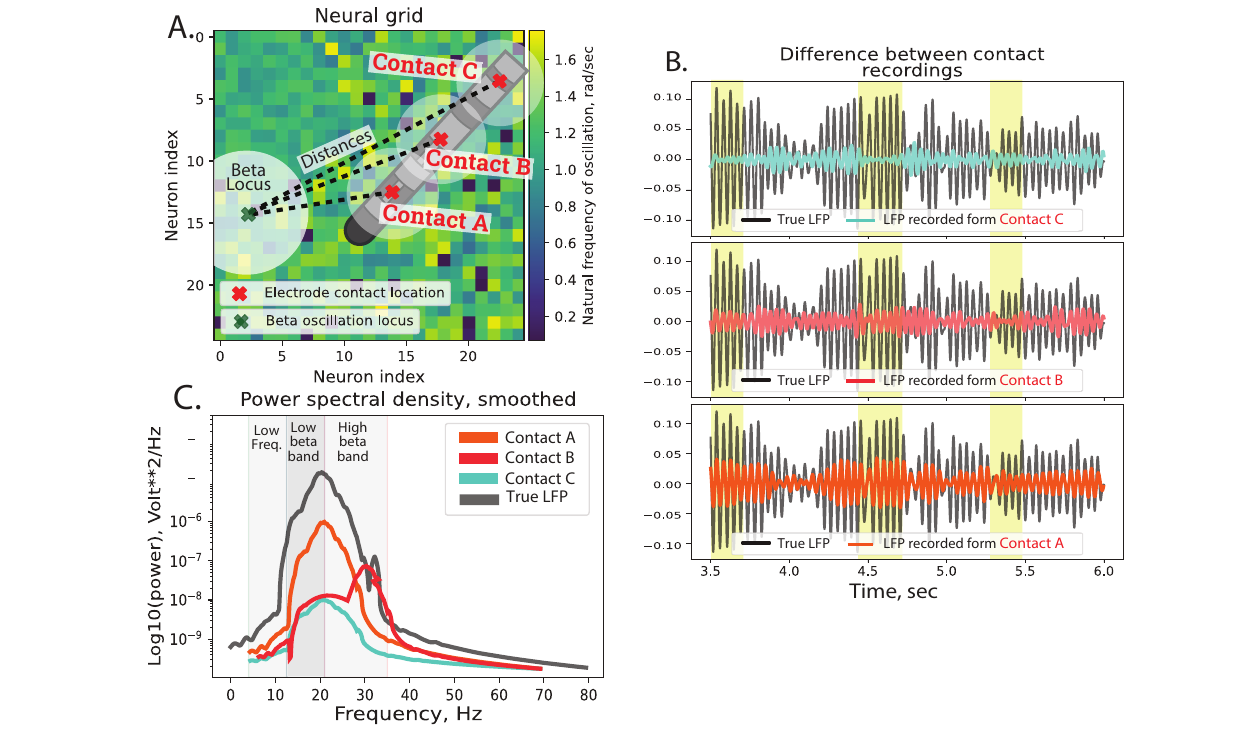}
    \caption{Example of how recorded LFP signal changes depending on the location of electrode recording contact w.r.t. beta oscillations locus. A. Electrode with 3 recording contacts, designated with A, B and C. Each contact have different distance to beta locus. B. Difference between recorded signals and true global LFP of the model. We marked important temporal periods with yellow. In this parts we can clearly see how different distance not only reduce the amplitude of the recorded signal, but also change the dynamics itself. For example, in the middle part of the signal, marked with yellow we can see that electrode contacts B and C recorded dynamics similar to global LFP, but with reduced amplitude, whereas contact A recorded no oscillations at all. C. Difference of PSD in recorded signals for the whole episode.}
    \label{fig:electrode_location_appendix}
\end{figure*}

Proximity of the active electrode contact to the maximum high beta oscillations locus lead to increase in the recorded LFP high beta power in the STN region and the a greater response to the stimulation of STN \cite{chen2022subthalamic}.
While the STN per se was originally assumed to be the principal target of therapeutic stimulation, several groups have shown that direct stimulation of numerous anatomical components of the STN region (e.g. fields of Forel, zona incerta, etc.) can result in similar clinical outcomes. Such conclusions have been based on retrospective studies of the anatomical location of therapeutic electrode contacts.
Besides STN, there is a difference in LFP recordings in different locations of GPi: oscillatory activity in GPi dynamically changes during volitional movement. A spatial topography of oscillatory activity in the pallidum. Still, although the functional meaning of these oscillations has not yet been fully elucidated, the fact they are spatially distributed in different regions of GPi, and are patient-specific \cite{aman2020directional}.
Also, the effect of HF-DBS may, depends on whether excitatory or inhibitory structures are stimulated \cite{popovych2014control}, which again depends on the spatial coupling profile between neurons and the location of the electrode.
%
Because of high-nonlinearity it is not only "the further relative distance, the weaker the amplitude of LFP", but the dynamics of recorded LFP change (Figure \ref{fig:electrode_location_appendix}).
%
Therefore, with an addition of the spatial domain the one more important task appears for aDBS - chose of the optimal stimulation target and the clinically most effective contacts, which is often a contact closer to the place of the highest beta oscillations as has been shown in the study with real patients \cite{aman2020directional}. 
Also there can be electrode shifts in a brain (see \ref{temp_f}) and initial inaccurate implantation of electrode, not as it has been planned \cite{chapelle2021early}, which makes the problem even more puzzled in spatial domain.

\textit{Partially observable/stimulated environment.}
Nowadays, in a number of works, a spatial domain is included into a model. Usually, a stimulation is assumed to affect all neurons equally, which highly simplify the work for aDBS \cite{daneshzand2018robust, su2023closed, krylov2020reinforcement, agarwal2023novel, wang2022adaptive, swann2018adaptive, liu2020neural, sui2022deep, herron2017cortical}. aDBS is considered to record and affect all environment, which is not the case in real settings. In realistic settings, aDBS can only affect and record from local neurons, with stimulation effectiveness decrease as a distance to the electrode increases. It can suppress oscillations locally, but not in far regions of neurons population. The same works for the recording contacts. 
The area of neurons that is activated with the electrode stimulation is called volume of tissue activated (VTA). And for STN it varies between 20\% to 60\%  of STN population neurons \cite{carlson2021computational, rajamani2024deep}.
Therefore, the locality of observation and stimulation is an important feature of any model \cite{fleming2020self}.
In the simplest case we consider, an isotropic medium, stimulation delivered using cylindrical DBS electrodes gives rise to a symmetric, omnidirectional VTA around the electrode \cite{cagnan2019emerging}. We set the electrode coordinates within the grid and incorporating a kernel that adjusts an observation and a stimulation intensity based on the distance from the electrode.
This feature introduce a severe complexity since the optimal locus of stimulation can be far from the electrode, which can force adaptive DBS to create more complicated policy of stimulation.
	
\textit{Electrode directed stimulation and multiple contacts.}
Recently, the segmented electrodes with directed stimulation were developed. The advent of segmented contact offers the possibility for directionally steering current to the specific STN region. \cite{yin2021local, krauss2021technology}. Current electrodes allow three radial directions of stimulation separated by 120 degrees \cite{cagnan2019emerging}. This will allow for more accurate stimulation and can decrease side effects of DBS even more \cite{cagnan2019emerging}.
Furthermore, the newly developed electrodes have multiple contacts, which allow them to shape the stimulation electric field. Which is relevant, since it is still unclear if one “optimal” stimulation target exists or if there may actually be multiple target regions \cite{butson2011probabilistic}.

In our environment, the directional stimulation is implemented as a change in the stimulation kernel shape, by cutting 2\textbackslash3 of its volume. Specifically, this is achieved using trigonometric functions to define the angular boundaries of the activation zone. The kernel is cut by applying a condition on the azimuthal angle $\psi$ in spherical coordinates, restricting it to a segment of $120^{\circ}$ corresponding to the active contact. This is mathematically represented as $\phi - \frac{\Delta\psi}{2} < \psi < \phi + \frac{\Delta\psi}{2}$, where $\Delta\psi = \frac{2\pi}{3}$ to simulate the $120$-degree directional steering. Here, $\phi$ is the angular position of the active contact, which can take values of $0^{\circ}$, $120^{\circ}$, or $240^{\circ}$ depending on the chosen radial direction.
Multiple stimulating and recording contacts can be created by simply setting more than one electrode contacts coordinates in the code of our model. The interactions functions and kernels of stimulation will be automatically adjusted.

\subsubsection{Temporal features} \label{temp_f}

For BCI the topic of neural activity drift is not new and a lot of papers note the importance of this phenomenon. If left unmitigated, these neural recording instabilities can lead the BCI to become uncontrollable, often within hours. Therefore the topic of BCI "stabilization" is highly popular in this field \cite{degenhart2020stabilization, ma2023using, adair2018evolving}.
A lot of PD computational models used for aDBS validation note the importance of temporal features and implement different types of them \cite{dovzhenok2013failure, fleming2020simulation, van2009mean, shukla2014modeling}. It should be noted that these features are of special importance for developing aDBS that can give a stable performance over time.

\textit{Electrode drift, neural drift and noisy observations.}

The importance of drift is outlined in the community \cite{krauss2021technology, schroll2014dysfunctional, rule2019causes, rubin2017computational} and implemented in some PD models used to train aDBS. In RTM models \cite{fang2023robust, zhu2021adaptive, daneshzand2018robust, fleming2020self} and in more simpler models \cite{cumin2007generalising, shukla2014modeling, zhou2017adaptive}. For example, it was shown that various aDBS algorithms perform differently in non-stationary environment and reduce their performance if non-stationary nature of signal does not take into account during developing of aDBS \cite{fang2023robust}.

Short-term and long-term synaptic plasticity have been studied in the BG \cite{rubin2017computational, popovych2014control}. It was shown that in real patients the structural connectivity is changing by itself and because of the action of DBS \cite{van2014neural}. Furthermore, it was shown that in the brain, the coupling strength would be governed by synaptic changes such as the diffusion of the neurotransmitter. These changes would occur over long timescales (namely slow varying changes in frequency) \cite{cumin2007generalising}.

There can be implemented different types of neural plasticity \cite{fung2013neural}. The most popular in spike-timing-dependent plasticity (STDP) \cite{popovych2014control} and structural plasticity \cite{manos2021long}. Structural plasticity allows neurons to establish new or delete preexisting synaptic connections; this occurs via the extension or retraction of axons and dendrites, or by modifying the number of axonal boutons or dendritic spines \cite{manos2021long}. The RTM models \cite{lindahl2016untangling, wang2018suppressing, farokhniaee2021cortical, fang2023robust} as well and Kuramoto models \cite{maistrenko2007multistability, cumin2007generalising} are developed with these plasticity types and used for aDBS development. 

This is also addressed as a micromotion: encapsulation of an implanted array of electrodes through formation of a glial sheath leads to recording inconsistencies, as the sheath can move an entire array. This movement is often labeled ‘micromotion,’ though this term is also applied to displacement of implants due to an accumulation of e.g., small head movements. Micromotion is a likely cause for the inability to record a neuron shape longer than a few weeks \cite{groothuis2014physiological}.

In the chronic state, there happen foreign body response - the impedance of implanted DBS electrodes changes over time as a result of the inflammatory response and gliosis around the electrode. This response consists of several phases that ultimately result in neuronal loss and the formation of a dense glial sheath that encapsulates the implant. During which glial encapsulation of the electrode, protein adsorption on electrode sites and the characteristics of the ionic environment at the electrode–electrolyte interface dictate the electrical characteristics of the electrode–tissue interface \cite{krauss2021technology, little2012brain, campbell2018chronically}. The low-conductivity giant cells in chronic stage restricted current spread they produced a magnitude of the potential at a given distance consistently less that in acute stage (after implantation) when the peri-electrode space was filled with extracellular fluid \cite{rosa2012neurophysiology}. For example, a study by Rousche and Norman reported a loss of viable electrodes as well as a large implant displacement due to a strong FBR around several electrodes \cite{groothuis2014physiological}. A few computational studies address this issue and check how aDBS will perform with varying electrode impedance \cite{fleming2020self}.
Drift for the recording can also happen because of unit drop-out, where a neuron dies or moves away from the electrode tip \cite{degenhart2020stabilization, groothuis2014physiological}.
In addition to a loss of signal, the relative movement of an implant in cortical tissue can contribute to local tissue damage and accelerate neuron loss near the electrode \cite{groothuis2014physiological}.
The source of noise in neurons - random opening of ion channels, the quintal releases of neural transmitters, the coupling of background neural activity, etc. \cite{lu2017desynchronizing}.

\textit{Pathological beta bursting and movement modulation.}
Beta activity is not continuously elevated, but fluctuates between long, greater than 400 ms, and short from 100 to 400 ms duration bursts of beta activity, with only long burst durations being positively correlated with motor impairment in PD. The proportion of long duration beta bursts was positively related to clinical impairment, while the proportion of short duration beta bursts was negatively related to clinical impairment in the unstimulated state. These results suggest that a long duration of uncontrolled synchronization has an important negative impact on motor performance, whereas we could speculate that short duration beta synchronization may impact positively \cite{tinkhauser2017modulatory}.
However not everything is so clear with beta modulation during movements \cite{vinding2020reduction}. It does not fall during long and cyclic movements, but fall because of brief movement. Also movement beta-bursting modulations is different for self-initiated and sign-initiated movements \cite{lofredi2019beta}.
The addition of beta-bursting introduce even more non-stationarity into the model, since beta bursting is assumed to be a transient non-stationary event. And it is also acknowledged in the community as an important feature of PD that can be used as a controlling signal \cite{tinkhauser2017modulatory, anderson2020novel, merk2022machine, vinding2020reduction, popovych2019adaptive, fleming2020simulation}. 
During these activities the modulation of amplitude and frequency of LFP happens, which suggests that amplitude and frequency modulation might both be important for coding motor state \cite{little2012brain}. A useful model must reliably indicate state information during more complex situations including voluntary and cued movements \cite{binns2024shared}. And this, in turn, directly affect aDBS policy and performance \cite{krauss2021technology}, since a long-term adaptive DBS need to self-optimize to allow for fluctuations in state over time \cite{tinkhauser2017modulatory}.

Several models used to train aDBS have implemented features of excessive beta bursting \cite{zhu2021adaptive, fleming2020simulation, ranieri2021data, lu2019application} and movement modulation \cite{binns2024shared}, as well as more long-term variations such as circadian rhythm and psychiatric state \cite{fang2023robust}. 


\subsection{Comparison of Kuramoto and RTM models}

The most widely used model in the field of aDBS frameworks is the RTM model and its variants \cite{terman2002activity}, but we opted against it. 
While RTM is considered somewhat biologically realistic, its units remain oscillators with slightly more complex waveforms than the sine functions in Kuramoto. 
Moreover, RTM models have neurobiological realism issues \cite{pascual2006computational}. 
The Kuramoto model being a flexible enough to approximate Hodgkin–Huxley (HH) model dynamics without losing overall system behavior \cite{hansel1993phase}, retains all key features used in other models while maintaining computational efficiency.

\begin{figure*}
    \centering
    \includegraphics[width=1.0\linewidth]{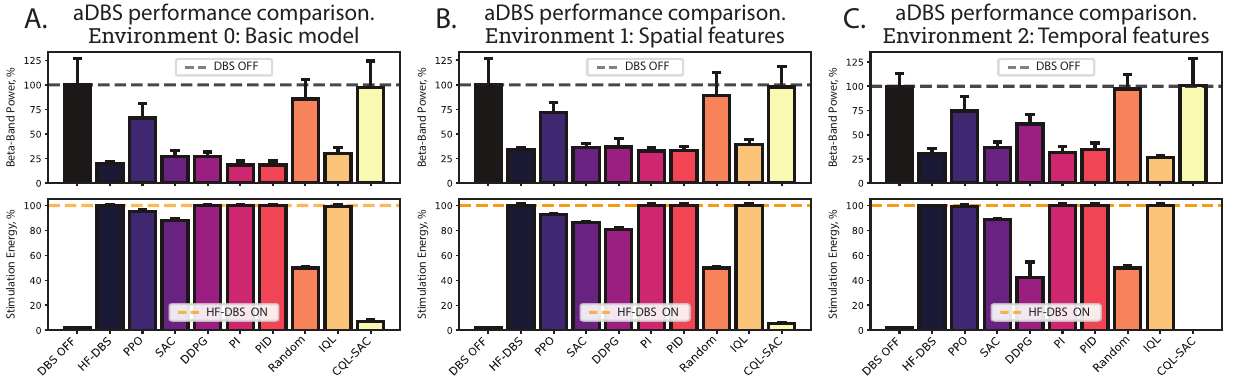}
    \caption{Performance of Different Control Strategies in three proposed Environments. The top panels show the efficiency of beta-band power suppression, expressed as a percentage relative to the average beta-band power in the uncontrolled environment (DBS OFF case, represented by the grey dashed line). The bottom panels display the energy consumption efficiency of different control strategies, measured as a percentage relative to the case of maximum energy consumption (as in the HF-DBS ON condition, indicated by the orange dashed line). Columns A, B, and C correspond to environment levels 0, 1, and 2, each incorporating different feature groups. The level 0 environment includes only bandwidth features, while level 1 adds spatial features. The level 2 environment incorporates features from all three groups. Error bars represent one standard deviation across 10 environment runs for env0 and env1 and 25 for env2 conducted during the validation procedure.}
    \label{fig:barplots}
\end{figure*}

\begin{figure*}
    \centering
    \includegraphics[width=1.0\linewidth]{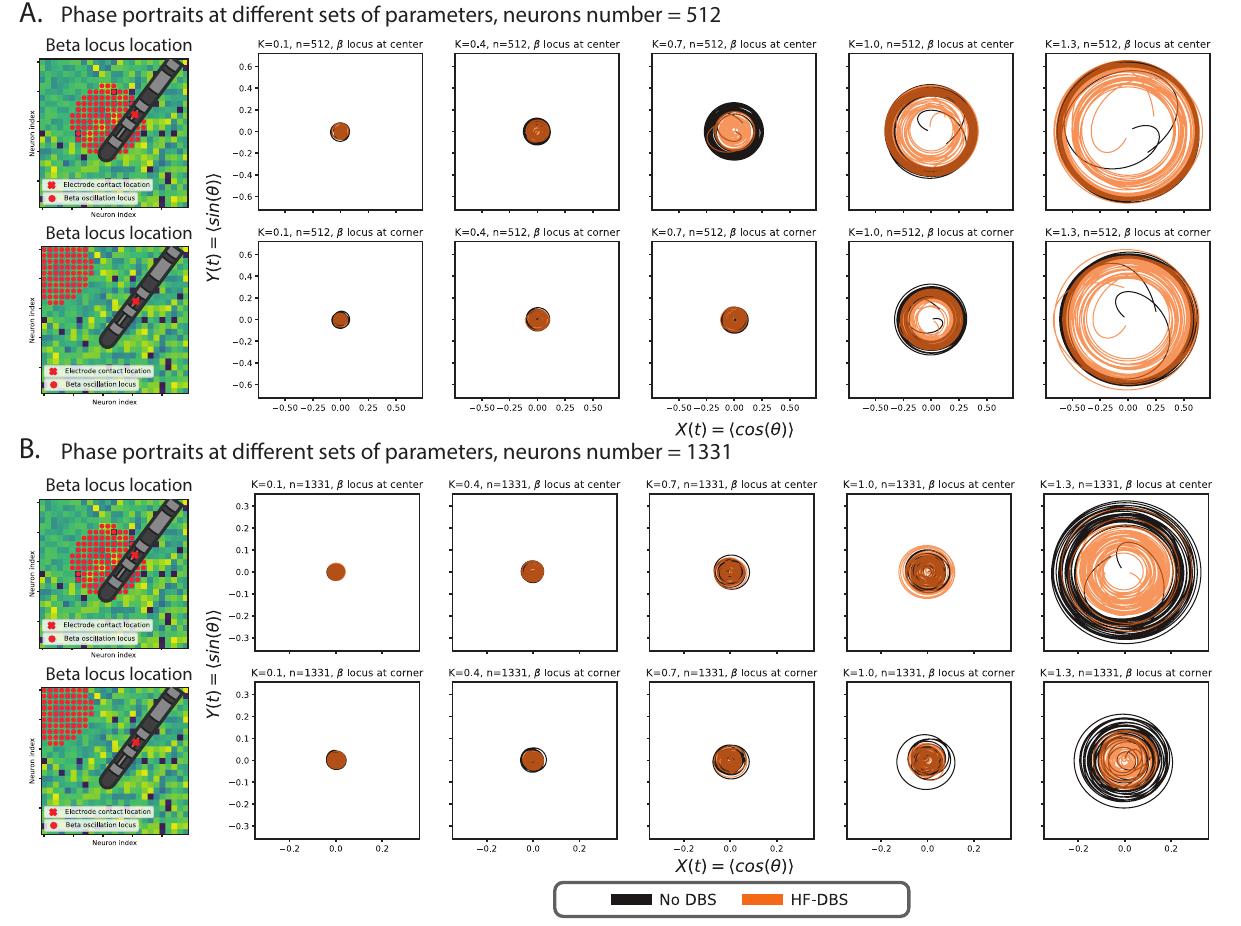}
    \caption{Phase portraits of LFP dynamics at different beta locus positions relative to the electrode for different coupling parameters and the number of neurons in the grid. A) The case for 512
    neurons in the grid. Left, sketches demonstrate a position of beta locus in the neural grid. Right, the corresponding phase portraits of LFPs at different coupling parameters. B) The same is also shown for the case of 1331
    neurons in the grid.}
    \label{fig:phase_plot}
\end{figure*}

\subsection{Model implementation details} \label{app: model_implement_details}

\subsubsection{Phase response curve} 
\begin{figure*}
    \centering
    \includegraphics[width=1.0\linewidth]{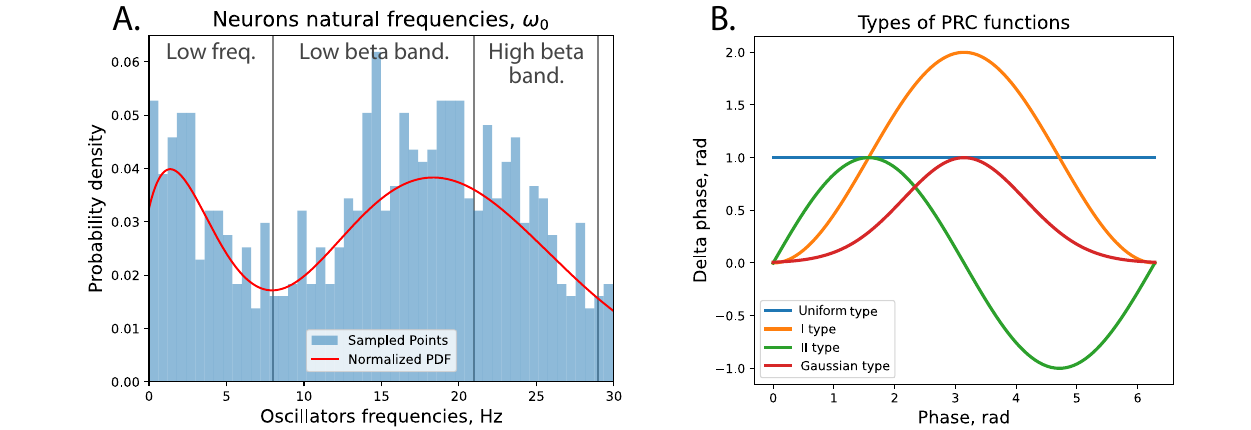}
    \caption{A) The distribution of natural frequencies of neurons. B) Types of phase response curve functions implemented in the proposed environment.}
    \label{fig:w0_prc}
\end{figure*}

PRC function describes how phase of the oscillator will change w.r.t. current phase of the oscillator itself.  The visualization of the functions is shown in Figure \ref{fig:w0_prc}. This function is discovered in real neurons and is required for imitating DBS effect on neurons \cite{smeal2010phase, danzl2007event, azodi2015phase}. In our model, 3 main types of PRC are available: type I $PRC_{I} = 1 - cos(\theta)$, type II $PRC_{II} = sin(\theta)$; 
and Gaussian PRC $PRC_{G} = e^{-(\frac{\theta - \mu}{2\sigma}^2)}$ (Figure \ref{fig:w0_prc}B).
Each function introduce additional non-linear response of the environment to the stimulation and make controlling task more complicated. In our experiments, to keep task more simple, we mostly used 'dummy' PRC curve, that maps any to effect of phase to $1$.

\subsubsection{Natural frequencies}
By controlling the shape of the distribution one can control which oscillatory mode will prevail in model's dynamics. In Figure \ref{fig:w0_prc}A we samples $\omega_0$ in such a way, that the majority of neurons have their natural frequencies around 17 Hz, which gives a pronounced peak. Also, we included other frequencies to keep the dynamics of the models variable - so-called, low frequencies between 4 and 8 Hz.
In our setting of the problem the low beta band is assumed to affect pathological state of PD patients. However, for example, one can shift the pathological oscillatory peak  in the model dynamics to high beta band (21 - 30 Hz) by moving the peak in the distribution to according value. After sampling from the distribution one should convert sampled values to rad/sec and used in the model.  

The values in beta locus were sampled from uniform distribution with minimum of 16 Hz and maximum of 18 Hz. The size of beta locus was set about 25 \% of overall neurons (Figure \ref{fig:w0_prc}A).

\begin{figure*}
    \centering
    \includegraphics[width=1.0\linewidth]{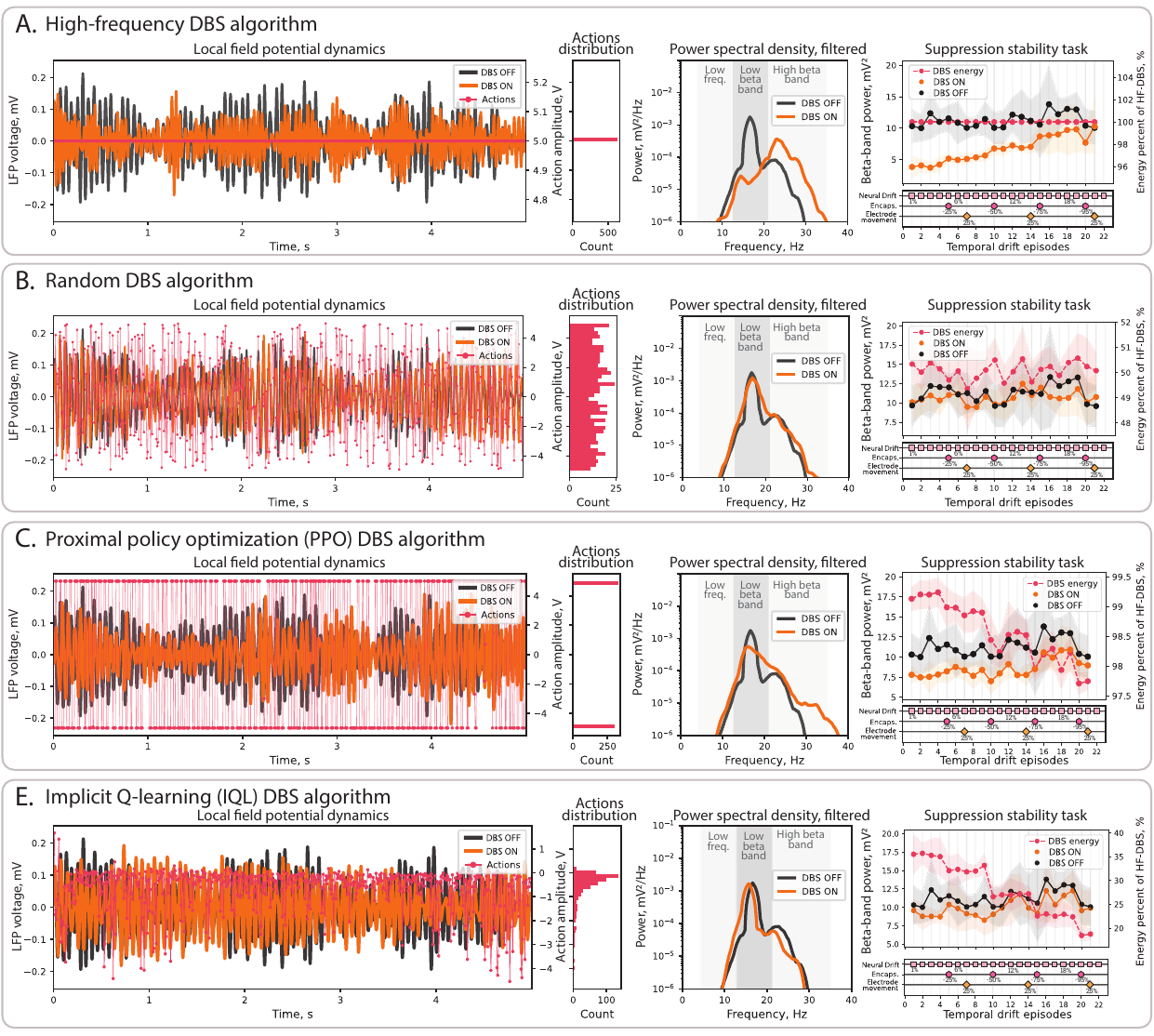}
    \caption{Detailed Analysis of Model Performance for Different DBS Algorithms. The figure presents performance results for A) HF-DBS, B) Random DBS, C) PPO-DBS, and D) DDPG-DBS algorithms. The leftmost panel illustrates beta oscillation suppression under each control strategy, with black and orange curves representing beta oscillations in the absence of DBS and under DBS control, respectively. Red markers indicate DBS pulse amplitudes at specific time points, with identical time lags between pulses. The second panel shows the distribution of DBS amplitudes over the considered time frame. The third panel displays the power spectral densities of LFP signals from the first panel, comparing No DBS and DBS conditions, where the suppression or shift of low beta-band frequencies represents the desired outcome. The rightmost panel depicts the dependence of each control strategy on the progressive injection of noise into the environment across successive episodes. Red polygons at the bottom indicate the time steps at which different noise sources were introduced: plasticity noise (red squares), reflecting changes in the natural frequencies of neurons; encapsulation noise, representing alterations in conductance between the DBS electrode and surrounding neurons (red hexagons); and electrode movement noise, simulating random shifts in the electrode's spatial position within the neural grid (red diamonds).}
    \label{fig:sup_agents_performance}
\end{figure*}


\subsection{Agent implementation details} \label{app:agents}
\subsubsection{Reward functions} \label{app: rewards}

For training RL agents we used other two popular reward functions. The second reward was previously introduced in \cite{krylov2020reinforcement} and amplifies the form of efficiency term, requiring an LFP to be minimally different from its average over the $k$ previous steps:
\begin{equation}
    r_2 = -(\lambda_{r_2}(LFP(t) - \langle LFP(t)\rangle_k)^2  + \kappa_{r_2}E_A).
\label{eq:R2}
\end{equation}

The third reward, recently introduced in \cite{gao2023offline}, in contrast, relax the requirements to efficiency by introducing a threshold value below which only energy costs are accounting:
\begin{equation}
    r_3 = -(\lambda_{r_3}\mathbb{I}(\beta_{BP}>\beta_{thr}) + \kappa_{r_3}E_A).
\label{eq:R3}
\end{equation}
The rationale is that complete suppression of low beta-band is unnecessary; a therapeutic effect is achieved by maintaining beta power below approximately 50\% of its pathological level (see Section \ref{sec:features}, Appendix \ref{app: features_justif}).

\subsubsection{Details of online agents training}
RL algorithms were implemented with pytorch \cite{paszke2017automatic} and stable baselines 3 libraries \cite{stable-baselines3}.

For three reward functions (Equations \ref{eq:R1}, \ref{eq:R2}, \ref{eq:R3}) we used the weights: $\lambda_{r_1} = 1\cdot10^{4} $ $ \kappa_{r_1} = 0.01 $, $\lambda_{r_2} = 1000 $ $ \kappa_{r_2} = 0.01 $, $\lambda_{r_3} = 1\cdot10^{4} $ $ \kappa_{r_3} = 0.1 $. 

\textbf{Details on how online agents were trained on 3 different environment levels.} For each of 3 levels of environment, we trained a separate agent. The length of the episode was set to 50 seconds (5555 steps).

For Env0, training was conducted with fixed spatial parameters: both the electrode and beta locus were positioned at the center of the neural grid. All spatial changes and temporal drifts were disabled. The stimulation kernel of aDBS was large enough to influence every neuron within the grid, and the model provided full observability with unitary conductances. To simplify the task, the Phase Response Curve (PRC) was set to uniform.

In Env1, every 5th episode introduced random shifts in the positions of the electrode's recording contact, stimulating contact, and beta locus. These shifts were approximately 20\% of the grid’s total length. Partial stimulation and observability were enabled to increase task complexity.

Env2 introduced three types of temporal drift events. Electrode drift occurred every 7 $\pm$ 2 episodes, with the electrode moving in a random direction but restricted from crossing grid boundaries. Electrode encapsulation drift occurred every 5 $\pm$ 1 episodes, reducing the conductance of both recording and stimulating contacts by 2\% per event. Conductance values were reset after 10 events. Neural drift was implemented each episode, with natural frequencies $\omega_n$ drifting by 2\% from their initial values and resetting after 7 episodes.

Parameter values for all three models used in training are listed in Appendix Table \ref{tab:configs}. Temporal drifts were integrated within the reset phase of the code to optimize computational efficiency while maintaining continuous differential equation calculations and consistent parameter settings throughout each episode.

\subsubsection{Details of offline agents training}

An offline dataset was collected based on the action history of DDPG, SAC, and PPO agents trained in an online RL setting (as described in the main text). Data was gathered over 50 episodes, with each episode consisting of 5,000 steps. At the start of each episode, one of the three policies was randomly selected to generate trajectories. For each of the three environment levels, an offline dataset was collected independently. The offline agents, IQL and CQL-SAC, were trained following the methodology in \cite{pan2024coprocessor}, using the same hyperparameters specified in that study. The validation procedure for offline agents was identical to that of the online agents, as described in the main text.

\subsubsection{Details of PI/PID controller implementation}

The parameters of PI/PID controller were tuned with Optuna hyperparameter optimization framework \cite{akiba2019optuna}.

For PI and PID controllers, we decided to define a bicriterial control task by a minimization of negative reward function, i.e. an error term corresponds to negative reward $e = -r_i,\ i=\{1,2,3\}$. Then, the PID control problem written as follows:
\begin{equation}
    e_{t}= K_pe_{t-1} + K_I\int_0^{t-1}e(\tau)d\tau + K_d\frac{de_{t-1}}{dt},
\label{eq:pid}
\end{equation}
where $K_p$, $K_I$, $K_d$ are proportional, integral, derivative gains respectively. The PI controller, therefore, included only proportional and integral terms, and the PID controller included all three. The parameters of PI/PID controller were tuned with Optuna hyperparameter optimization framework \cite{akiba2019optuna} by running PI/PID controllers on environments of different levels during 1000 steps. A total number of trials per each reward function was 70. 

\subsection{Suppression stability task: details} \label{app:stability}
For the suppression stability test, electrode conductance was reduced by 25\% every 5th episode, reaching 95\% reduction by the 22nd episode, affecting only 3\% of the neuronal population. Electrode movement occurred every 7th episode, randomly shifting the electrode by about 25\% of the grid size. Neural drift occurred continuously, shifting natural frequencies by 1\% each episode while preserving the initial distribution. We ran 10 environments for 22 episodes each in parallel to gather statistics.
\begin{table*}[ht]
\centering
\caption{Specific configuration parameters of environments for training online RL agents. \\\textbf{*} - denotes the initial values: they will change as the environment evolves, because of the temporal drift.}
\label{tab:configs}
\begin{tabular}{p{6.3cm}p{3.3cm}p{3.3cm}p{3.3cm}}
\toprule
\textbf{Parameters} &      \textbf{Env0} &       \textbf{Env1} &       \textbf{Env2} \\
\midrule
Number of neurons &       512 &        512 &        512 \\
Grid size & [8, 8, 8] &  [8, 8, 8] &  [8, 8, 8] \\
Random seed &        10 &         10 &         10 \\
Natural frequencies distribution mean, deg. &        17 &         17 &         17 \\
Natural frequencies distribution sd, deg. &         1 &          1 &          1 \\
Electrode kernel size scaling &       0.1 &        0.1 &        0.1 \\
Beta locus size, \% &      0.55 &       0.55 &      0.55* \\
Transient state duration, sec &       2.0 &        2.0 &        2.0 \\
Electrode stimulation duration, sec &    0.0015 &     0.0015 &     0.0015 \\
Electrode pause duration, sec &    0.0075 &     0.0075 &     0.0075 \\
DBS amplitude bounds &   [-5, 5] &    [-5, 5] &    [-5, 5] \\
Electrode PRC scaling &       1.0 &        1.0 &        1.0 \\
Electrode PRC type &     dummy &      dummy &      dummy \\
Verbose dt, sec &    0.0005 &     0.0005 &     0.0005 \\
Total episode duration, sec &      50.0 &       50.0 &       50.0 \\
Observation window duration, sec &      1.17 &       1.17 &       1.17 \\
Initialization distribution type &    normal &     normal &     normal \\
Initialization mean, Hz &     $\pi$ &     $\pi$ &      $\pi$ \\
Initialization sd, Hz &       0.6 &        0.6 &        0.6 \\
Spatial features &    False  &      True  &      True  \\
Distance scaling  &       0.1 &        0.1 &        0.1 \\
Naive DBS stimulation type &    False  &     False  &     False  \\
Directed stimulation TURN ON &         - &          - &          - \\
Neural spatial interaction kernel type &       cos &        cos &        cos \\
Stimulation contact coordinate & [4, 3, 4] & [4, 3, 4]* & [4, 3, 4]* \\
Recording contact coordinate & [1, 1, 1] & [1, 1, 1]* & [1, 1, 1]* \\
Beta locus center coordinate & [4, 4, 4] & [4, 4, 4]* & [4, 4, 4]* \\
Recording contact kernel type &     naive &   gaussian &   gaussian \\
Temporal drift TURN ON &    False  &     False  &      True  \\
Random period of update of temporal features &         - &          - &      True  \\
Electrode drift frequency &         - &          - &          7 \\
Plasticity drift frequency &         - &          - &          1 \\
Plasticity percent, \% &         - &          - &          2 \\
Reset plasticity drift process, episode &         - &          - &          7 \\
Electrode encapsulation drift frequency &         - &          - &          5 \\
Encapsulation percent, \% &         - &          - &          2 \\
Movement modulation drift frequency &         - &          - &          - \\
\bottomrule
\end{tabular}
\end{table*}
\begin{table*}
\centering
\caption{Evaluation results of baseline aDBS algorithms across three environment variations using three basic reward functions. Details on the evaluation environment parameters are provided in Appendix \label{app: rewards} and in Section \ref{section:metrics}. Two key metrics were used for assessment: beta-band power (restricted to the low-beta range, 12.5–21 Hz over the entire episode) and total stimulation energy. Evaluations were conducted six times for Env0 and Env1 and 25 times for Env2. The best-performing model is highlighted in bold, selected as a balanced compromise between minimizing beta-band power and reducing energy consumption. High-frequency DBS (HF-DBS) represents the maximum possible energy usage, while the DBS OFF condition indicates the highest beta-band power, reflecting neural activity without stimulation.}
\label{tab:vertical_results}
\begin{tabular}{lllllll}
\toprule
 &     \multicolumn{6}{c|}{Env0 (Basic model)} \\
\midrule
 &     \multicolumn{2}{c|}{Reward \#1}            
 &     \multicolumn{2}{c|}{Reward \#2}
 &     \multicolumn{2}{c|}{Reward \#3}
\\
\midrule
&  \multicolumn{1}{p{2.2cm}|}{Beta-band power, $\%$} 
&  \multicolumn{1}{p{2.2cm}|}{Stimulation energy, mJ} 
&  \multicolumn{1}{p{2.2cm}|}{Beta-band power, $mV^2$} 
&  \multicolumn{1}{p{2.2cm}|}{Stimulation energy, mJ} 
&  \multicolumn{1}{p{2.2cm}|}{Beta-band power, $mV^2$} 
&  \multicolumn{1}{p{2.2cm}|}{Stimulation energy, mJ} \\
\midrule
DBS OFF &              11.83 $\pm$3.2 &                     0 &           11.83 $\pm$3.2 &                     0 &           11.83 $\pm$3.2 &                     0 \\
HF-DBS &               2.34 $\pm$0.2 &              5555 $\pm$0 &           2.34 $\pm$0.23 &              5555 $\pm$0 &            2.34 $\pm$0.2 &              5555 $\pm$0 \\
PI &                2.2 $\pm$0.2 &              5555 $\pm$0 &             2.2 $\pm$0.2 &              5555 $\pm$0 &            2.22 $\pm$0.2 &              5555 $\pm$0 \\
PID &                2.2 $\pm$0.2 &              5555 $\pm$0 &            2.23 $\pm$0.2 &              5555 $\pm$0 &            2.23 $\pm$0.2 &              5555 $\pm$0 \\
Random policy &              10.15 $\pm$2.3 &             2775 $\pm$19 &           10.15 $\pm$2.3 &             2775 $\pm$19 &           10.15 $\pm$2.3 &             2775 $\pm$19 \\
PPO &               7.85 $\pm$1.8 &             5274 $\pm$23 &             8.0 $\pm$3.8 &             2153 $\pm$51 &            10.1 $\pm$2.4 &              270 $\pm$25 \\
SAC &               \textbf{3.24 $\pm$0.7} &             \textbf{4877 $\pm$27} &           10.31 $\pm$5.5 &              461 $\pm$87 &            10.1 $\pm$3.7 &               9 $\pm$0.0 \\
DDPG &               3.18 $\pm$0.6 &            5549 $\pm$1.6 &            10.1 $\pm$5.6 &             485 $\pm$132 &           10.22 $\pm$3.6 &                 1 $\pm$0 \\

IQL &                3.6 $\pm$0.7 &             5512 $\pm$26 &              \textbf{8.97 $\pm$5} &            \textbf{1880 $\pm$320} &            11.9 $\pm$2.8 &              149 $\pm$48 \\
CQL-SAC  &               11.5 $\pm$3.2 &              390 $\pm$10 &            12.0 $\pm$2.6 &              423 $\pm$38 &            11.5 $\pm$4.8 &              398 $\pm$45 \\
\midrule
 &     \multicolumn{6}{c|}{Env1 (Spatial features)} \\
\midrule
DBS OFF &                9.1 $\pm$2.5 &                     0 &             9.1 $\pm$2.5 &                     0 &             9.1 $\pm$2.5 &                     0 \\
HF-DBS &               3.09 $\pm$0.2 &              5555 $\pm$0 &            3.09 $\pm$0.2 &              5555 $\pm$0 &            3.09 $\pm$0.2 &              5555 $\pm$0 \\
PI &               3.06 $\pm$0.2 &              5555 $\pm$0 &            2.97 $\pm$0.1 &              5555 $\pm$0 &            3.05 $\pm$0.2 &              5555 $\pm$0 \\
PID &                3.0 $\pm$0.2 &              5555 $\pm$0 &             3.0 $\pm$0.2 &              5555 $\pm$0 &             3.0 $\pm$0.2 &              5555 $\pm$0 \\
Random policy &               8.09 $\pm$2.2 &             2770 $\pm$34 &            8.09 $\pm$2.2 &             2770 $\pm$34 &            8.09 $\pm$2.2 &             2770 $\pm$34 \\
PPO &               6.53 $\pm$0.9 &             5167 $\pm$25 &            9.29 $\pm$4.1 &              698 $\pm$93 &            9.86 $\pm$3.8 &                58 $\pm$3 \\
SAC &              3.31 $\pm$0.18 &            4807 $\pm$31  &            9.56 $\pm$3.7 &               122 $\pm$5 &            9.88 $\pm$3.7 &                12 $\pm$0 \\
DDPG &               \textbf{3.36 $\pm$0.4} &             \textbf{4474 $\pm$54} &             9.5 $\pm$3.4 &                51 $\pm$5 &            9.77 $\pm$3.8 &               0.5 $\pm$0 \\
IQL &               3.55 $\pm$0.5 &            5548 $\pm$8.6 &             \textbf{7.5 $\pm$1.3} &            \textbf{1937 $\pm$136} &             \textbf{8.8 $\pm$2.0} &             \textbf{219 $\pm$7.9} \\
CQL-SAC &                8.9 $\pm$1.9 &               292 $\pm$6 &             9.1 $\pm$2.1 &               253 $\pm$9 &             9.2 $\pm$3.7 &              229 $\pm$16 \\
\midrule
 &     \multicolumn{6}{c|}{Env2 (Temporal features)} \\
\midrule
DBS OFF &               11.3 $\pm$1.5 &                     0 &            11.3 $\pm$1.5 &                     0 &            11.3 $\pm$1.5 &                     0 \\
HF-DBS &                3.4 $\pm$0.7 &              5555 $\pm$0 &             3.4 $\pm$0.7 &              5555 $\pm$0 &             3.4 $\pm$0.7 &              5555 $\pm$0 \\
PI &                3.6 $\pm$0.7 &              5555 $\pm$0 &             3.7 $\pm$0.5 &              5555 $\pm$0 &             4.3 $\pm$1.8 &              5555 $\pm$0 \\
PID &               4.91 $\pm$1.6 &              5555 $\pm$0 &             3.9 $\pm$0.8 &              5555 $\pm$0 &             4.0 $\pm$1.2 &              5555 $\pm$0 \\
Random policy &              10.99 $\pm$1.7 &             2773 $\pm$12 &           10.99 $\pm$1.7 &             2773 $\pm$12 &           10.99 $\pm$1.7 &             2773 $\pm$12 \\
PPO &               9.27 $\pm$1.8 &              5511 $\pm$7 &            8.42 $\pm$1.7 &              5512 $\pm$5 &           11.43 $\pm$2.3 &              742 $\pm$22 \\
SAC &               \textbf{4.17 $\pm$0.6} &             \textbf{4928 $\pm$37} &           11.81 $\pm$2.4 &              222 $\pm$60 &           11.47 $\pm$1.5 &            3033 $\pm$442 \\
DDPG &              10.62 $\pm$1.7 &            4978 $\pm$148 &           13.28 $\pm$3.0 &                42 $\pm$5 &            \textbf{6.91 $\pm$1.1} &            \textbf{2352 $\pm$693} \\
IQL &               3.01 $\pm$0.2 &              5553 $\pm$1 &           \textbf{10.07 $\pm$1.6} &            \textbf{1971 $\pm$153} &            12.1 $\pm$2.2 &              181 $\pm$12 \\
CQL-SAC &               12.2 $\pm$2.9 &              265 $\pm$19 &            11.8 $\pm$2.2 &              235 $\pm$18 &            11.4 $\pm$3.2 &              358 $\pm$49 \\
\bottomrule
\end{tabular}
\end{table*}



\end{document}